\documentclass{emulateapj}
\usepackage{graphicx}
\usepackage{natbib}
\shorttitle{PS1 Rapidly-Evolving Transients}
\shortauthors{M. R. Drout et al.}

\begin{document}

\title{Rapidly-Evolving and Luminous Transients from Pan-STARRS1}

\author{M. R. Drout\altaffilmark{1}, R. Chornock\altaffilmark{1}, A. M. Soderberg\altaffilmark{1}, N. E. Sanders\altaffilmark{1}, R. McKinnon\altaffilmark{1,2}, A. Rest\altaffilmark{3}, R. J. Foley\altaffilmark{4,5}, D. Milisavljevic\altaffilmark{1}, R. Margutti\altaffilmark{1}, E. Berger\altaffilmark{1}, M. Calkins\altaffilmark{1}, W. Fong\altaffilmark{1}, S. Gezari\altaffilmark{6}, M. E. Huber\altaffilmark{7}, E. Kankare\altaffilmark{8}, R. P. Kirshner\altaffilmark{1}, C. Leibler\altaffilmark{9}, R. Lunnan\altaffilmark{1}, S. Mattila\altaffilmark{8}, G. H. Marion\altaffilmark{10},  G. Narayan\altaffilmark{11}, A. G. Riess\altaffilmark{2,12}, K. C. Roth\altaffilmark{13}, D. Scolnic\altaffilmark{12}, S. J. Smartt\altaffilmark{14}, J. L. Tonry\altaffilmark{7}, W. S. Burgett\altaffilmark{7}, K. C. Chambers\altaffilmark{7}, K. W. Hodapp\altaffilmark{7}, R. Jedicke\altaffilmark{7}, N. Kaiser\altaffilmark{7}, E. A. Magnier\altaffilmark{7}, N. Metcalfe\altaffilmark{15}, J. S. Morgan\altaffilmark{7}, P. A. Price\altaffilmark{16}, C. Waters\altaffilmark{7}}

\altaffiltext{1}{Harvard-Smithsonian Center for Astrophysics, 60 Garden Street, Cambridge, MA 02138}
\altaffiltext{2}{Department of Physics, Yale University, New Haven, CT 06520-8121, USA}
\altaffiltext{3}{3 Space Telescope Science Institute, 3700 San Martin Drive, Baltimore, MD 21218, USA}
\altaffiltext{4}{Astronomy Department, University of Illinois at Urbana-Champaign, 1002 West Green Street, Urbana, IL 61801, USA}
\altaffiltext{5}{Department of Physics, University of Illinois at Urbana-Champaign, 1110 West Green Street, Urbana, IL 61801, USA}
\altaffiltext{6}{Department of Astronomy, University of Maryland, College Park, MD 20742-2421, USA}
\altaffiltext{7}{Institute for Astronomy, University of Hawaii at Manoa, Honolulu, HI 96822, USA}
\altaffiltext{8}{Finnish Centre for Astronomy with ESO (FINCA), University of Turku, V\"ais\"al\"antie 20, FI-21500 Piikki\"o, Finland}
\altaffiltext{9}{Department of Astronomy \& Astrophysics, University of California, Santa Cruz, CA 95060, USA}
\altaffiltext{10}{Department of Astronomy, University of Texas at Austin, Austin, TX 78712, USA}
\altaffiltext{11}{National Optical Astronomy Observatory, 950 North Cherry Ave., Tucson, AZ 85719 USA}
\altaffiltext{12}{Department of Physics and Astronomy, Johns Hopkins University, 3400 North Charles Street, Baltimore, MD 21218, USA}
\altaffiltext{13}{Gemini Observatory, 670 North Aohoku Place, Hilo, HI 96720, USA}
\altaffiltext{14}{Astrophysics Research Centre, School of Mathematics and Physics, QueenÕs University Belfast, Belfast, BT7 1NN, UK}
\altaffiltext{15}{Department of Physics, Durham University, South Road, Durham DH1 3LE, UK}
\altaffiltext{16}{Department of Astrophysical Sciences, Princeton University, Princeton, NJ 08544, USA}

\begin{abstract}
In the past decade, several rapidly-evolving transients have been discovered whose timescales and luminosities are not easily explained by traditional supernovae (SN) models.  The sample size of these objects has remained small due, at least in part, to the challenges of detecting short timescale transients with traditional survey cadences.  Here we present the results from a search within the Pan-STARRS1 Medium Deep Survey (PS1-MDS) for rapidly-evolving and luminous transients.  We identify 10 new transients with a time above half-maximum (t$_{1/2}$) of less than 12 days and $-$16.5 $>$ M $>$ $-$20 mag. This increases the number of known events in this region of SN phase space by roughly a factor of three.  The median redshift of the PS1-MDS sample is z$=$0.275 and they all exploded in star forming galaxies.  In general, the transients possess faster rise than decline timescale and blue colors at maximum light (g$_{\rm{P1}}$ $-$ r$_{\rm{P1}}$ $\lesssim$ $-$0.2).  Best fit blackbodies reveal photospheric temperatures/radii that expand/cool with time and explosion spectra taken near maximum light are dominated by a blue continuum, consistent with a hot, optically thick, ejecta.  We find it difficult to reconcile the short timescale, high peak luminosity (L $>$ 10$^{43}$ erg s$^{-1}$), and lack of UV line blanketing observed in many of these transients with an explosion powered mainly by the radioactive decay of $^{56}$Ni.  Rather, we find that many are consistent with either (1) cooling envelope emission from the explosion of a star with a low-mass extended envelope which ejected very little ($<$0.03 M$_\odot$) radioactive material, or (2) a shock breakout within a dense, optically thick, wind surrounding the progenitor star.  After calculating the detection efficiency for objects with rapid timescales in the PS1-MDS we find a volumetric rate of 4800 $-$ 8000 events yr$^{-1}$ Gpc$^{-3}$ (4 $-$ 7\% of the core-collapse SN rate at z$=$0.2). 
\end{abstract}

\email{mdrout@cfa.harvard.edu}

\section{Introduction}\label{Sec:Intro}

Modern supernova (SN) searches are expanding our knowledge of the phase space occupied by cosmic explosions (e.g., peak luminosity, timescale).  High cadence surveys have led to the discovery of a variety of rapidly-evolving (t$_{1/2} \lesssim$ 12 days) and luminous ($-$15 $>$ M $>$ $-$20) transients.  These include the type Ic SNe 2005ek \citep{Drout2013} and 2010X \citep{Kasliwal2010}, the type Ib SN 2002bj \citep{Poznanski2010}, the type IIn SN PTF09uj \citep{Ofek2010}, and the type Ibn SN 1999cq \citep{Matheson2000}, with the latter two showing signs of circumstellar interaction (see \citealt{Filippenko1997} for a review of traditional SN classifications).  

The properties of these events are varied, and individual events have been hypothesized to be due to a wide number of explosion mechanisms/progenitors, including the explosion of a stripped massive star \citep{Drout2013,Kleiser2014,Tauris2013}, the detonation of a helium shell on a white dwarf \citep[e.g.][]{Shen2010,Perets2010}, the shock breakout from a dense circumstellar shell \citep{Ofek2010}, and a super-Eddington tidal disruption flare \citep{Cenko2012,Strubbe2009}.

Since their discovery, the overall number of these rapidly-evolving transients has remained small.  This is due to a relatively unknown combination of the true rate of such events and the inefficiency with which optical SN searches detect transients with very short timescales. Despite some success discovering transients at early times, many searches are still optimized for the detection of type Ia SN (t$_{1/2} \sim$ 20 $-$ 30 days). As a result, many rapidly-evolving transients are discovered around maximum light, and decay timescales are used as a proxy for overall timescale or rise time in explosion models.  Since rapidly-evolving transients probe the extremes of both explosion parameters and progenitor configurations, such assumptions can have an effect on the theoretical interpretation of these transients.

As such, obtaining a larger sample of rapidly-evolving events with well constrained rise and decline timescales would be beneficial for two reasons. First, the assumptions which go into many simplified analytic models for SN (e.g.\ \citealt{Arnett1982}) break down at such short timescales. More detailed modeling of a larger number of transients is necessary to determine whether these events are extreme cases of progenitors/explosions we already know, or if they represent entirely new classes of transients.  Second, the true rate at which these rapidly-evolving transients occur will influence our understanding of various stages of stellar evolution.  Although rates have been estimated by several previous works \citep{Drout2013,Perets2011,Poznanski2010}, they are difficult to constrain when the sample size of known objects from any given survey is typically $\sim$ 1. 

The Pan-STARRS1 Medium Deep Survey (PS1-MDS) is well suited for this task because of its rapid cadence and multiple band coverage to a significant depth ($\sim$ 24 mag).  In this paper we present the results from a search within the PS1-MDS survey data for rapidly-evolving and luminous transients with well constrained rise times.  The search was designed to identify transients which persist for multiple rest-frame days, but have t$_{1/2} \lesssim$ 12 days. We have identified 10 such objects in the duration of the survey (Oct 2009 - March 2014).  This represents a significant increase ($\times$ 3) in the total number of transients that have been found in this portion of SN phase space.

The paper is structured as follows: In Section~\ref{Sec:Obs} we give a brief overview of the PS1-MDS, describe the process by which we selected transients, and summarize our observations.  In Sections~\ref{Sec:Overview}, ~\ref{Sec:Photometry} and ~\ref{Sec:Spectra} we give an overview of our selected sample and describe their photometric and spectroscopic properties.  In Section~\ref{Sec:Hosts} we examine the host galaxies of the entire sample.  In Section~\ref{Sec:Rates} we calculate volumetric and relative rates for these objects based on four years of the PS1-MDS and, finally, in Section~\ref{Sec:Discussion} we discuss likely explosion mechanisms and progenitor systems for these events.

\section{Observations and Sample Selection}\label{Sec:Obs}

\subsection{PS1-MDS Overview}

PS1 is a wide-field imaging system dedicated to survey observations.  Located on Haleakala, Hawaii, it possesses a 1.8-m diameter primary mirror and a 3$^{\circ}$.3 diameter field of view \citep{Kaiser2010}.  The imager consists of an array of sixty 4800 $\times$ 4800 pixel detectors with a pixel scale of 0.$''$258, providing an instantaneous field of view of 7.1 deg$^2$  \citep{Tonry2009}.  Observations are obtained with a set of five broadband filters (g$_{P1}$r$_{P1}$i$_{P1}$z$_{P1}$y$_{P1}$, hereafter grizy$_{\rm{P1}}$) which are similar, although not identical, to those used by the Sloan Digital Sky Survey (SDSS; \citealt{Ahn2012}).  Details of the filters and photometry system are given in \citet{Tonry2012} and \citet{Stubbs2010}.

The PS1-MDS consists of 10 fields, each a single PS1 imager footprint, distributed throughout the sky.  Approximately 25\% of PS1 observing time is dedicated to revisiting the PS1-MDS fields on a nightly basis in griz$_{P1}$ to a 5-$\sigma$ depth of $\sim$ 23.3 mag with a nominal cadence of 3 days in any given filter.  In optimal observing conditions, g$_{\rm{P1}}$ and r$_{\rm{P1}}$ are observed on the same night with i$_{\rm{P1}}$ and z$_{\rm{P1}}$ observations following on consecutive evenings.  y$_{P1}$ is observed during times of full moon to a depth of $\sim$ 21.7 mag.

Initial reduction and processing of all PS1-MDS images are carried out with the PS1 Image Processing Pipeline (IPP; \citealt{Magnier2006,Magnier2008}).  This includes standard reductions, astrometric solution, and stacking of nightly images as well as source detection and photometry.  The nightly PS1-MDS stacks are then transferred to the Harvard FAS Research Computing cluster where difference images are produced and potential transients identified by the {\tt photpipe} pipeline \citep{Rest2005,Rest2013}.  Potential transients are visually inspected for possible promotion to status of transient alert. This visual inspection acts to filter out false positives due to time varying trails from saturated sources and poorly subtracted galaxy cores (with obvious large-scale dipole patterns). Once a transient reaches alert status it is flagged for potential spectroscopic follow-up.  Spectra are acquired for $\sim$10\% of PS1-MDS alerts, with final selections left to the spectroscopic observers.

\subsection{Rapid Transient Selection Criteria}\label{sec:selection}

As part of the normal operation of the PS1-MDS described above, we identified a number of transients that evolved on rapid timescales and reached peak luminosities associated with SN ($-$16 $>$ M $>$ $-$21 mag).  Objects identified in this way (i.e.\ flagged by a human operating the pipeline) are certainly useful for determining the properties of rapidly-evolving transients.  However, calculating volumetric rates based on the number of transients identified in a survey requires a \emph{well defined} set of selection criteria.  To systematically identify objects that lie in this portion of transient phase space, we initiated a search within the $\sim$5000 transient sources discovered with the {\tt photpipe} pipeline during the duration of the PS1-MDS. As a first cut, we required that the transient satisfy the following three criteria in a minimum of two photometric bands (all times are observer frame)\footnote{A given photometric band must satisfy all three criteria in order to count towards the two required bands.}.

\begin{enumerate}
\item The transient must rise by $\gtrsim$ 1.5 mag in the 9 days immediately prior to observed maximum light.
\item The transient must decline by $\gtrsim$ 1.5 mag in $\sim$ 25 days post observed maximum.  
\item The transient must be present ($>$ 3 $\sigma$) in at least three sequential observations.  This criteria selects a unique population of transients than those discussed in \citet{Berger2013} (which possess timescales shorter than 1 day).
\end{enumerate}

\begin{figure*}[!ht]
\begin{center}
\includegraphics[width=0.245\textwidth]{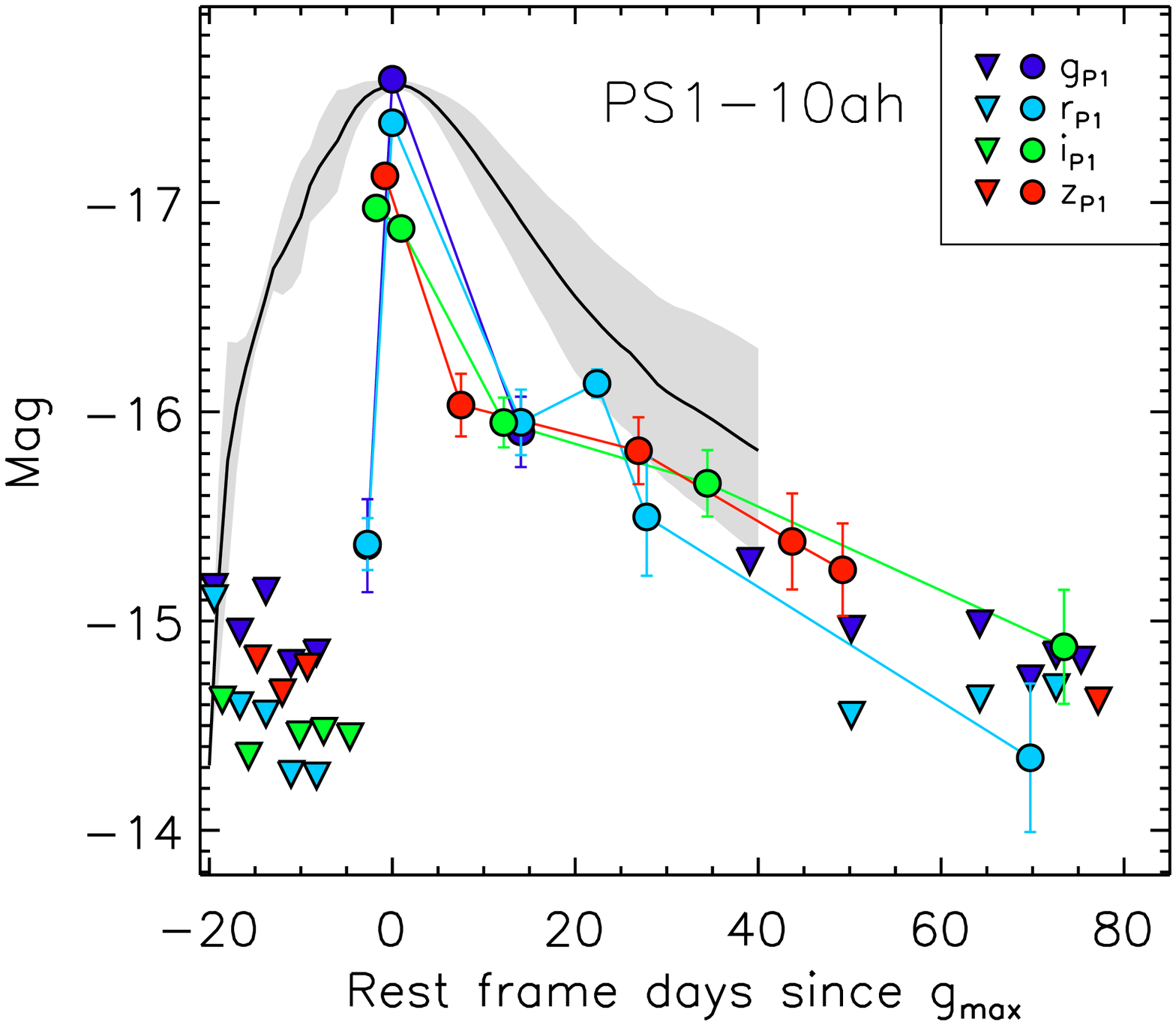}
\includegraphics[width=0.245\textwidth]{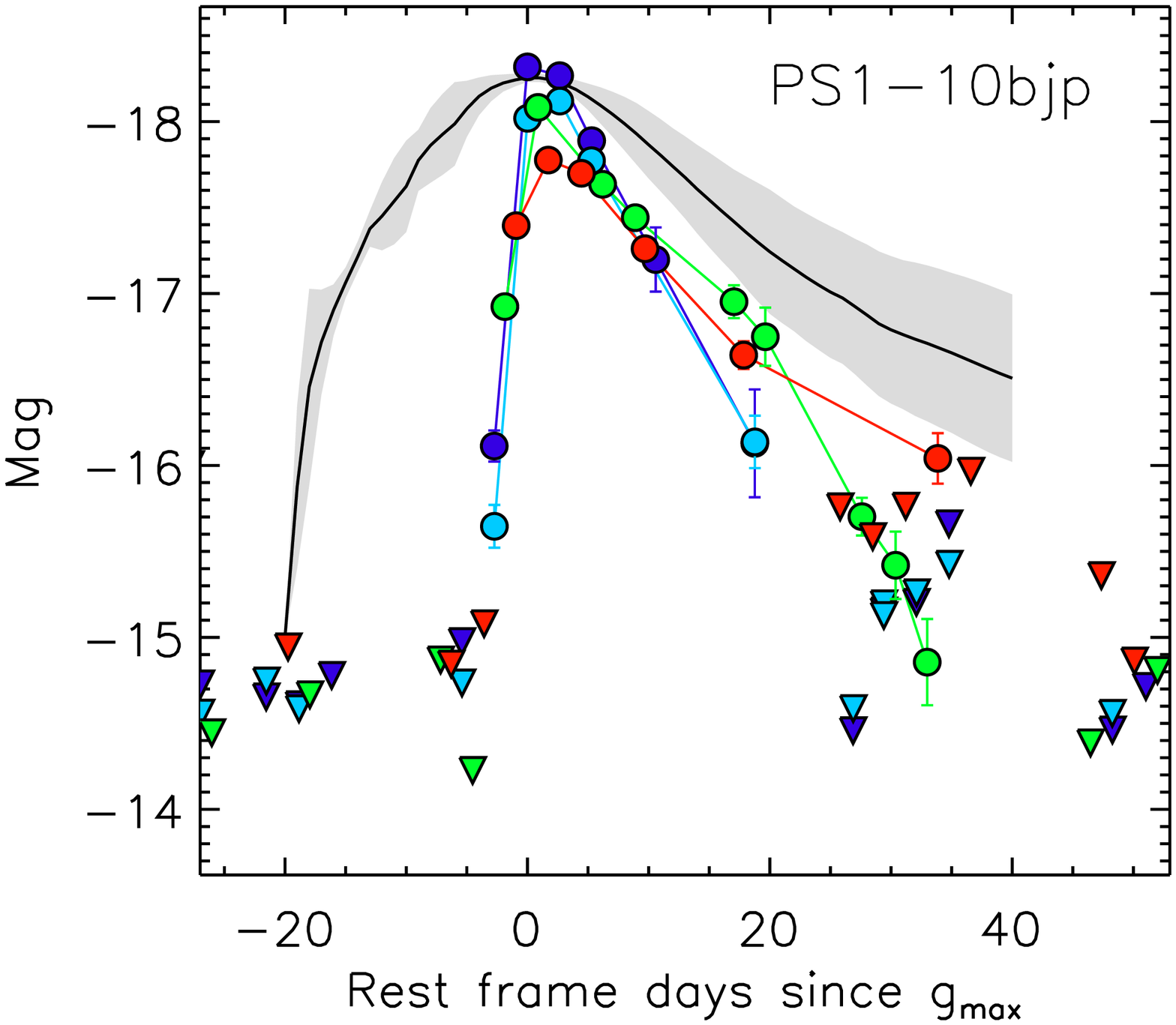}
\includegraphics[width=0.245\textwidth]{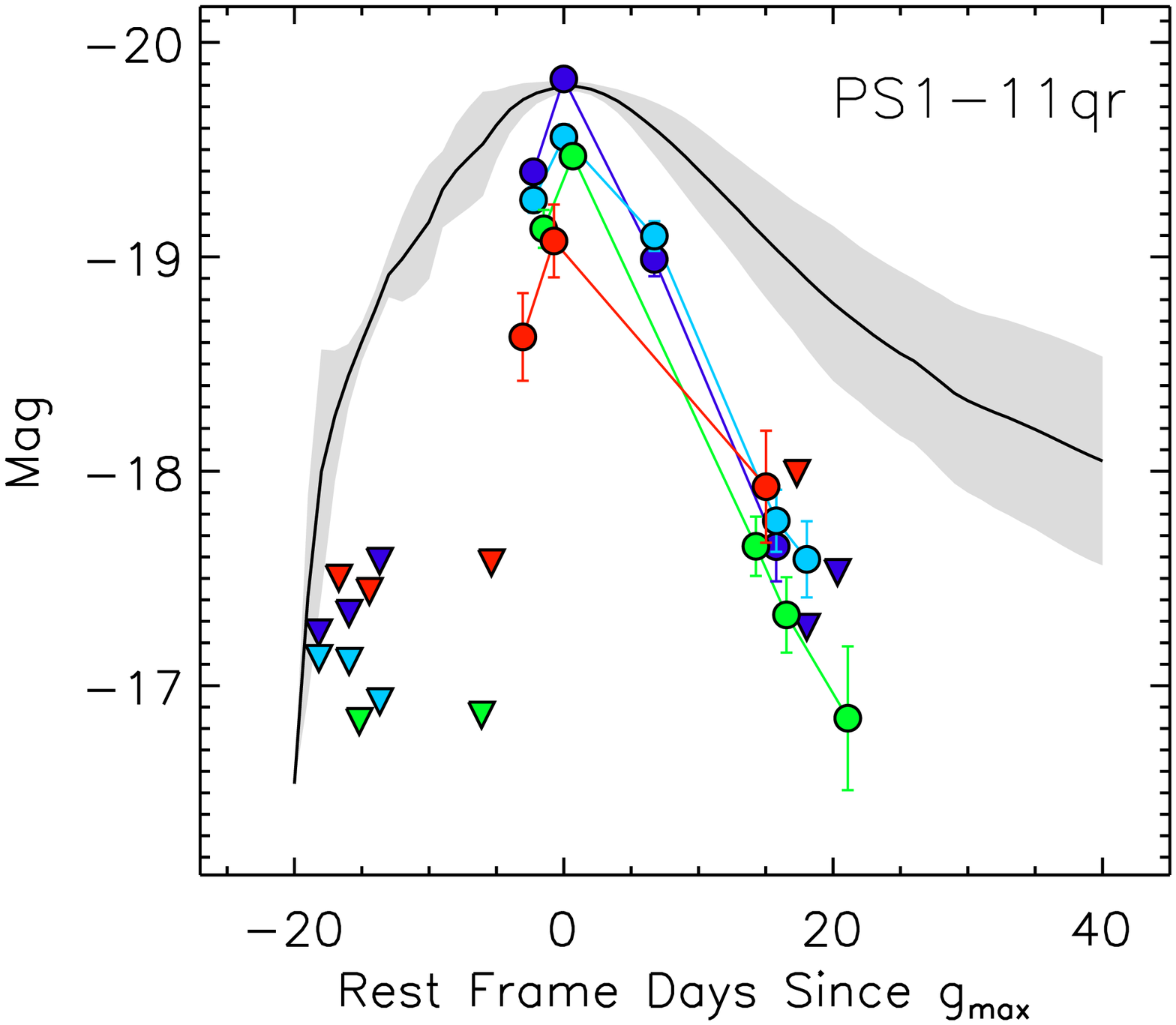}
\includegraphics[width=0.245\textwidth]{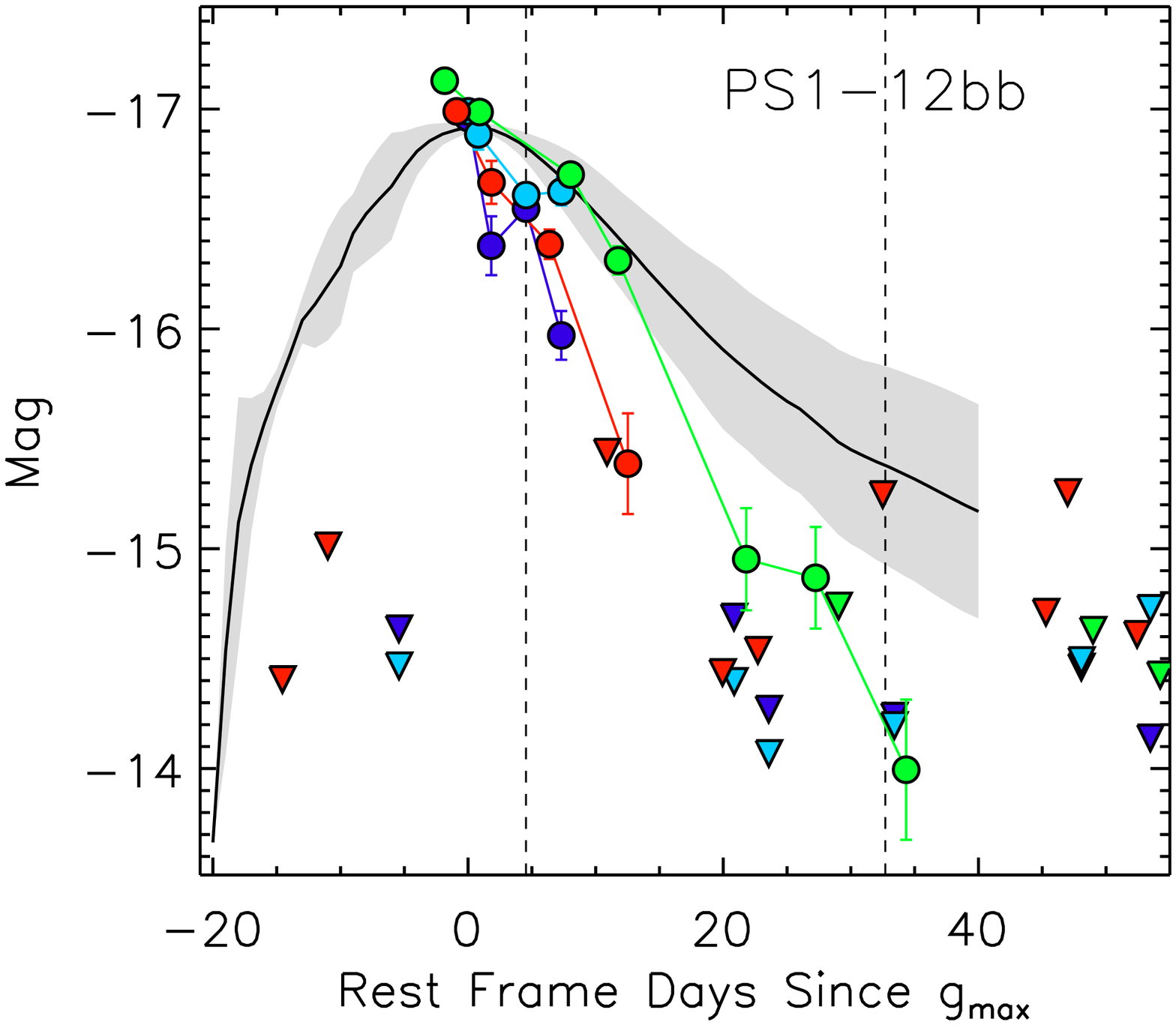}
\includegraphics[width=0.245\textwidth]{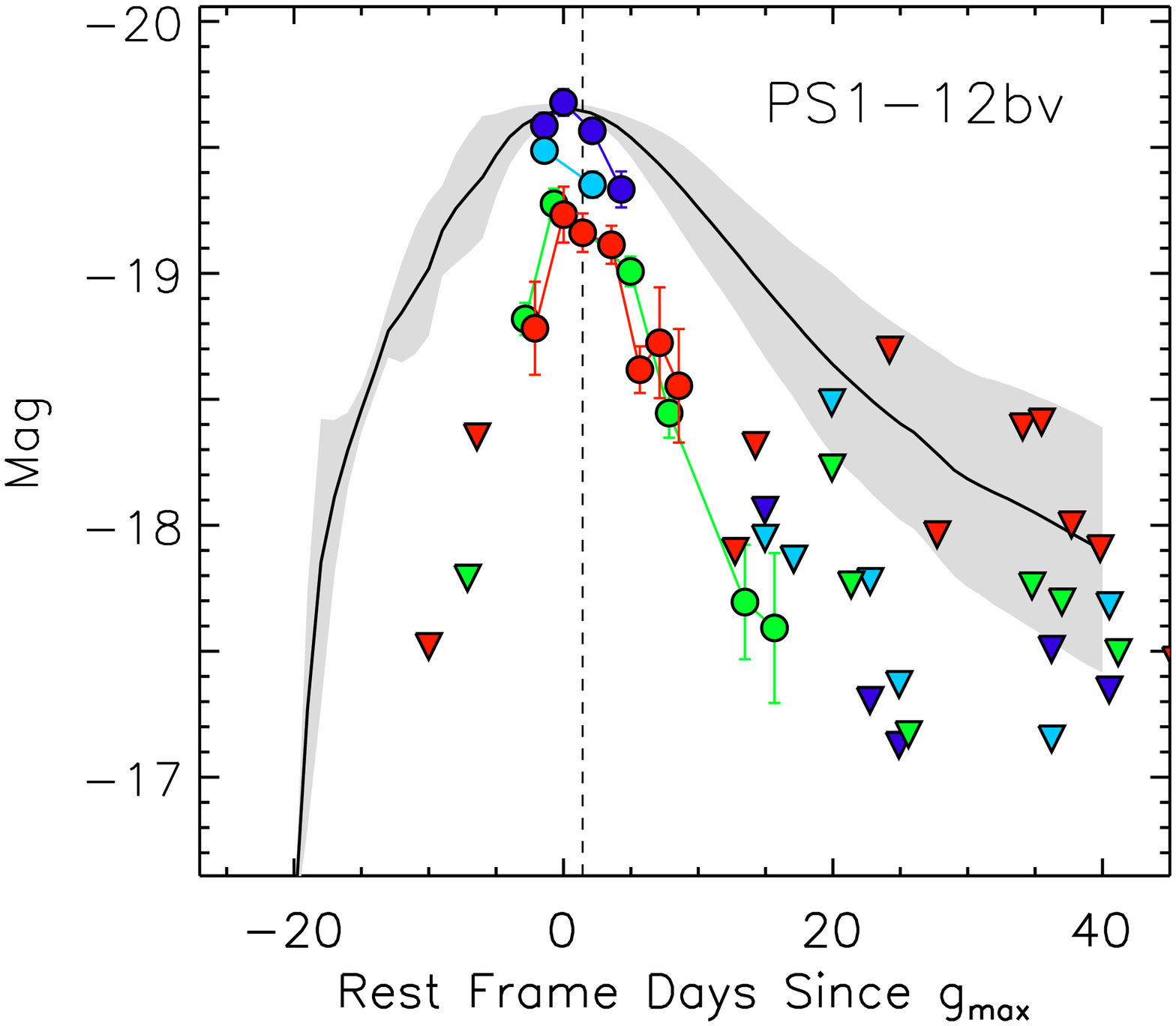}
\includegraphics[width=0.245\textwidth]{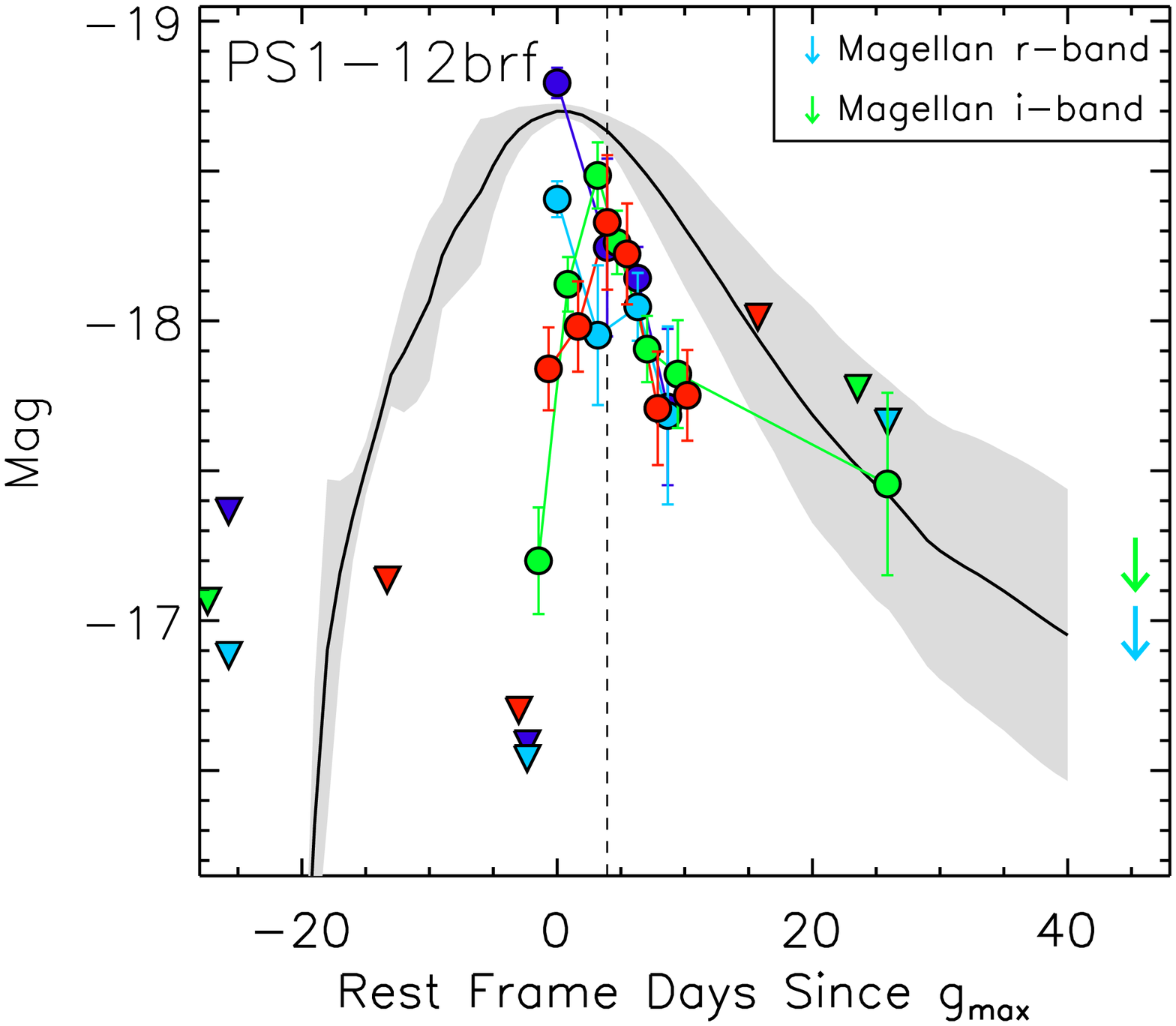}
\end{center}
\caption{PS1 absolute magnitude, rest-frame, light curves for gold sample transients. Circles represent griz$_{\rm{P1}}$ detections and triangles represent 3$\sigma$ upper limits.  Vertical dashed lines indicate epochs when spectroscopic observations were acquired.  The grey shaded region is the R$-$band type Ibc template from \citet{Drout2011}, normalized to the peak magnitude of the PS1-MDS transient. \label{fig:GoldPhot1}}
\end{figure*}

\begin{figure*}[!ht]
\includegraphics[width=0.245\textwidth]{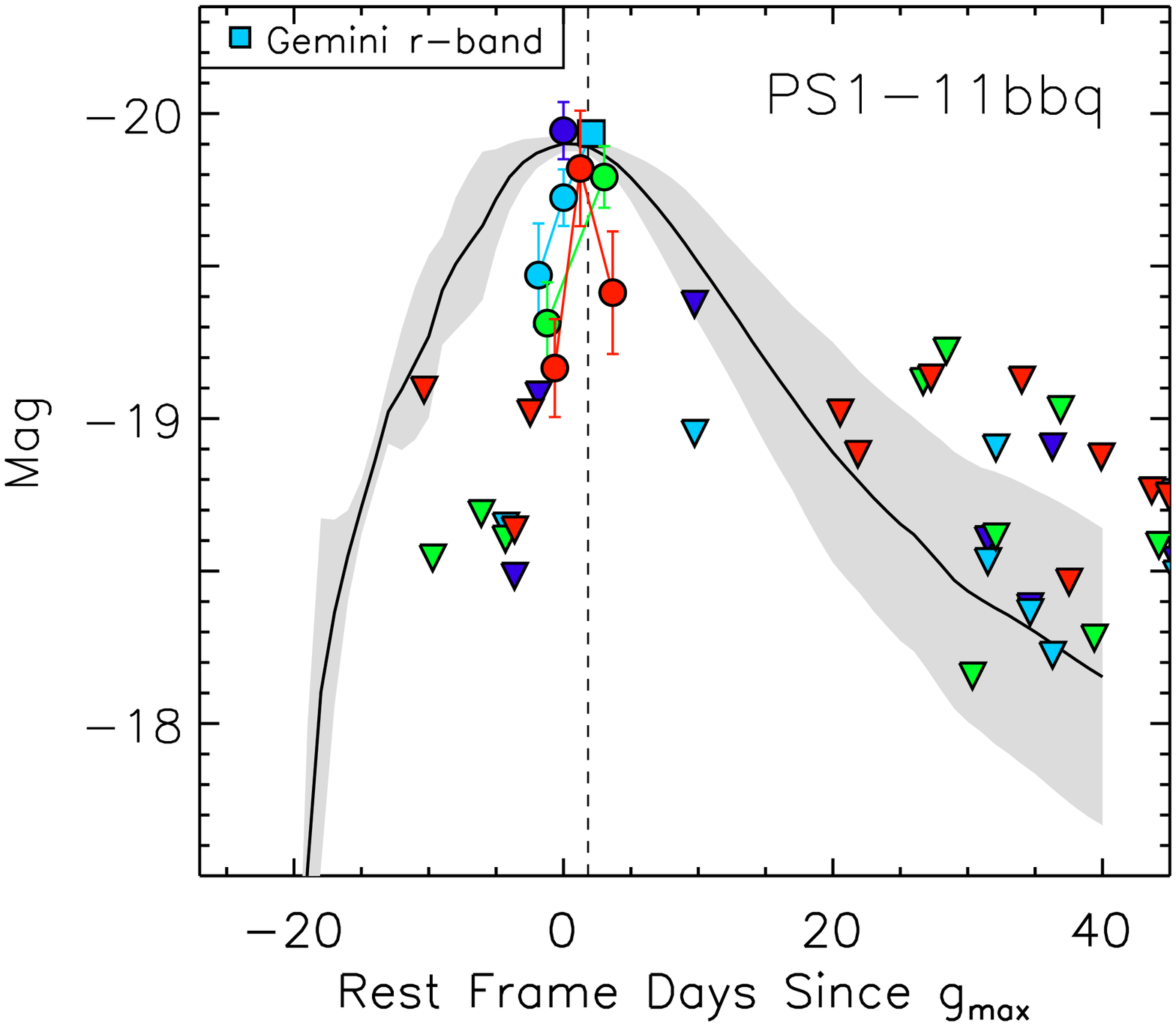}
\includegraphics[width=0.245\textwidth]{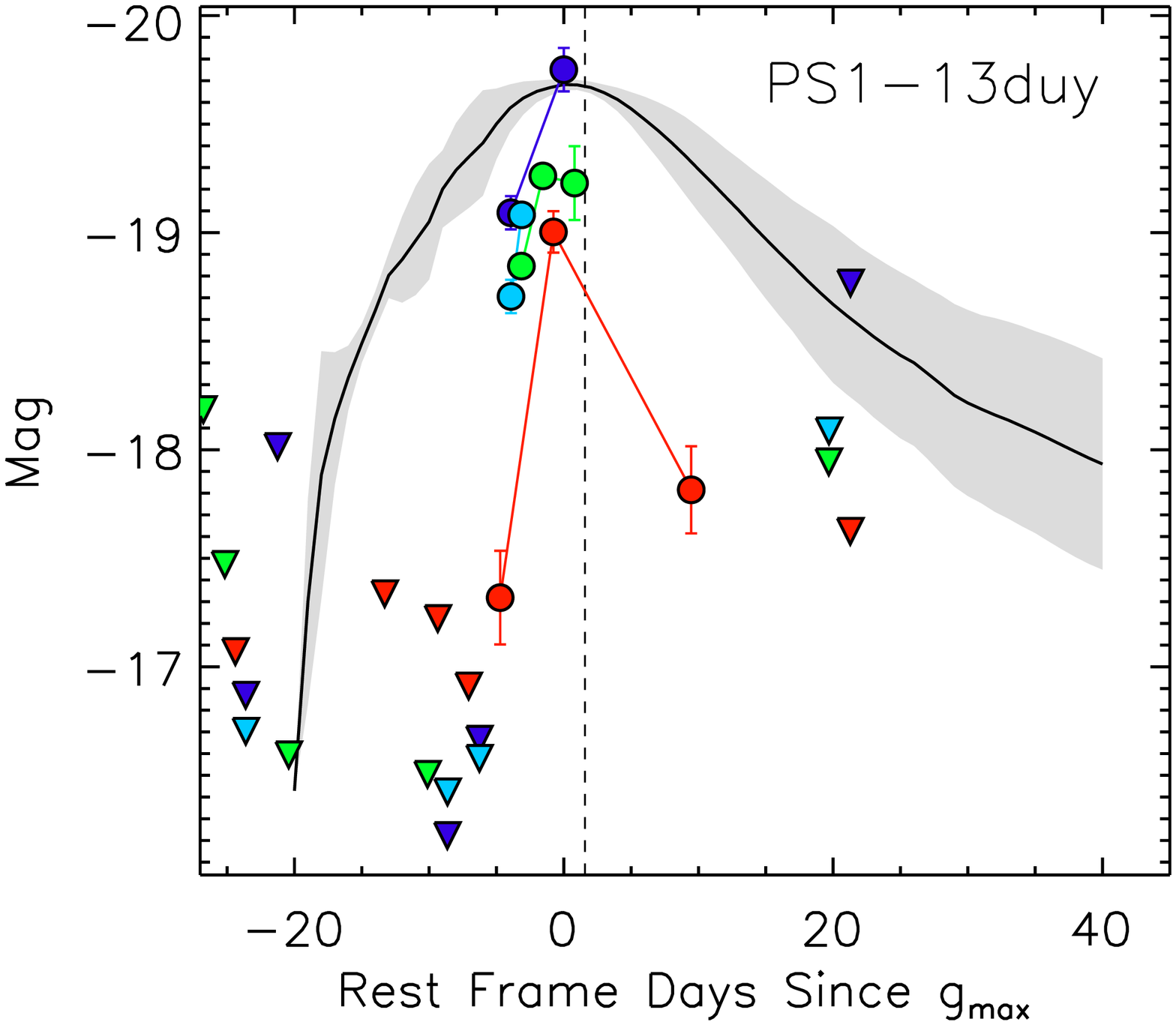}
\includegraphics[width=0.245\textwidth]{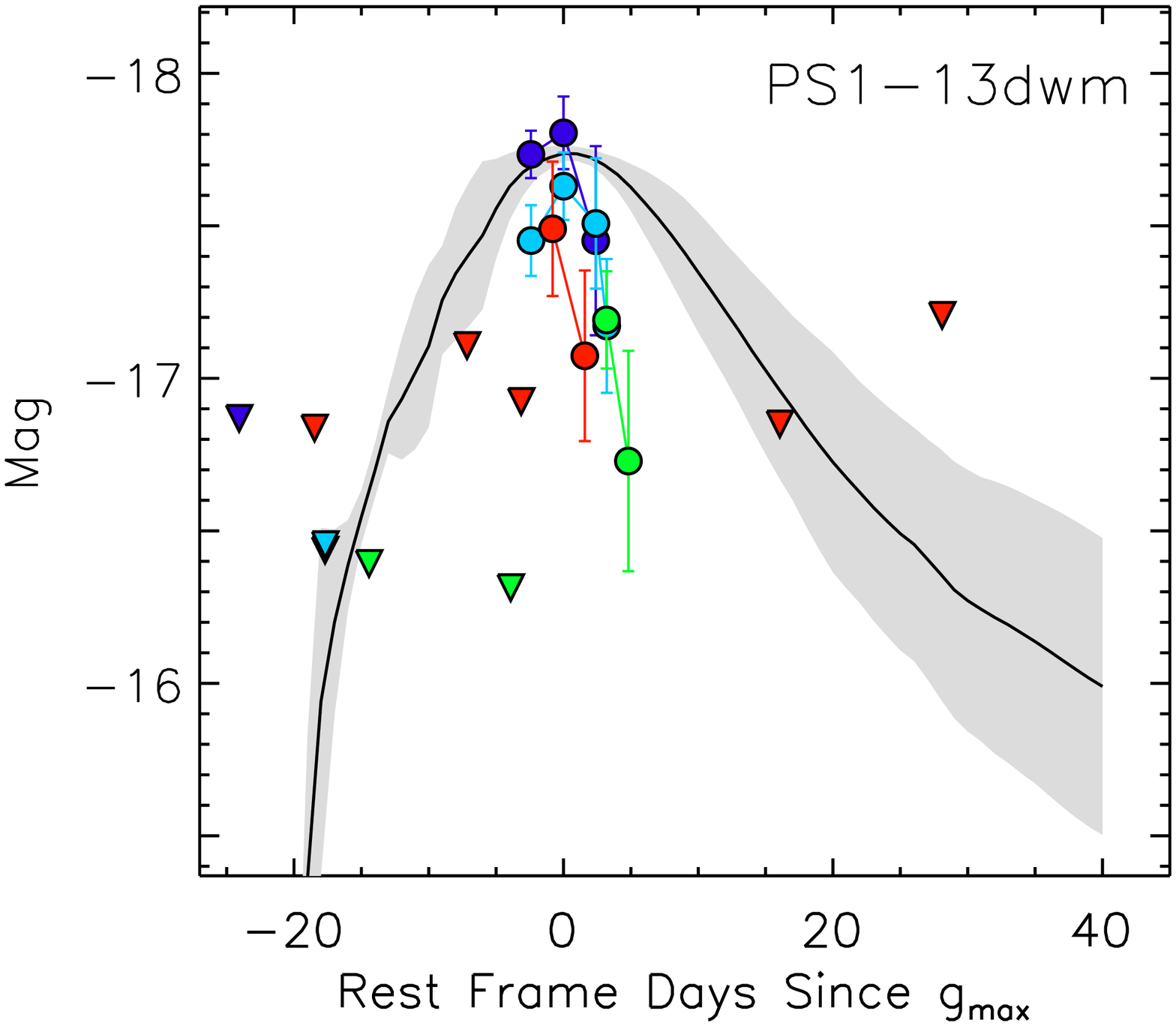}
\includegraphics[width=0.245\textwidth]{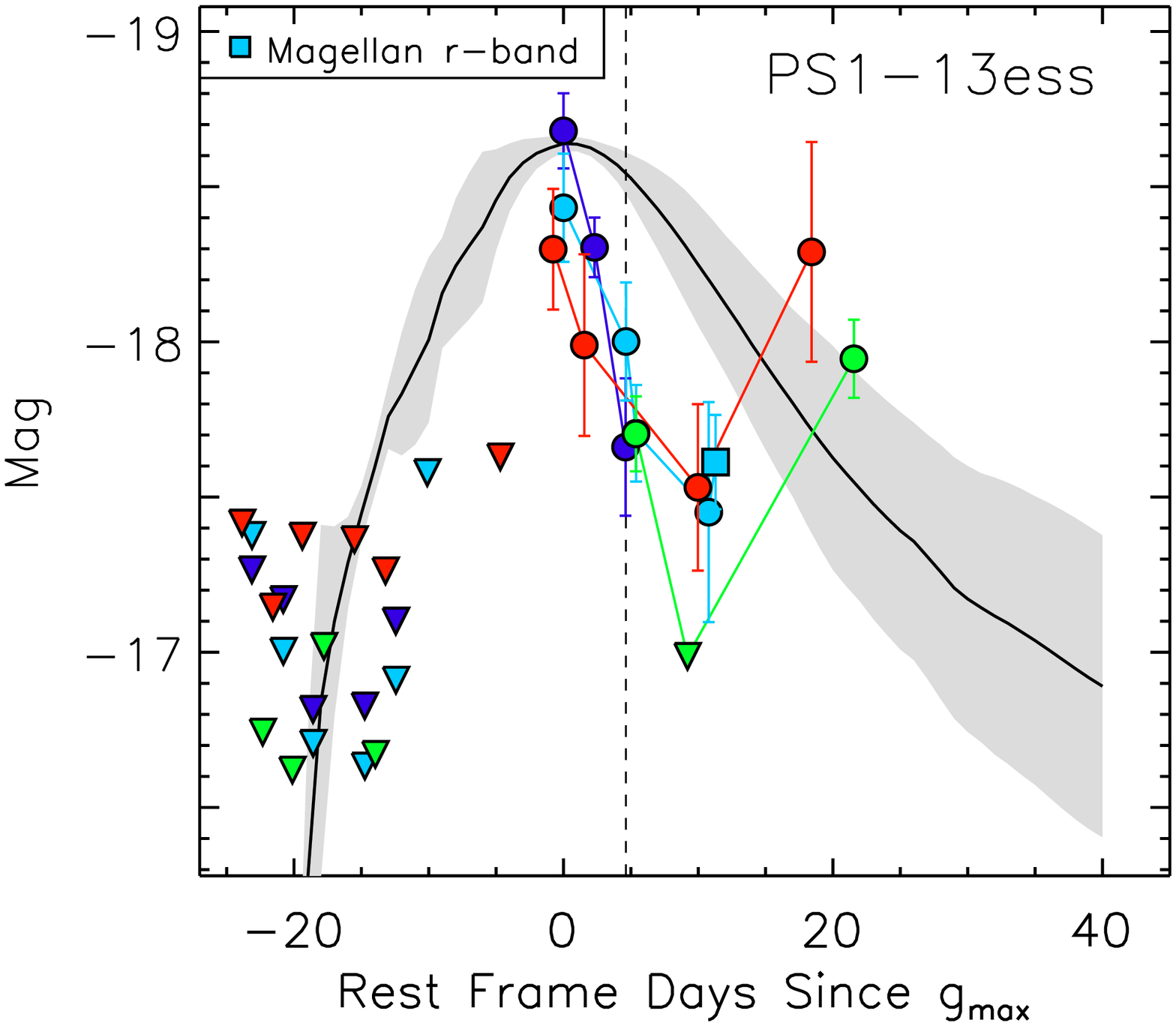}
\caption{Same as Figure~\ref{fig:GoldPhot1} for silver sample objects. \label{fig:GoldPhot2}}
\end{figure*}

\begin{figure*}[!ht]
\includegraphics[width=0.245\textwidth]{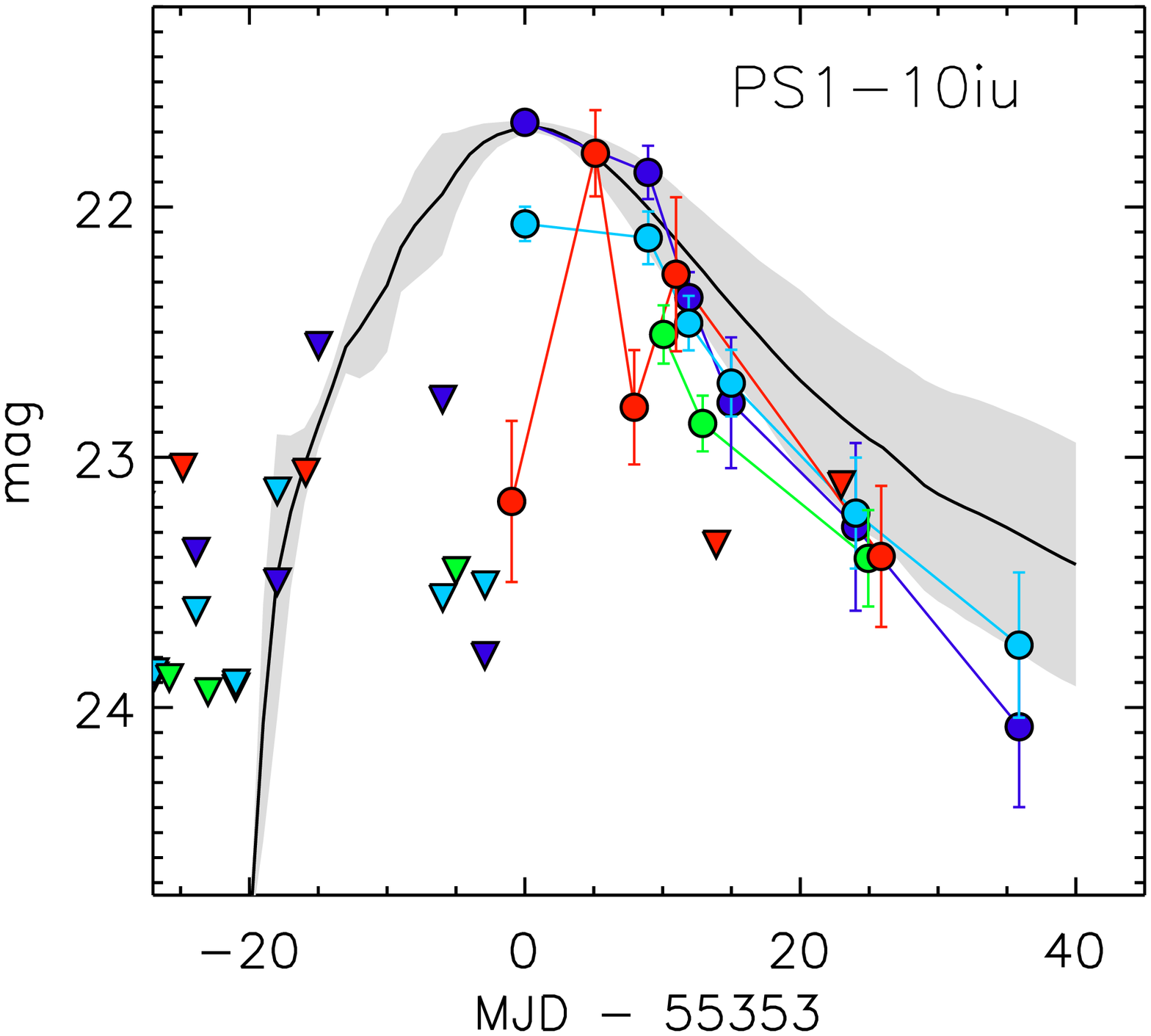}
\includegraphics[width=0.245\textwidth]{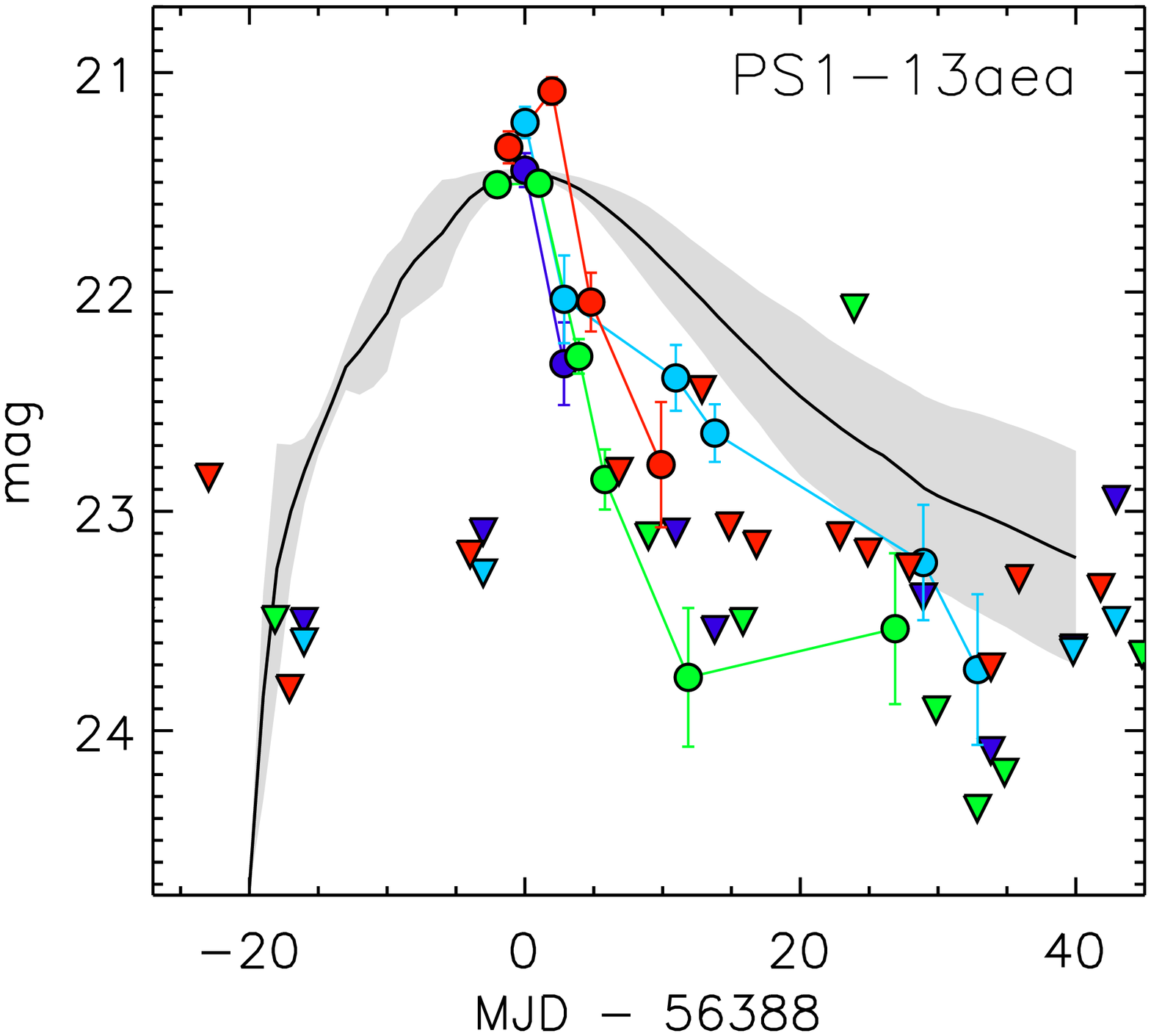}
\includegraphics[width=0.245\textwidth]{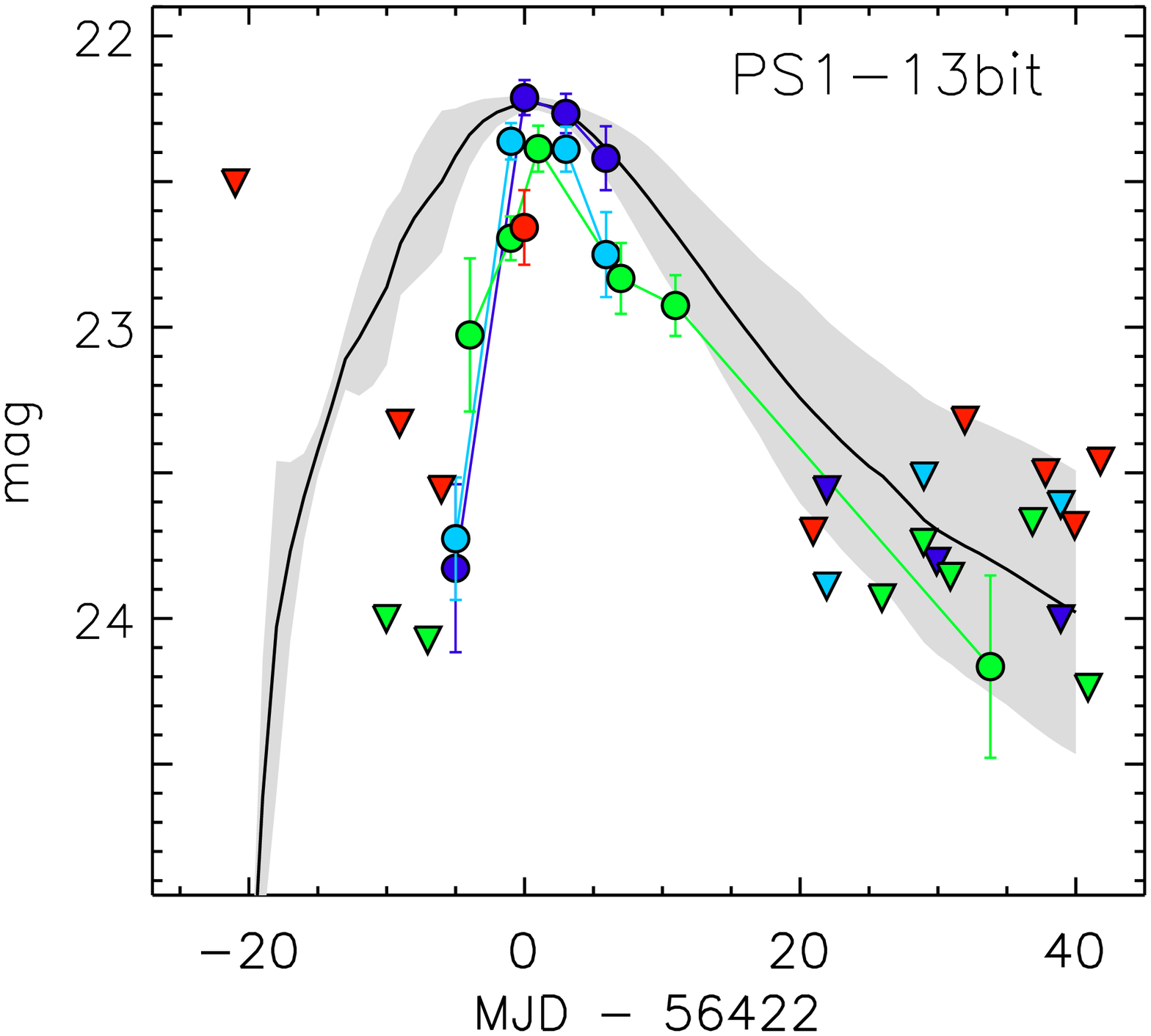}
\includegraphics[width=0.245\textwidth]{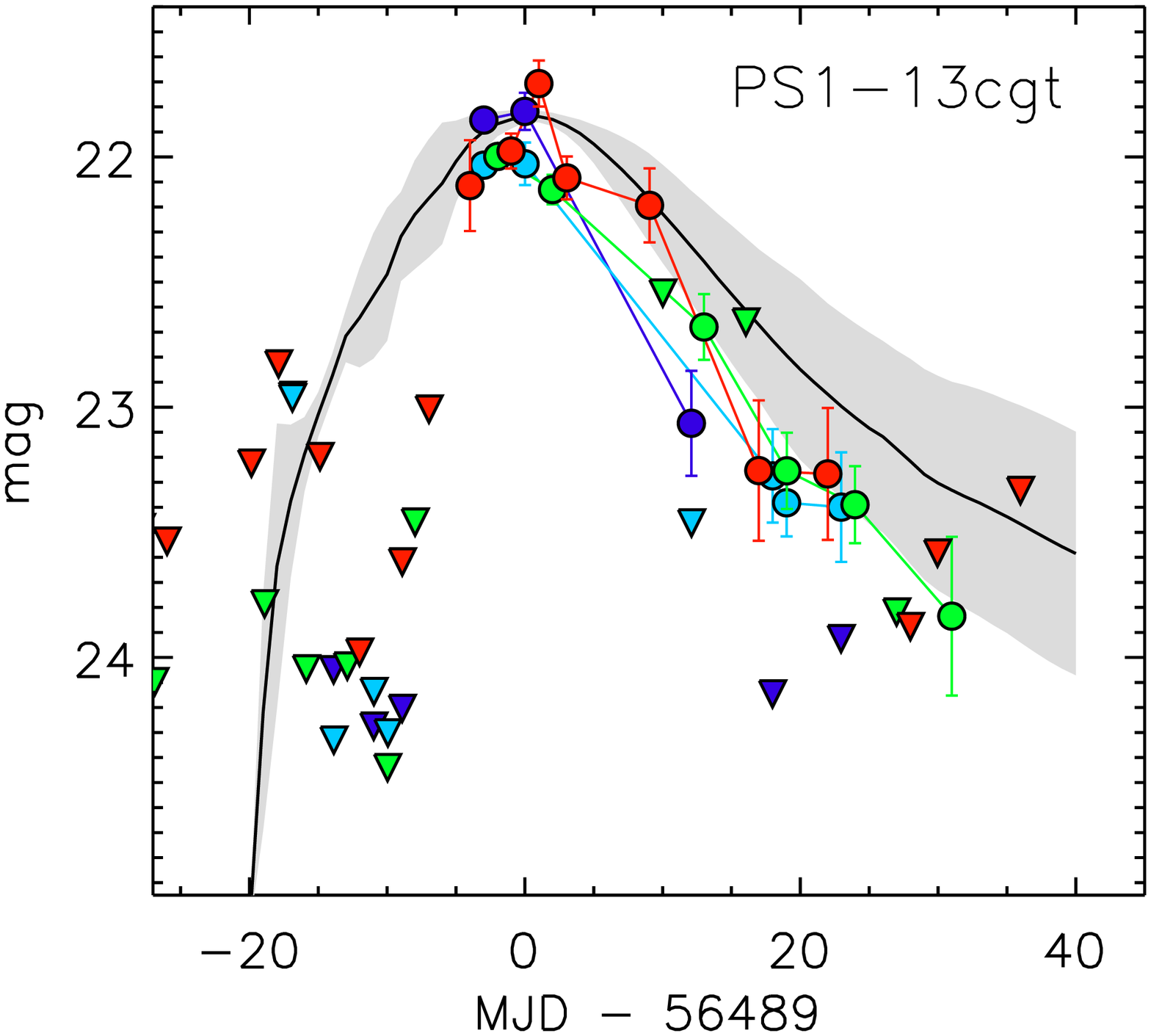}
\caption{PS1 apparent magnitude, observer-frame,  light curves for our bronze (non-spectroscopic) sample. Symbols have the same meaning as Figure~\ref{fig:GoldPhot1}. \label{fig:SilverPhot}}
\end{figure*}

These criteria are motivated by the observed timescales of known SN and are intended to exclude the bulk of type Ia, type Ib/c and type II SN.   In V-band, typical type Ia and type Ib/c SN will rise by 0.6 $-$ 0.75 mag in the 10 days immediately prior to maximum light and decline by 1.0 $-$ 1.2 mag in the 25 days post-maximum \citep{Riess1999,Drout2011,Li2011}.  At higher redshifts PS1 will probe bluer wavelengths (which typically evolve more rapidly) but this will compete with the effects of time dilation.  Many type II SN would easily pass the rise time requirement, but they decline by $\lesssim$0.5 mag in the 25 days post maximum \citep{Li2011,Hamuy2003}.  The most rapid type Ic SN (e.g.\ SN\,1994I; \citealt{Richmond1996}) and most rapid SN1991bg-like type Ia SN \citep{Taubenberger2008} would barely pass all three of our requirements if caught at maximum light.  However, any given PS1 band is only observed every three days and we require the same band to pass both our rise and decline cuts. We therefore expect our selected objects to evolve more rapidly than these events\footnote{Consider a transient that barely passes our selection criteria (i.e.\ rises 1.5 mag in the 9 days prior to maximum light and declines 1.5 mag in the 25 days post maximum). If instead of observing exactly at maximum light we observe at +1 day then the transient will appear to rise slightly \emph{less} than 1.5 mag in the 10 days prior to \emph{observed} maximum.  As a result, it will no longer pass our rise time selection criteria.}.
  
A number of known low-luminosity transients (nova, CVs, M dwarf flares, etc.) could pass these selection criteria if located within the Milky Way or a nearby galaxy. Thus, all objects that passed these initial criteria were then examined by eye in order to select objects that were likely of extragalactic origin.  Objects were removed from consideration if their was a clear stellar source at the location of the transient.  In several cases, spectra were obtained of rapidly-evolving transients which revealed Balmer lines in emission at zero redshift, indicating the transient was likely galactic. Our search identified a handful ($\sim$10) of these low luminosity events.  If a transient was either host-less, or exploded near a low signal to noise, unresolved, source (such that the stellar versus galactic nature of the host could not be robustly determined) it was kept in the sample and is discussed below.

In the end, by combining the results from this systematic search with results from the normal operation of the PS1-MDS pipeline, we identify 14 transients of interest.  For ten of these transients we have obtained spectra of their underlying hosts, confirming their extragalactic origin.  In Section~\ref{Sec:Overview} we provide an overview of the selected objects and divide them into three groups based on the quality of their light curves and constraints available for their distance.  In the rest of this section we describe the observations obtained for all 14 transients.

\subsection{PS1 Transient Photometry}

The 14 objects in our sample were discovered in the PS1-MDS imaging spanning 2010 $-$ 2013.  Discovery dates, coordinates, and other basic information are listed in Table~\ref{tab:basic}.  Detailed information on the production of final PSF, template subtracted, griz$_{P1}$ photometry is given in \citet{Rest2013} and \citet{Scolnic2013}.  For transients discovered before Oct.\ 2011, object specific deep templates were constructed from pre-explosion images.  For transients discovered after this time, template subtraction was performed using the PS1 ``deep stacks'' constructed from high quality PS1 images obtained between 2010 and 2011.

In Figures~\ref{fig:GoldPhot1}, \ref{fig:GoldPhot2} and \ref{fig:SilverPhot} we plot the griz$_{P1}$ light curves for all 14 transients.  This includes 3$\sigma$ pre- and post-explosion limits. To give perspective on the rapid timescale of these events, we have also plotted the type Ibc template light curve (grey area; normalized to the PS1 transient peak magnitude) from \citet{Drout2011}.  Photometry is listed in Table~\ref{tab:Photom}.

\subsection{Other Transient Photometry}

In addition to the PS1 photometry, we also obtained one epoch of $r-$band imaging for PS1-11bbq with Gemini GMOS \citep{Hook2004}, one epoch of $ri-$band imaging for PS1-12brf  with Magellan IMACS \citep{Dressler2006} and one epoch of $r-$band imaging for PS1-13ess with Magellan IMACS. This additional photometry was obtained at +2, +45 and +12 rest-frame days for the three objects, respectively. The images were processed using standard tasks in IRAF\footnote{IRAF is distributed by the National Optical Astronomy Observatory, which is operated by the Association for Research in Astronomy, Inc.\, under cooperative agreement with the National Science Foundation.} and calibrated using PS1 magnitudes of field stars.  We subtracted contributions from the host galaxies using PS1 template images and the ISIS software package as described in \citet{Chornock2013}.  These points are also shown (squares) in Figures~\ref{fig:GoldPhot1} and \ref{fig:GoldPhot2}, and listed in Table~\ref{tab:Photom}.

\subsection{Galaxy Photometry}

For our entire sample we compile griz-band photometry for any underlying galaxy/source.  When possible, we utilize the SDSS DR9 Petrosian magnitudes which account for galaxy morphology.  For cases where the underlying galaxy/source was too faint for a high signal-to-noise SDSS detection, we perform aperture photometry on the PS1 deep template images, choosing an aperture to encompass all of the visible light. Cross-checks show the PS1 and SDSS magnitudes are consistent. In Figures~\ref{fig:GoldStamp} and \ref{fig:SilverStamp} we show the environments immediately surrounding the transients. 

\subsection{Optical Spectroscopy}\label{Sec:SpecObs}

Spectroscopic follow-up for the PS1-MDS is carried out on a number of telescopes, with the MMT, Magellan and Gemini bearing most of the load.  Spectra are acquired for roughly 10\% of the transients identified by the {\tt photpipe} pipeline, with final selections left to the observer.  We obtained spectra of five transients while they were active, including two observations of PS1-12bb.  The epochs on which these spectra were taken are indicated by a dashed vertical line in the appropriate panels of Figures~\ref{fig:GoldPhot1} and~\ref{fig:GoldPhot2}.  Host galaxy spectra were also obtained for six transients.  A summary of our spectroscopic data is given in Table~\ref{tab:Spectra}.

Initial reduction (overscan correction, flat fielding, extraction, wavelength calibration) of all long slit spectra was carried out using the standard packages in IRAF.  Flux calibration and telluric correction were performed using a set of custom idl scripts (see, e.g., \citealt{Matheson2008,Blondin2012}) and standard star observations obtained the same night as the science exposures.  Spectra obtained with the Hectospec multi-fiber spectrograph \citep{Fabricant2005} were reduced using the IRAF package ``hectospec'' and the CfA pipeline designed for this instrument.

\section{Sample Overview}\label{Sec:Overview}

The 14 rapidly-evolving transients we identify in the PS1-MDS can be usefully split into three groups based on (1) the quality of their observed light curves and (2) constraints available on their distances.  For the rest of the manuscript these sub-groups will be designated ``gold'', ``silver'', and ``bronze''.

\subsection{Gold and Silver Samples}

\begin{figure*}[!ht]
\begin{center}
\includegraphics[width=0.19\textwidth]{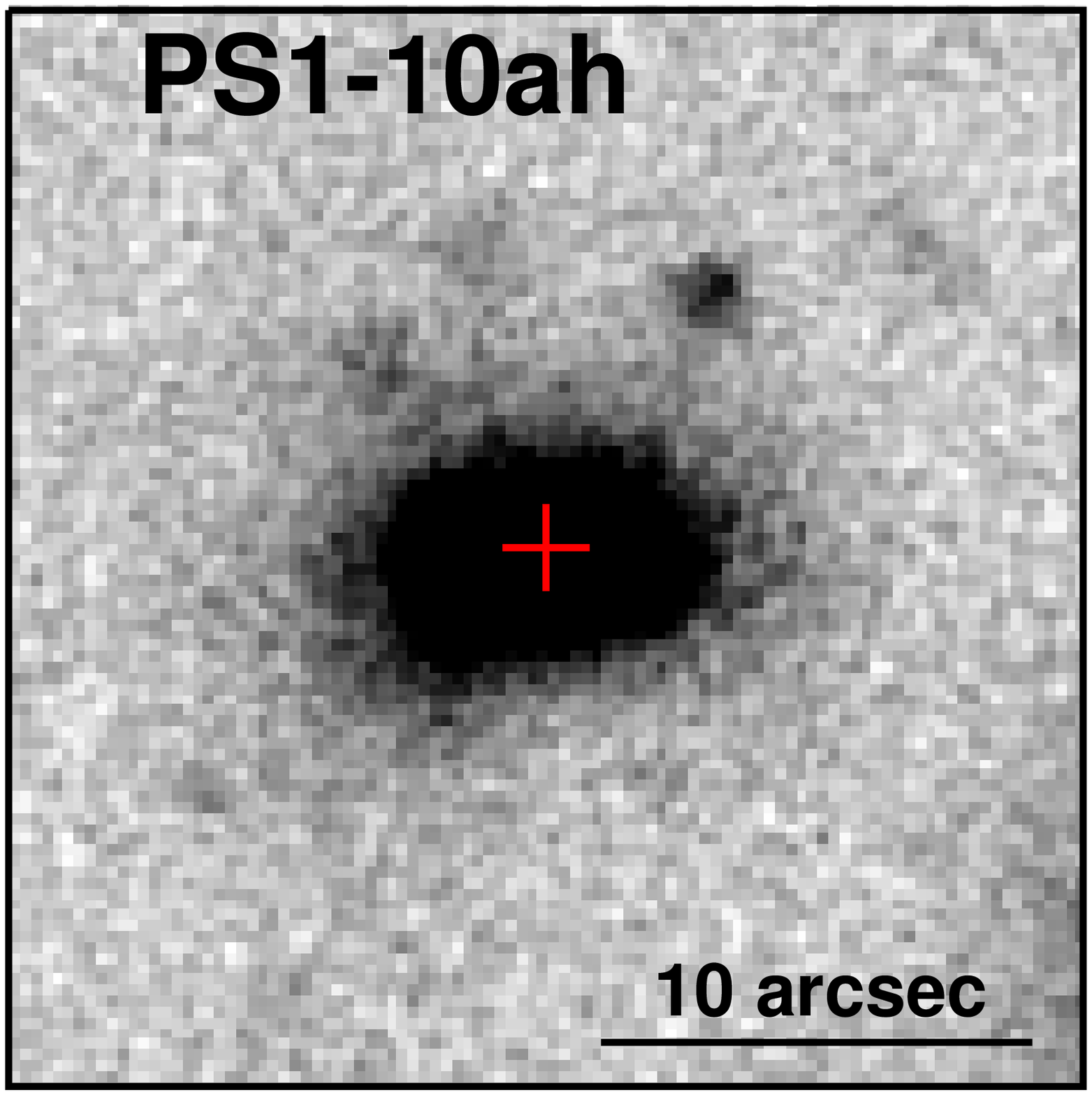}
\includegraphics[width=0.19\textwidth]{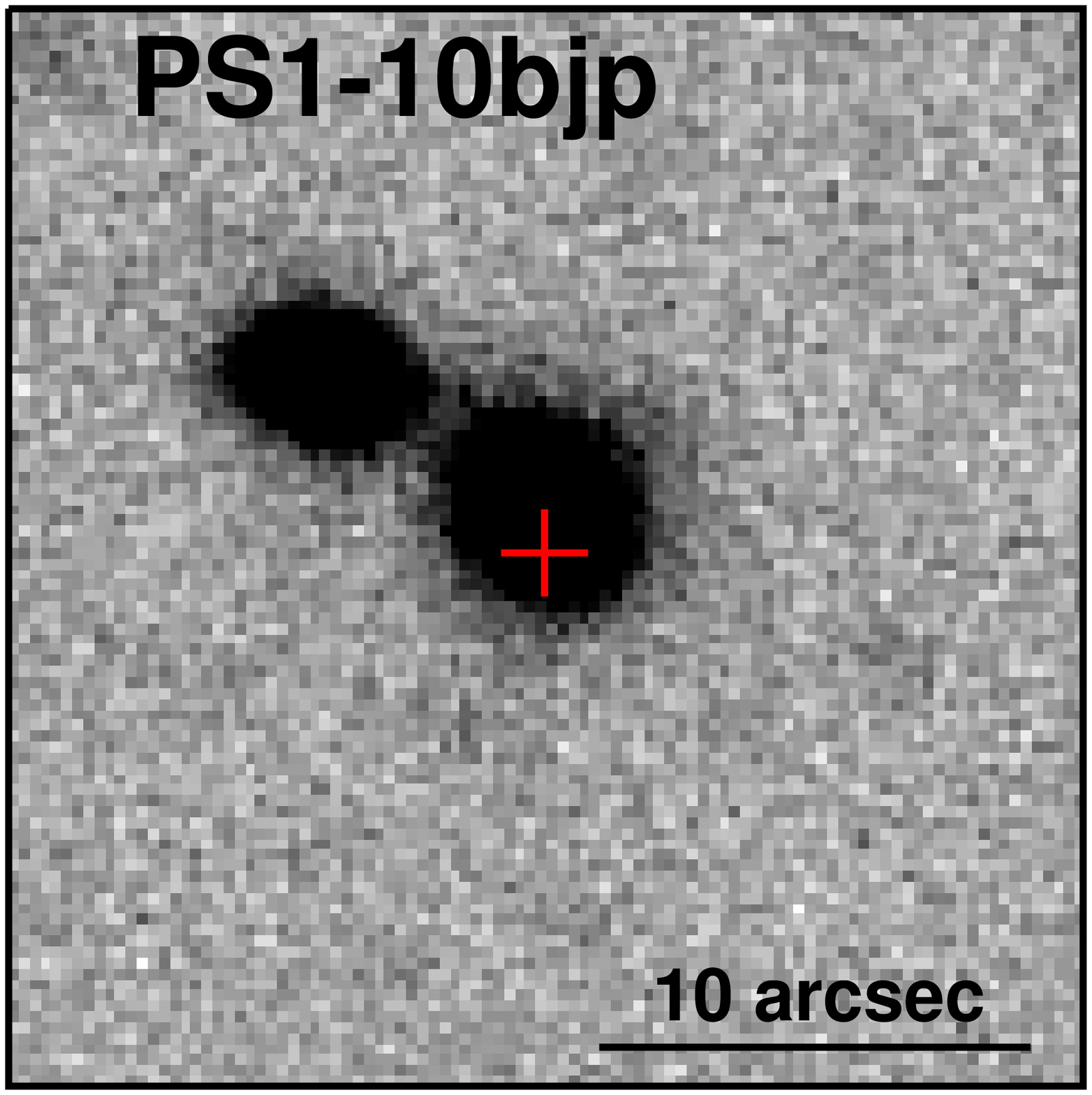}
\includegraphics[width=0.19\textwidth]{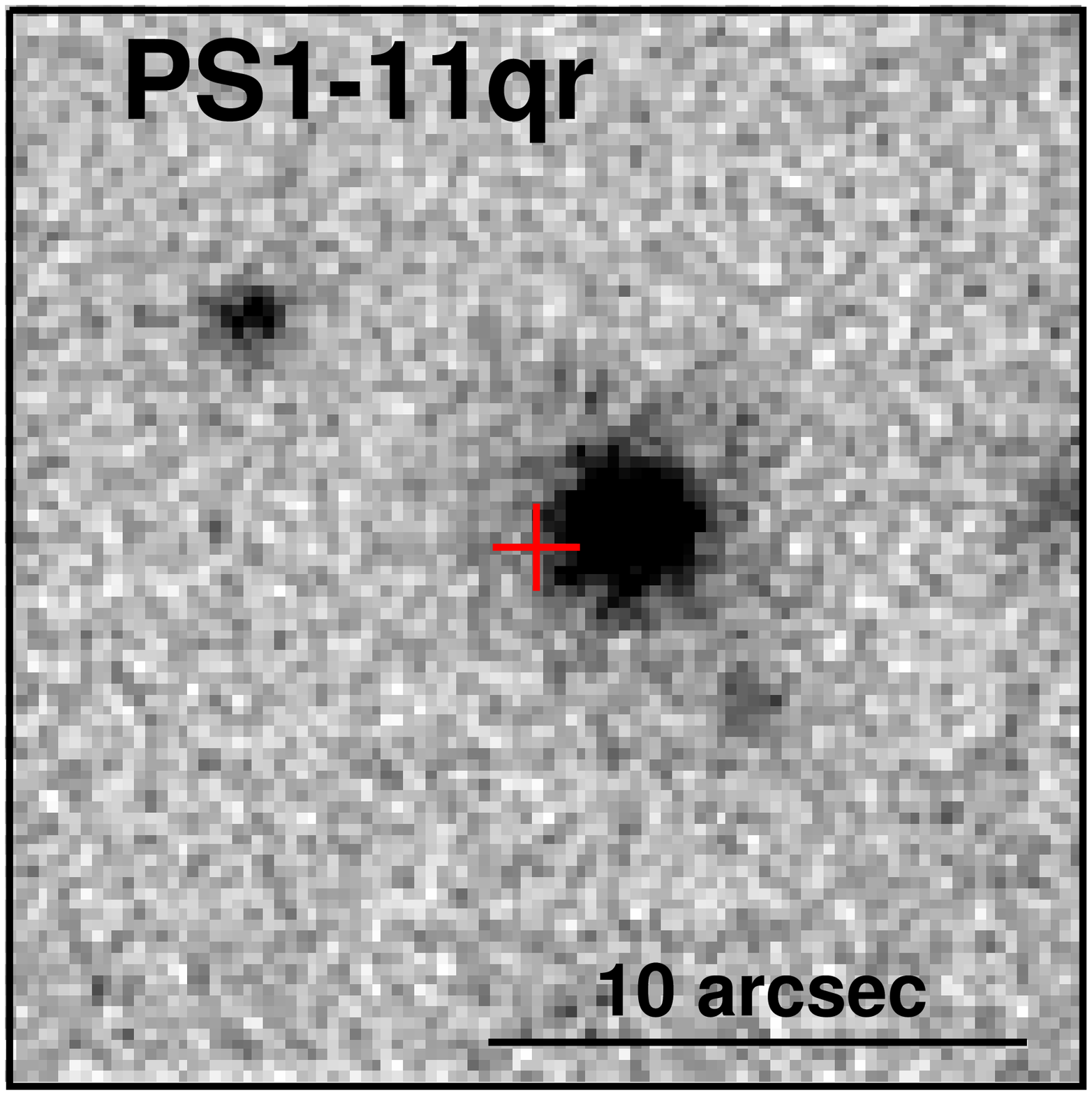}
\includegraphics[width=0.19\textwidth]{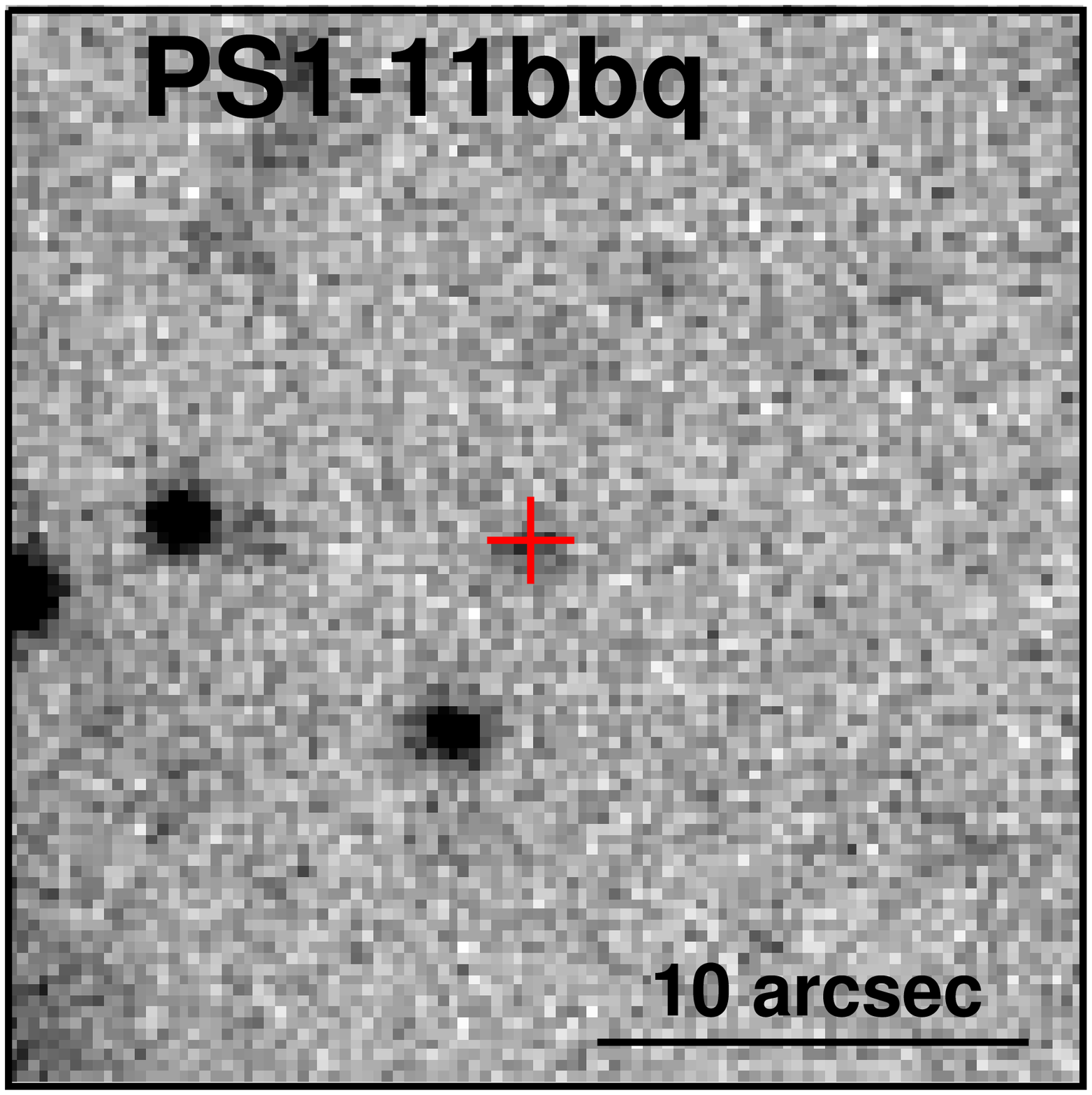}
\includegraphics[width=0.19\textwidth]{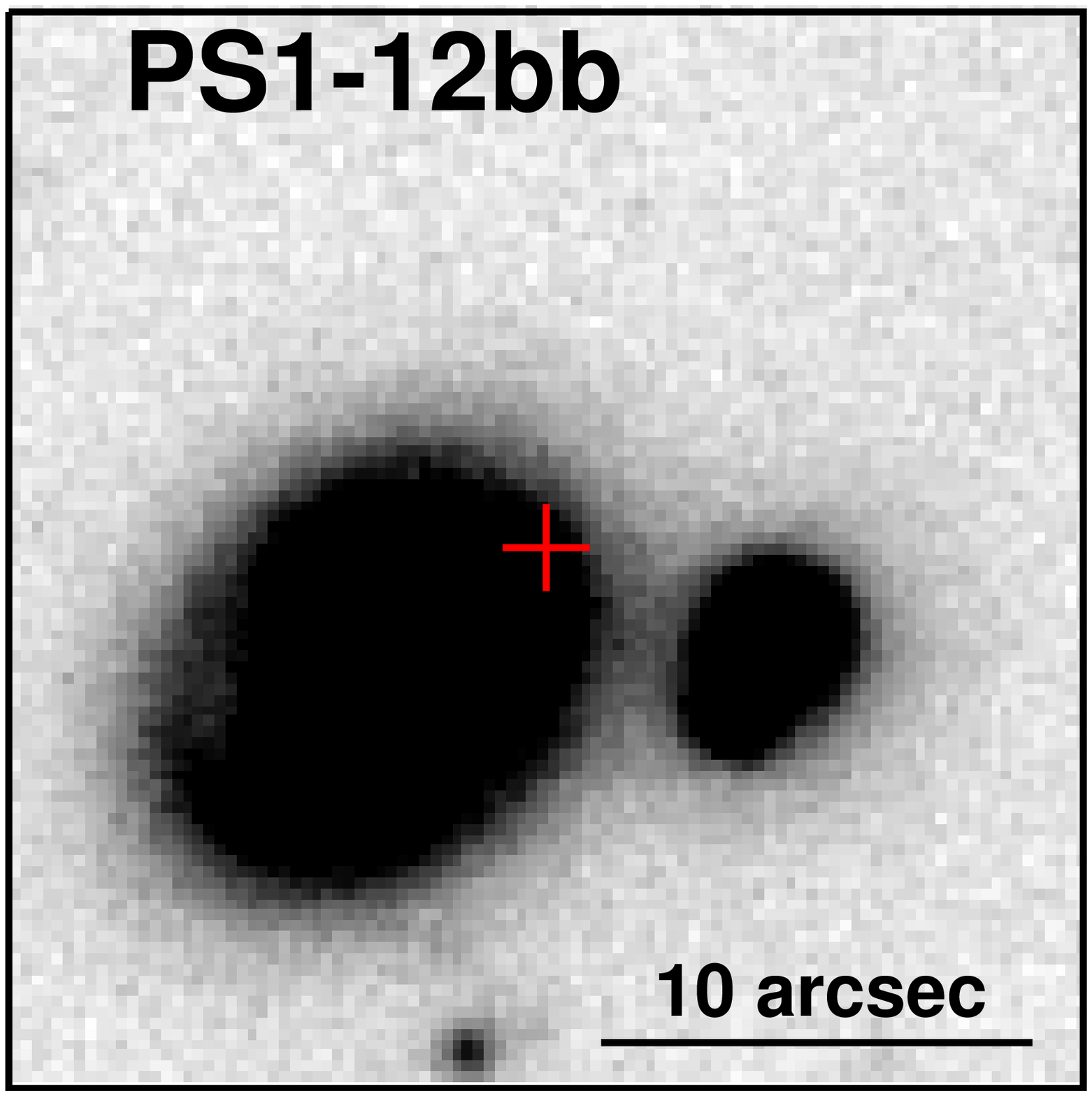}
\includegraphics[width=0.19\textwidth]{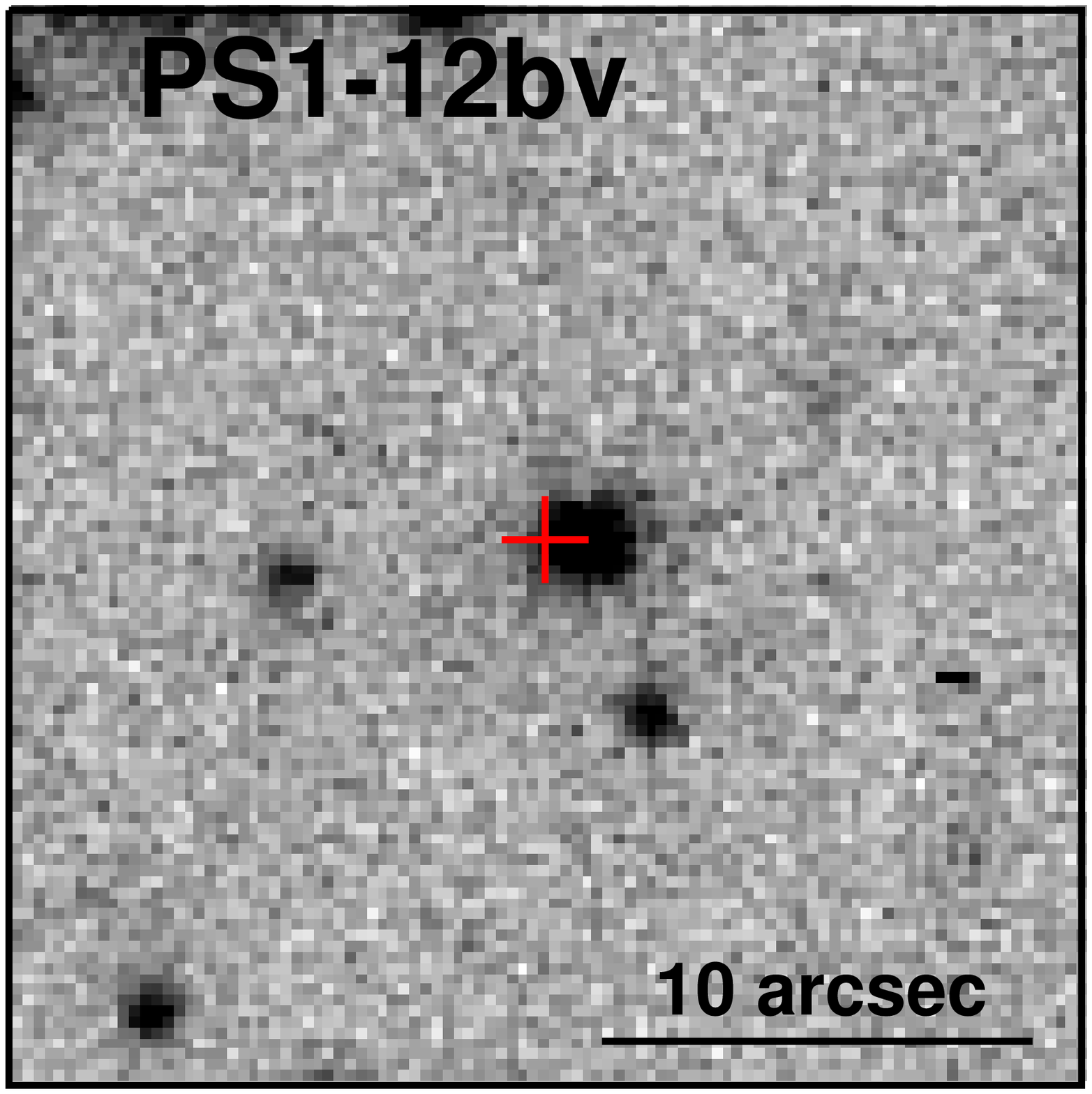}
\includegraphics[width=0.19\textwidth]{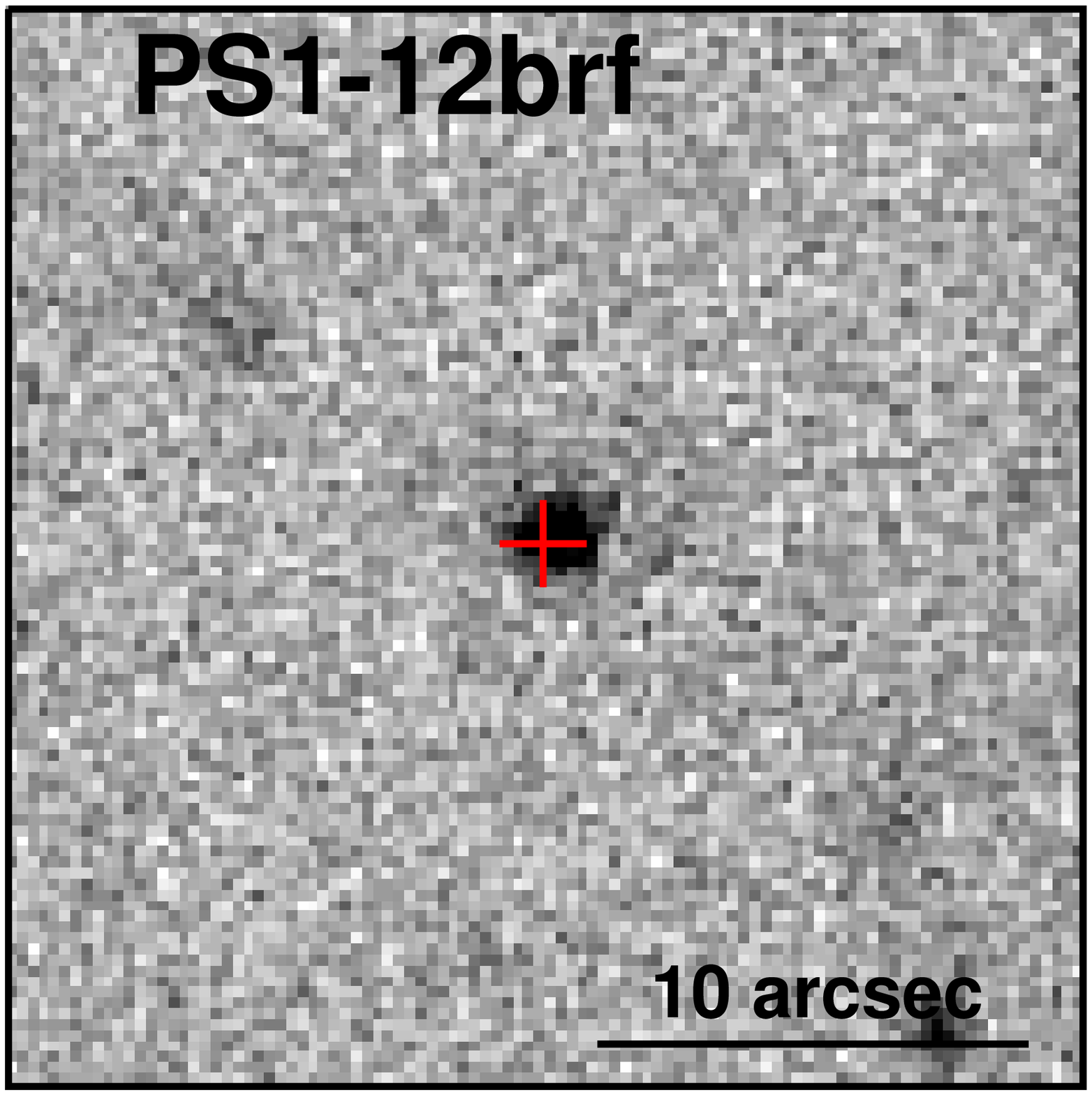}
\includegraphics[width=0.19\textwidth]{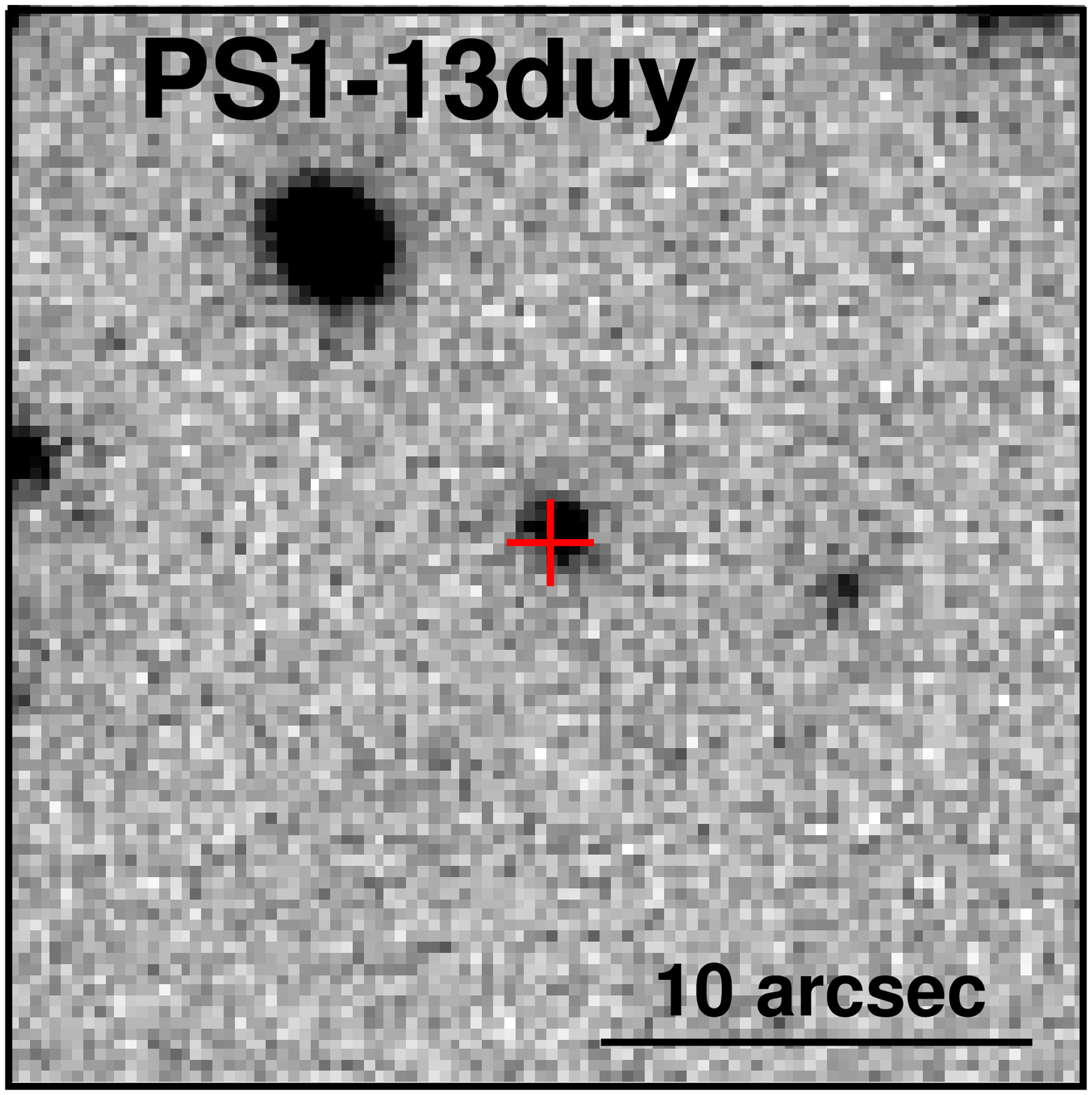}
\includegraphics[width=0.19\textwidth]{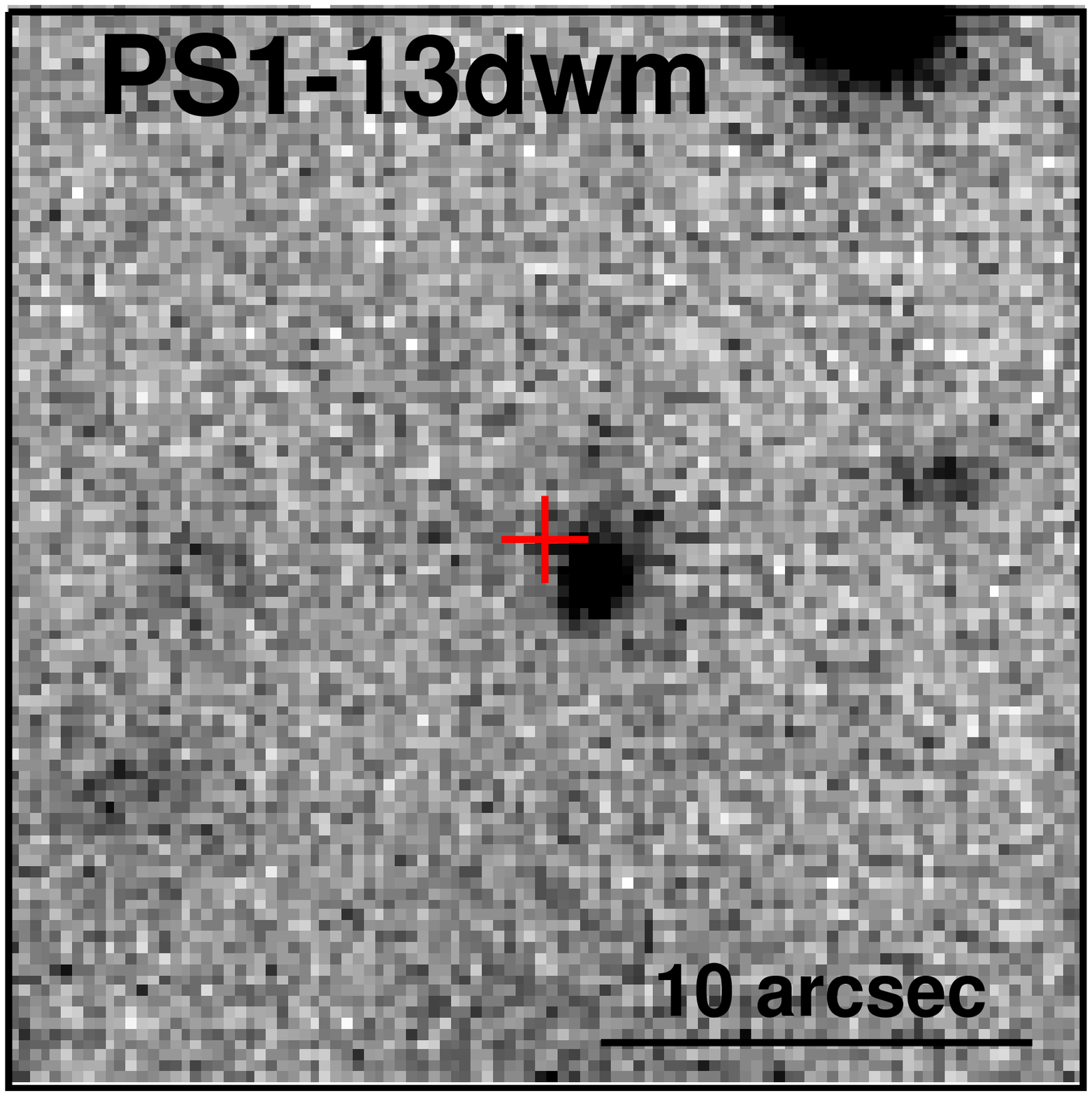}
\includegraphics[width=0.19\textwidth]{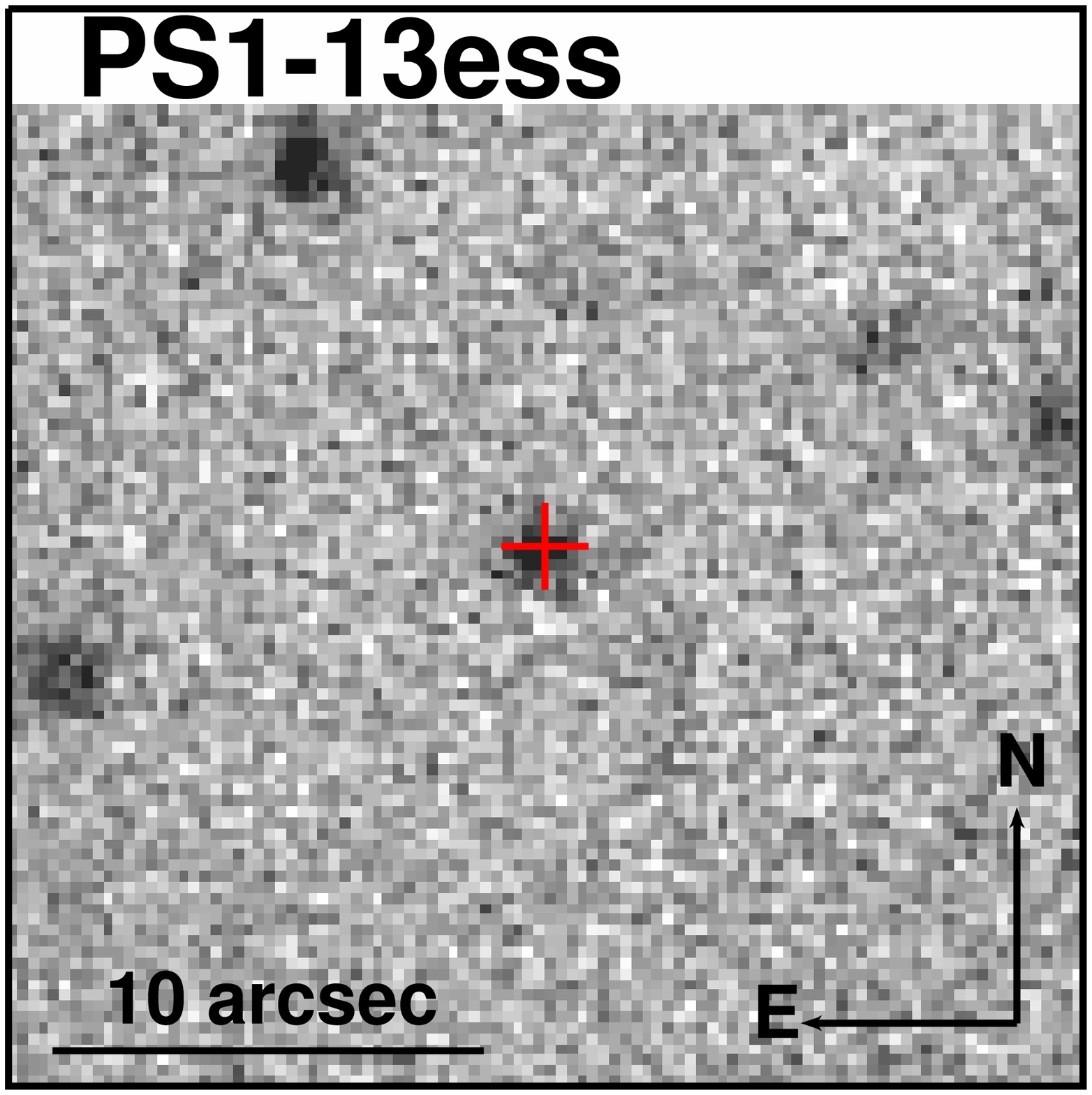}
\end{center}
\caption{Gold and silver sample explosion environments. Images are from PS1 templates.  Transient explosion sites are marked with red crosses. \label{fig:GoldStamp}}
\end{figure*}

\begin{figure*}[ht!]
\begin{center}
\includegraphics[width=\textwidth]{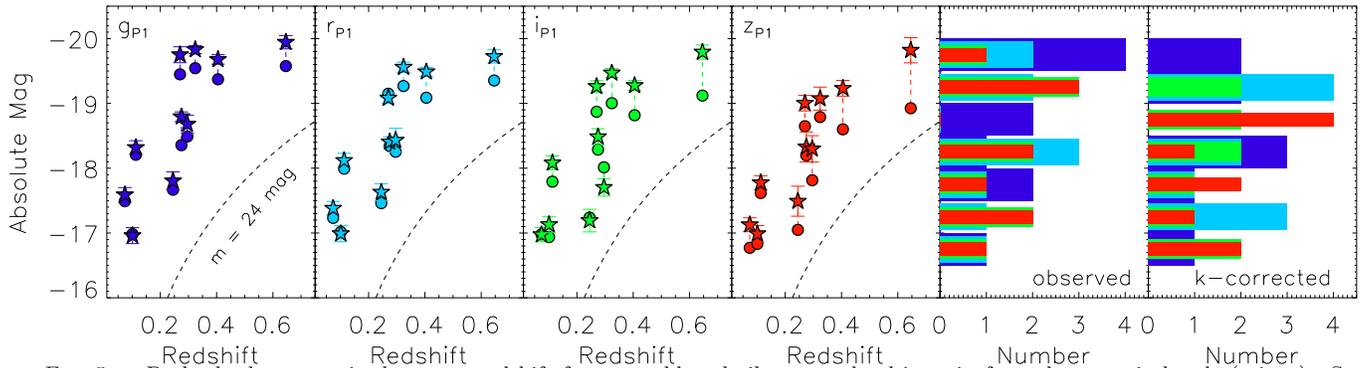}
\caption{Peak absolute magnitude versus redshift for our gold and silver sample objects in four photometric bands (griz$_{\rm{P1}}$).  Stars represent our observed values corrected for distance and extinction while circles have been roughly k-corrected to rest-frame bandpasses (see Section~\ref{sec:kcor}). Dashed lines represent a constant apparent magnitude of 24 mag.  Histograms for both sets of magnitudes are given in the two right most panels (blue $=$ g$_{\rm{P1}}$, cyan $=$ r$_{\rm{P1}}$, green $=$ i$_{\rm{P1}}$, red $=$ z$_{\rm{P1}}$). \label{fig:AbsMag}}
\end{center}
\end{figure*}

Together our gold and silver samples contain 10 objects.  For these events, we obtained spectroscopic redshifts of their host galaxies, allowing us to constrain their true luminosity scale.

Our gold sample is composed of six objects. These are our highest quality events, all of which satisfy the photometric selection criteria listed in Section~\ref{sec:selection}.  The absolute magnitude, rest-frame, light curves for these events are shown in Figure~\ref{fig:GoldPhot1}.  

Our silver sample is composed of four objects.  These are events that were noted as rapidly-evolving during the normal operations of the PS1-MDS but which possess sparser light curves (Figure~\ref{fig:GoldPhot2}).  In all cases the observed light curves are sufficient to characterize them as rapidly-evolving, but sparse enough such that they fail the systematic selection criteria described above.  For instance, PS1-13ess only has one band with a deep limit in the 9 days prior to observed maximum (as opposed to the requisite two).  For a majority of this manuscript, the silver objects will be analyzed with our gold sample, as they further inform the properties of rapidly-evolving transients.  However, in Section~\ref{Sec:Rates} when calculating volumetric rates, only objects which pass the well defined set of selection criteria outlined above will be considered.

In Figure~\ref{fig:GoldStamp} we show the 25$'$$\times$25$'$ region surrounding the gold and silver transients, all of which have an associated host. Narrow emission and absorption lines were used to measure the redshift to each host.  These range from z $=$ 0.074 (PS1-10ah) to z = 0.646 (PS1-11bbq) with a median redshift of z $=$ 0.275. In Table~\ref{tab:basic} we list the redshift, luminosity distance, and Milky Way reddening in the direction of each transient \citep{Schlafly2011}.  Throughout this paper we correct only for Milky Way extinction.  All calculations in this paper assume a flat $\Lambda$CDM cosmology with H$_0$ $=$ 71 km s$^{-1}$ Mpc$^{-1}$, $\Omega$$_m$ $=$ 0.27, and $\Omega$$_\Lambda$ $=$ 0.73.

In Figure~\ref{fig:AbsMag} we plot peak absolute magnitude versus redshift in griz$_{\rm{P1}}$ for the gold/silver transients.  Stars represent our observed magnitudes corrected for distance and MW extinction.  Circles represent absolute magnitudes that have been k-corrected to the rest-frame griz$_{\rm{P1}}$ bandpasses based on best fit blackbodies (see Section~\ref{Sec:Photometry}).  We see that our sample spans a wide range of absolute peak magnitude ($-$17 $>$ M $>$ $-$20).

\subsection{Bronze Sample}

\begin{figure}[!ht]
\begin{center}
\includegraphics[width=0.19\textwidth]{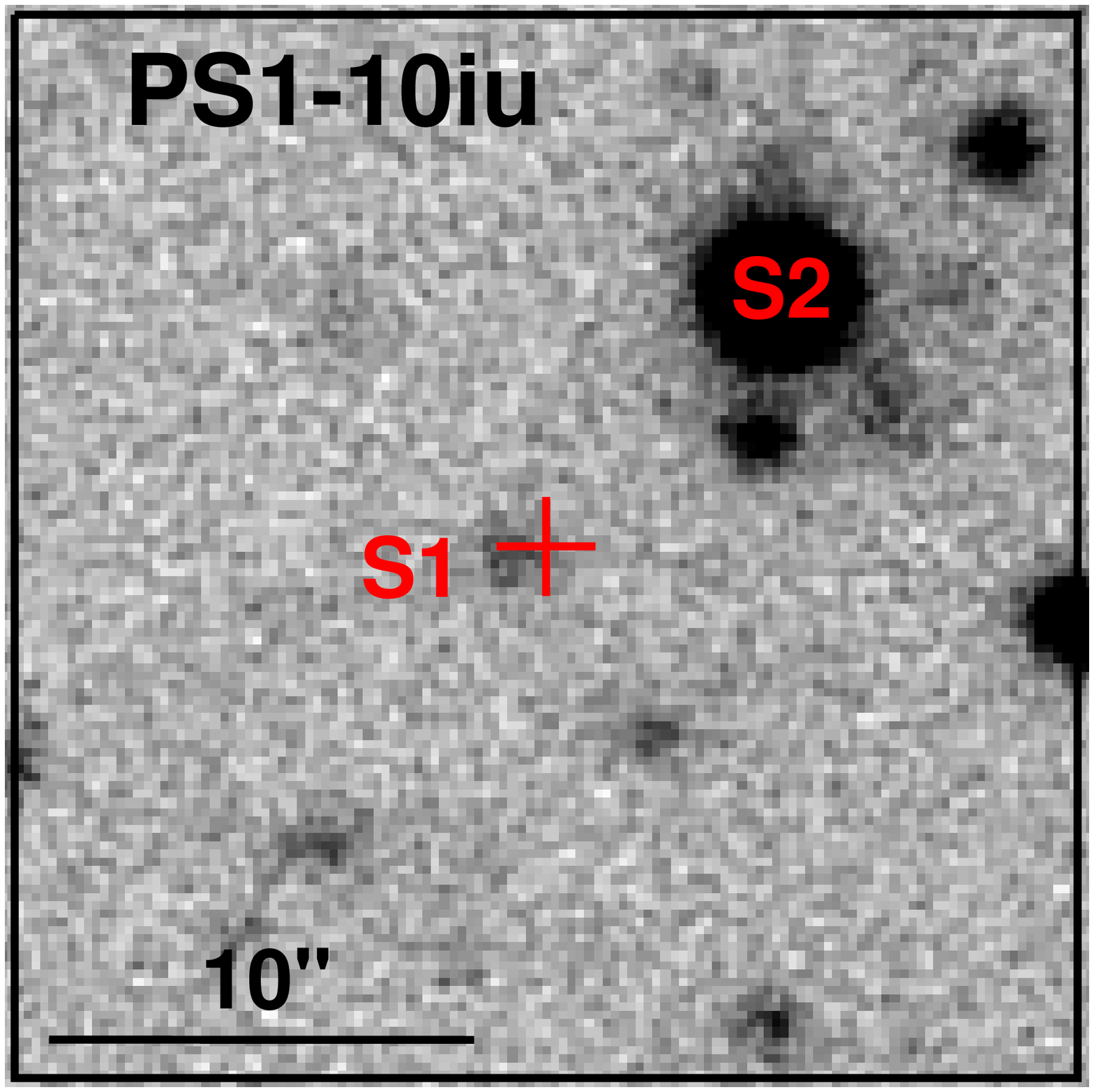}
\includegraphics[width=0.19\textwidth]{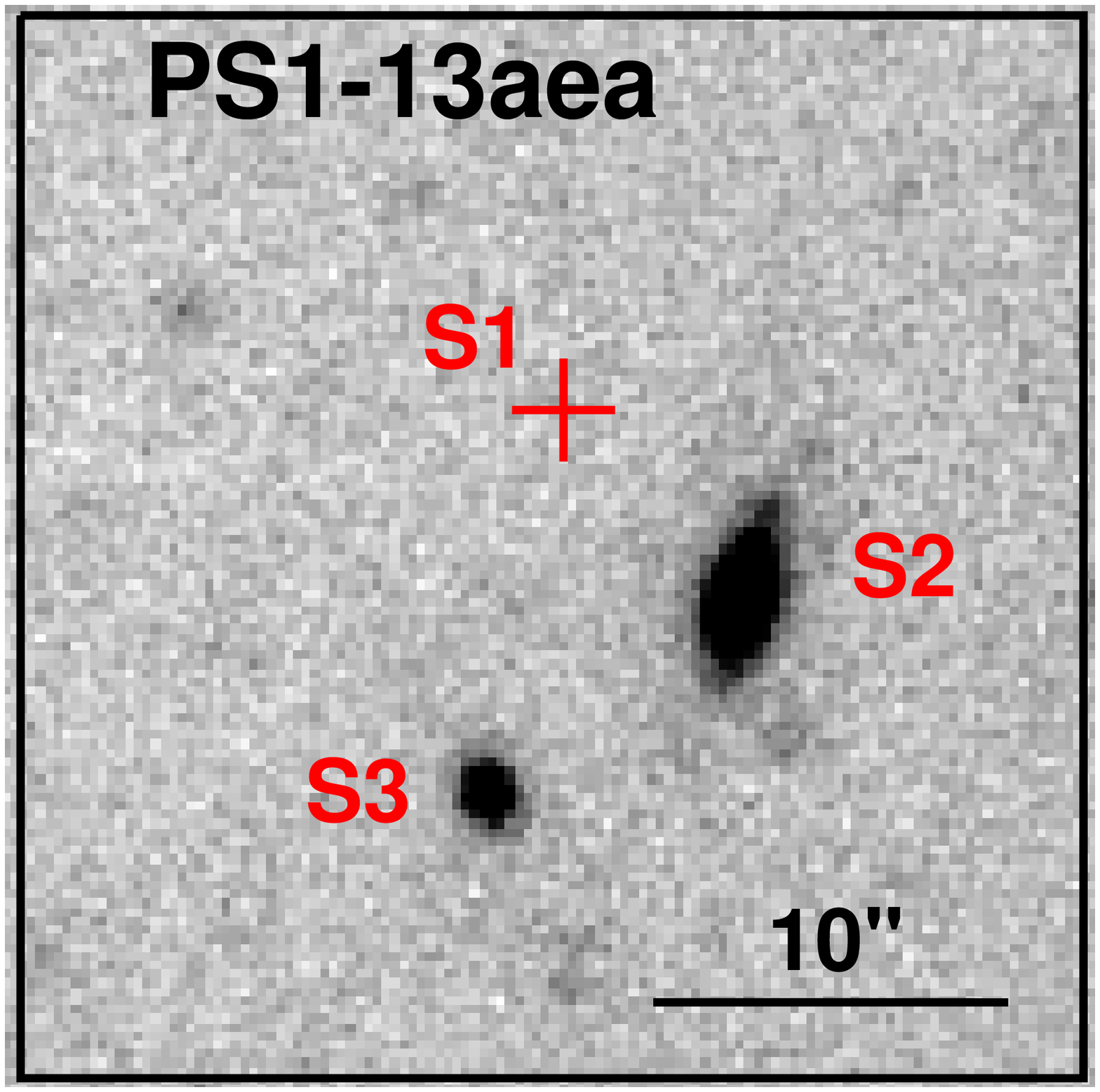}
\includegraphics[width=0.19\textwidth]{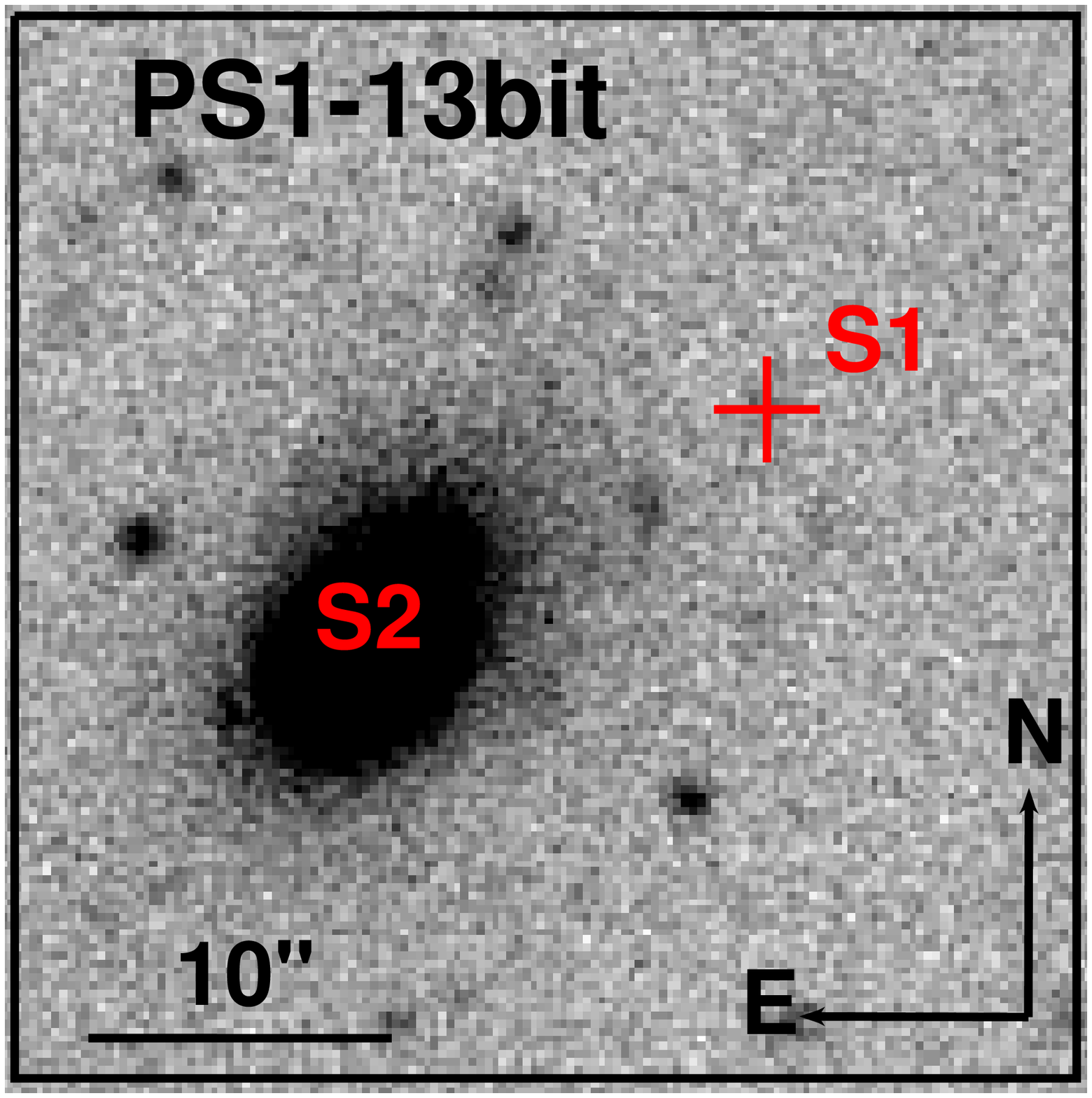}
\includegraphics[width=0.19\textwidth]{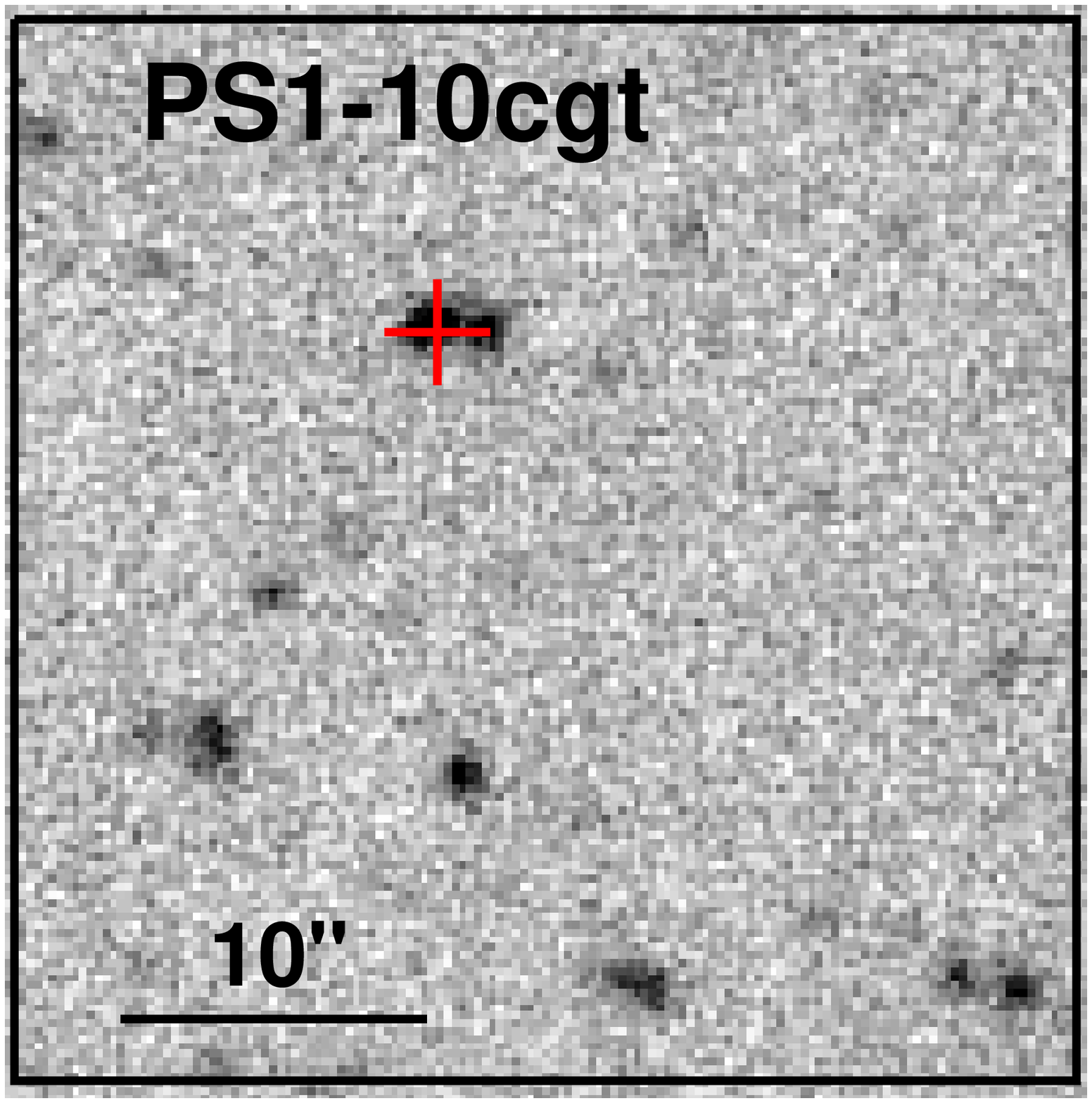}
\includegraphics[width=0.45\textwidth]{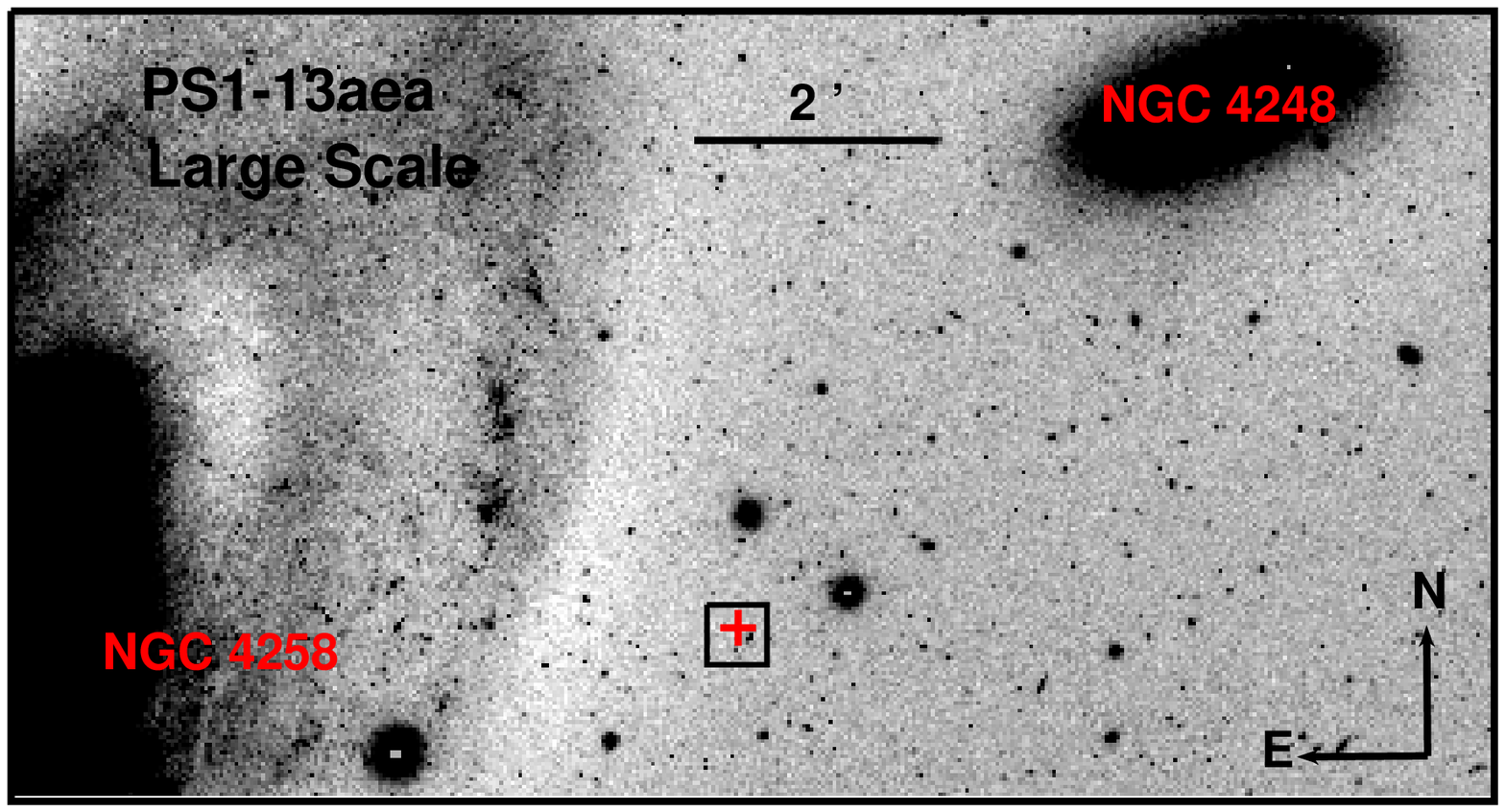}
\end{center}
\caption{\emph{Top and Middle Rows:} Bronze sample explosion environments. All images are from PS1 templates.  Nearby galaxies are  labeled in red.  The location of the transient is indicated by a red cross. \emph{Bottom Row:} The large scale environment around PS1-13aea, which exploded in the region between NGC\,4258 and NGC\,4248. \label{fig:SilverStamp}}
\end{figure}

Our bronze sample is composed of four objects.  These events have light curves that were flagged by the selection criteria in Section~\ref{sec:selection} but for which we were unable to spectroscopically confirm the extragalactic nature of their hosts.  The 25$'$$\times$25$'$ region surrounding our bronze sample is shown in Figure~\ref{fig:SilverStamp}.  All four transients have a faint (25  mag $<$ m$_i$ $<$ 22 mag) underlying source visible in the PS1 deep stacks, although in several cases there is ambiguity about the true host (for instance, PS1-13aea exploded in the region between NGC4258 and NGC4248; see Figure~\ref{fig:SilverStamp}).  In Appendix A we quantify the most likely hosts and discuss the consequences of each probable host for the nature of these transients.  We find it likely that a subset of our bronze sample events are extragalactic, and therefore additional rapidly-evolving and luminous transients.  However, due to the uncertainty in the distance to any individual bronze event, for a majority of this manuscript we will focus on the properties of the gold and silver transients.  

\section{Photometric Properties}\label{Sec:Photometry}

\begin{figure}[!ht]
\begin{center}
\includegraphics[width=\columnwidth]{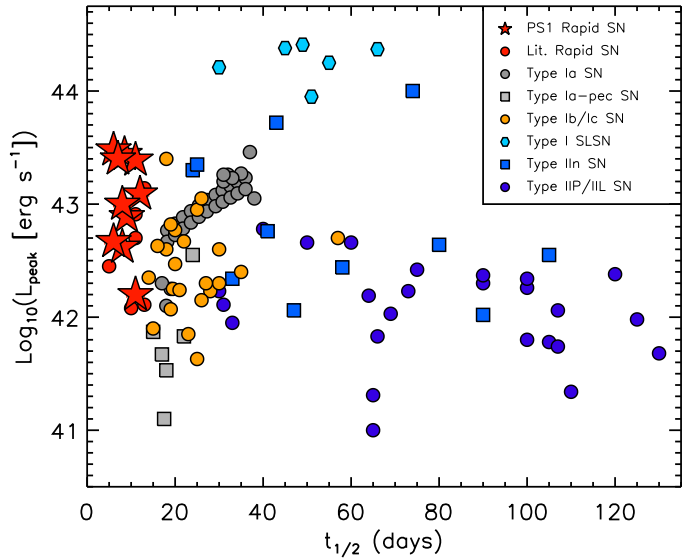}
\caption{The phase space of SN: peak luminosity versus rest-frame time above half-maximum for a variety SN.  The PS1-MDS transients described in this paper are shown as red stars.  They span an order of magnitude in peak luminosity and significantly increase the number of  known transients with short characteristic times.  Other events from the literature are shown as colored circles/squares: type Ia SN \citep{Nugent2002,Taubenberger2008,Scalzo2012}, type Iax \& ``calcium-rich'' Ia \citep{Foley2013a,Kasliwal2012,Perets2010}, type Ib/Ic \citep{Drout2011,Valenti2008,Campana2006,Taubenberger2006,Bersten2012,Olivares2012,Cobb2010,Valenti2012b}, type IIP/IIL \citep{Arcavi2012,Hamuy2003,Andrews2011,Botticella2009}, type IIn \citep{Kiewe2012,Margutti2014}, type I SLSN \citep{Quimby2011,Chomiuk2011,Lunnan2013}, and other rapidly-evolving events \citep[red;][]{Drout2013,Kasliwal2010,Poznanski2010,Ofek2010,Matheson2000}.  \label{fig:PhaseSpace}}
\end{center}
\end{figure}

The gold and silver events presented in this work increase the number of known transients with t$_{1/2} \lesssim 12$ days by approximately a factor of three.  This is illustrated in Figure~\ref{fig:PhaseSpace} where we plot peak pseudo-bolometric luminosity versus t$_{1/2}$ for the objects presented in this work (red stars) and other rapid SN (red circles) in comparison to type Ia, Ia-x, ``Calcium-rich'' Ia, Ib/c, IIn, IIP, and type-I SLSN (see caption for references).  The sample of PS1-MDS transients significantly increases the population of known objects at the shortest timescales and spans a wide range of peak luminosities.  In the this section we discuss the photometric properties of the PS1-MDS identified transients.

\subsection{Timescales}

In Table~\ref{tab:PhotomProps} we summarize the photometric properties of the gold/silver transients, including their observed/absolute magnitudes and several measurement of their rise and decline timescales.  The t$_{1/2\rm{,rise}}$ and t$_{1/2\rm{,decline}}$ timescales as well as the number of magnitudes the light curves decline in the first 15 days post-maximum ($\Delta$m$_{15}$) were determined by linearly interpolating our observed light curve\footnote{Uncertainties were estimated using a Monte Carlo approach to produce 1000 realizations of our light curve, drawing from the uncertainties from of each data point.}. All values are determined with respect to the time of \emph{observed} maximum, and are therefore influenced by the three day cadence in each band. However, this should not strongly affect our qualitative assessment of these objects. The quoted rise time values encompass the entire range of values permitted by the observed photometry/limits in each band. 

\begin{figure}[ht!]
\begin{center}
\includegraphics[width=\columnwidth]{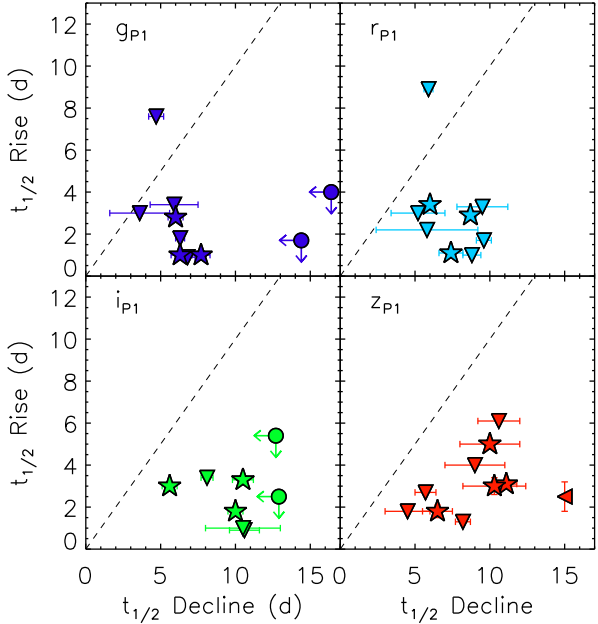}
\end{center}
\caption{t$_{1/2\rm{,rise}}$ versus t$_{1/2\rm{,decline}}$ for the gold and silver transients in four photometric bands (griz$_{\rm{P1}}$).  Stars represent objects with measured rise and decline timescales, triangles represent objects with an upper limit in one of the two timescales, and circles with arrows represent objects with upper limits for both timescales.  In general, the PS1-MDS rapidly-evolving transients rise significantly faster than they decline. \label{fig:CharTimes}}
\end{figure}

In general the transients have faster rise than decline timescales.  This is shown in Figure~\ref{fig:CharTimes} where we plot the t$_{1/2\rm{,rise}}$ versus t$_{1/2\rm{,decline}}$ timescales.  Most events fall well below the dashed line denoting equality, indicating that decline timescale is \emph{not} a good proxy for overall timescale in these events.  In fact, for six events, our data do not exclude a transient that rose on a timescale $\lesssim$1 day.  In contrast, the typical $\Delta$m$_{15}$ values for these transients (1 $-$ 2 mags) are slightly more rapid than typical stripped envelope core-collapse SN ($\Delta$m$_{15} \sim$ 0.5 $-$ 1.0; \citealt{Drout2011}) but slower than the rapidly declining type I SN\,2005ek, SN\,2010X, and SN\,2002bj \citep{Drout2013,Kasliwal2010,Poznanski2010}, which possess $\Delta$m$_{15}$ $\gtrsim$ 3 mag.

Three of the transients (PS1-10ah, PS1-12brf, PS1-13ess) show evidence for a change in decline timescale.  PS1-10ah initially declines very rapidly, with a linear decline rate of $\sim$0.15 mag day$^{-1}$, which then shallows to $\sim$ 0.02 mag day$^{-1}$ around 10 days post-maximum.  PS1-13ess shows an initial decline very similar to PS1-10ah, but then \emph{rises} again as evidenced by our final i$_{\rm{P1}}$ and z$_{\rm{P1}}$ data points.  This may liken it to the double peaked type IIb SN\,1993J (see Section~\ref{Sec:Discussion}). The other objects in our sample show no discernible change in slope out to $\sim$ 10 $-$ 20 days post-maximum.

\begin{figure}[!ht]
\begin{centering}
\includegraphics[width=\columnwidth]{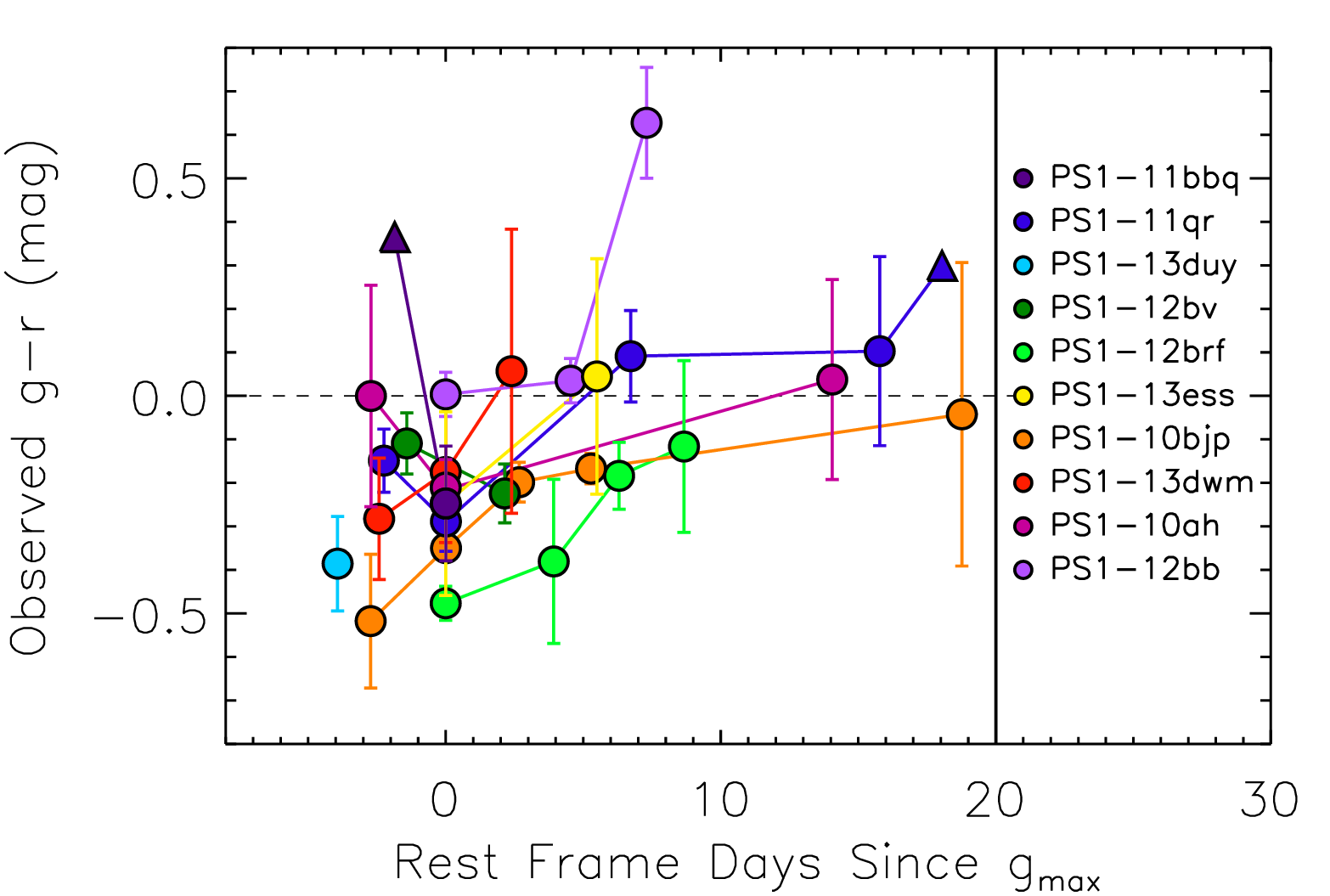}
\includegraphics[width=\columnwidth]{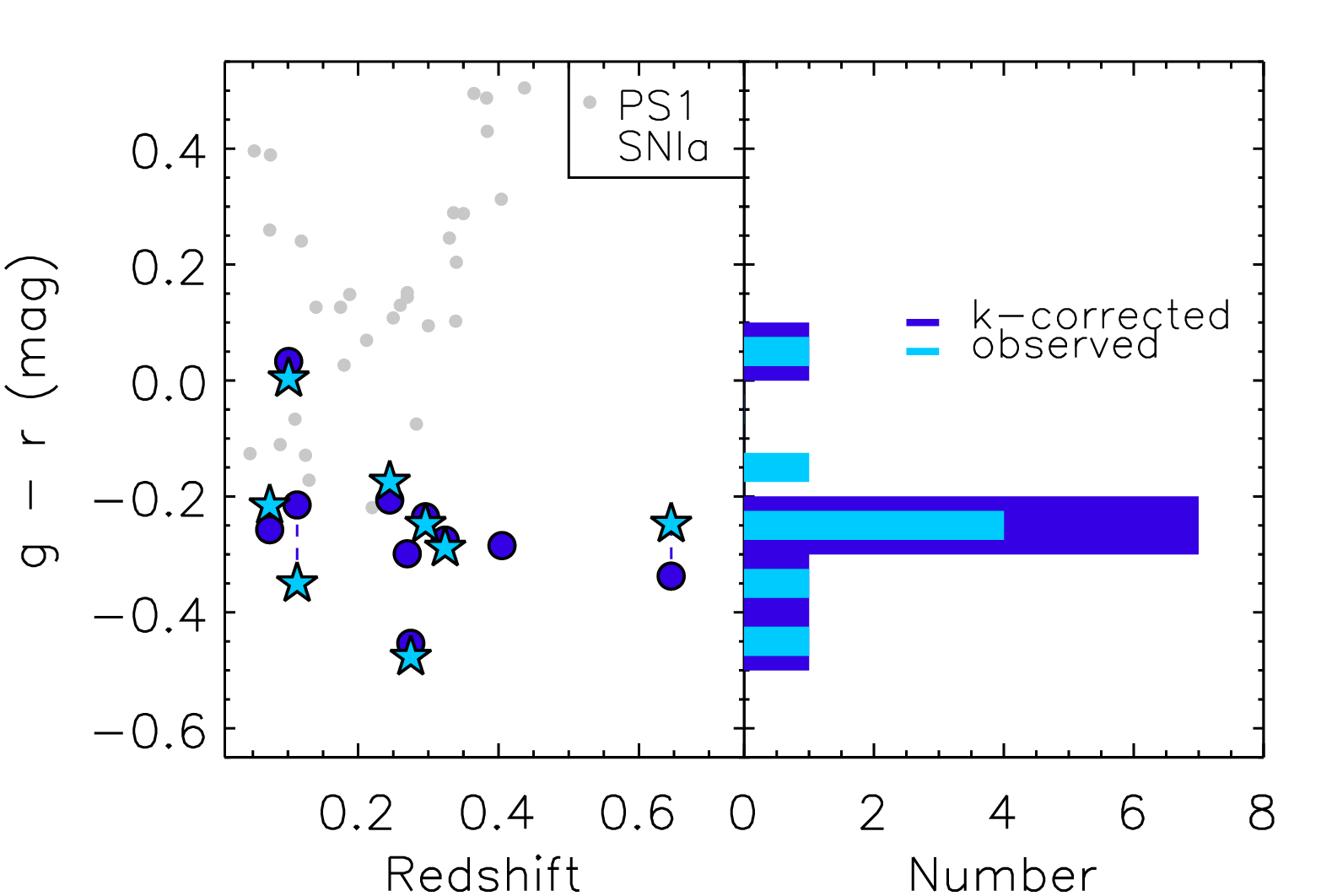}
\end{centering}
\caption{\emph{Top Panel:} Observed g$_{\rm{P1}}$ $-$ r$_{\rm{P1}}$ color evolution for the gold and silver transients.  Triangles mark lower limits.  Colors are initially blue and redden slightly with time.  The g$_{\rm{P1}}$ and r$_{\rm{P1}}$ bandpasses probe slightly different wavelength ranges for each transient. \emph{Bottom Panel:} g$_{\rm{P1}}$ $-$ r$_{\rm{P1}}$ colors at maximum light versus redshift.  Observed values are shown as cyan stars and values k-corrected to rest-frame bandpasses are shown as blue circles (see Section~\ref{sec:kcor}).  Histograms of both are given in the right panel.  A majority of our objects show blue colors with $-$ 0.2 $>$ g$_{\rm{P1}}$ $-$ r$_{\rm{P1}}$ $>$ $-$0.3. Grey dots are a random selection of PS1-MDS type Ia SN. \label{fig:grcolor}}
\end{figure}

\subsection{Colors and SEDs}

In the normal operation mode of the PS1-MDS only g$_{P1}$ and r$_{P1}$ observations are acquired on the same day, with i$_{P1}$ and z$_{P1}$ observations following on consecutive evenings.  In the upper panel of Figure~\ref{fig:grcolor} we plot the g$_{P1}$ $-$ r$_{p1}$ colors from every epoch on which both bands were observed.  The colors are generally blue (g$_{P1}$ $-$ r$_{p1}$ $\lesssim$ 0.0) and redden slightly with time.  In the lower panel of Figure~\ref{fig:grcolor} we plot g$_{\rm{P1}}$ $-$ r$_{\rm{P1}}$ color at maximum as a function of transient redshift.  A majority of objects possess $-$0.2 $>$ g$_{P1}$ $-$ r$_{P1}$  $>$ $-$0.3. Only PS1-12bb shows significantly redder colors with g$_{P1}$ $-$ r$_{p1}$  $=$ 0.0. For comparison, we also show a randomly selected sample of PS1 type Ia SN (grey dots).  The type Ia SN are significantly redder than our rapidly-evolving transients, especially at redshifts greater than z$=$0.2 where the g$_{\rm{P1}}$ and r$_{\rm{P1}}$ filters probe the portion of the type Ia spectrum where line blanketing greatly reduces the UV flux. 

\begin{figure}[!ht]
\includegraphics[width=\columnwidth]{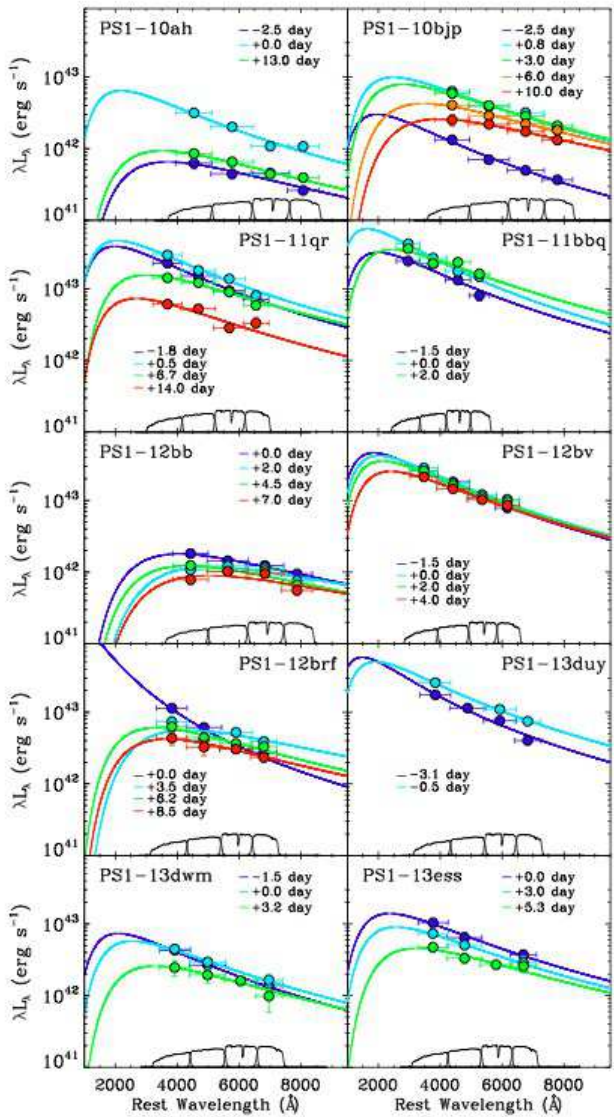}
\caption{Multi-epoch SEDs for the gold and silver transients.  Best fit blackbody curves are shown as solid lines. The griz$_{\rm{P1}}$ bandpasses are plotted on the bottom axis for each event. (A higher resolution version of this figure is available though the journal publication.)  \label{fig:GoldSED}}
\end{figure}

In order to gain a more complete view of the SED evolution of our sample, we interpolate our griz$_{P1}$ light curves to a set of common epochs.  The uncertainly involved in such an interpolation is accentuated due to the rapidly-evolving nature of our sample, and we take care to only interpolate to epochs for which there was a cluster of griz$_{P1}$ observations within a few days.  In Figure~\ref{fig:GoldSED} we plot the SEDs.  For a majority of the events, no turnoff is seen in the SED at blue wavelengths.  This is true even for our highest redshift object, PS1-11bbq, for whom the g$_{\rm{P1}}$ extends out to a rest fame wavelength of $\sim$2500 \AA.

\subsection{Temperature and Radius Evolution}\label{sec:kcor}

\begin{figure}[!ht]
\includegraphics[width=\columnwidth]{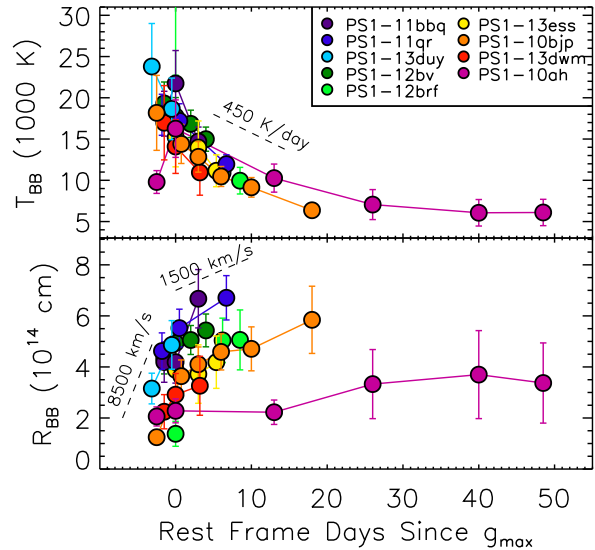}
\caption{Best fit blackbody temperatures (top panel) and radii (bottom panel) as a function of time.  In general, temperatures evolve from around 20,000 K near maximum to 7,000 K at later times. Exceptions include PS1-12brf (whose initial temperature is around 50,000 K), PS1-10ah (whose temperature rise to maximum light) and PS1-12bb (whose best-fit temperature remains constant around 8000 K; not plotted).  Radii are a few time 10$^{14}$ cm, and expand with time. \label{fig:TempRad}}
\end{figure}

Also shown in Figure~\ref{fig:GoldSED} are the best fit blackbody curves for each epoch.  For nine (of ten) transients the best fit blackbodies yield temperatures and radii that cool/expand with time, as one expects for an explosion with expanding ejecta (e.g. SN). The best fit blackbody temperatures and radii for these events are plotted as a function of time in Figure~\ref{fig:TempRad}.  On average, the temperatures cool from $\sim$20,000 K near maximum to $\sim$7,000 K at $\sim$20 days, and the radii expand from $\sim$10$^{14}$ cm to a few $\times$10$^{14}$ during the time period over which they are constrained. Dashed lines in Figure~\ref{fig:TempRad} show fiducial cooling rates and photosphere expansion speeds, which are based on our best observed object: PS1-10bjp (orange).  The presence of extragalactic extinction along the line of sight would slightly increase these temperatures and radii, but would not affect our conclusion that their evolution is consistent with explosions that possess an expanding ejecta.

The temperature/radius evolution of three transients warrant further note: (1) PS1-12brf displays an initial best fit blackbody temperature of $\sim$50,000 K.  This exact value is uncertain because our data lies far down the Rayleigh-Jeans tail, but from Figure~\ref{fig:GoldPhot1} we see that it possesses extreme colors near maximum light.  This is also the only transient for which our data shows longer wavelengths reaching maximum at later times. (2) PS1-10ah shows tentative evidence for a temperature that \emph{rises} to maximum and then cools. (3) PS1-12bb is our only object that does not show a characteristic cooling/expanding behavior.  The best fit blackbody temperature shows little variation and is consistent with $\sim$ 7,000 K from 0 $<$ t $<$ 20 days. 

Finally, by assuming the spectra of our objects can be approximated by these best fit blackbodies, we can perform rough k-corrections on our observed magnitudes to rest-frame griz$_{\rm{P1}}$ bandpasses.  This will allow us to directly compare the magnitudes observed for objects with differing redshifts. We perform synthetic PS1 photometry on these blackbodies and find the ``rest'' griz$_{P1}$ peak magnitudes listed in Table~\ref{tab:PhotomProps}. These ``k-corrected'' peak magnitudes are shown as circles in Figure~\ref{fig:AbsMag} and are used to produce ``rest-frame'' g$_{\rm{P1}}$ $-$ r$_{\rm{P1}}$ colors, which are plotted in Figure~\ref{fig:grcolor}.

\subsection{Pseudo-Bolometric Light Curves} \label{Sec:Bolo}

\begin{figure}[!ht]
\includegraphics[width=\columnwidth]{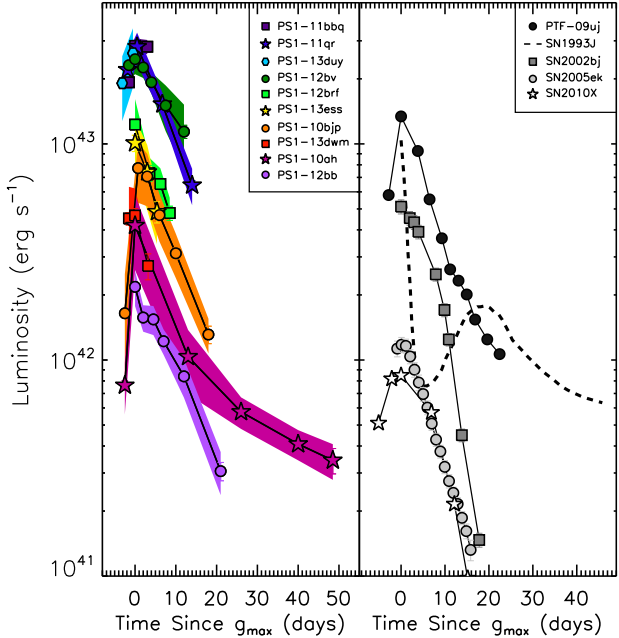}
\caption{\emph{Left Panel:} Pseudo-bolometric light curves for the gold and silver transients. \emph{Right Panel:} Pseudo-bolometric light curves for other rapidly-evolving transients from the literature: the type Ic SN 2005ek \citep{Drout2013} and 2010X \citep{Kasliwal2010}, the type Ib SN\,2002bj \citep{Poznanski2010}, the type IIb SN\,1993J \citep{Schmidt1993}, and the type IIn PTF09uj \citep{Ofek2010}.  \label{fig:BoloLC}}
\end{figure}

To constrain the total energy radiated in these explosions, we construct pseudo-bolometric light curves from our multi-band data.  We first perform a trapezoidal interpolation to our observed griz$_{P1}$ light curves and then account for missing IR flux by attaching a blackbody tail based on the best fit values described above.  In order to account for missing UV flux and the varied wavelength coverage of our observations (due to differing redshifts) we then extend the same blackbody from the edge of the observed g$_{P1}$ band back to 2500 \AA\ (rest-frame). This represents the blue edge of the wavelength range covered by our highest redshift event, and the approximate range of our spectra in Section~\ref{Sec:Spectra}. 

Using this formulation, our peak pseudo-bolometric luminosities span a range of approximately 2$\times$10$^{42}$ ergs s$^{-1}$ $<$ L $<$ to 3$\times$10$^{43}$ ergs s$^{-1}$, and are plotted in Figure~\ref{fig:PhaseSpace}.  If we had instead utilized a UV bolometric correction that integrated the entire best fit blackbody, these values would be approximately a factor of 2 higher.  In the left panel of Figure~\ref{fig:BoloLC} we plot our derived pseudo-bolometric light curves. The number of epochs for which we can constrain the luminosity is limited for the silver transients. The energy radiated by the six gold transients between $-$3 and +20 days ranged from 2 $\times 10^{48}$ ergs (PS1-12bb) to 2 $\times 10^{49}$ ergs (PS1-11qr).

\begin{figure}[ht!]
\begin{center}
\includegraphics[width=\columnwidth]{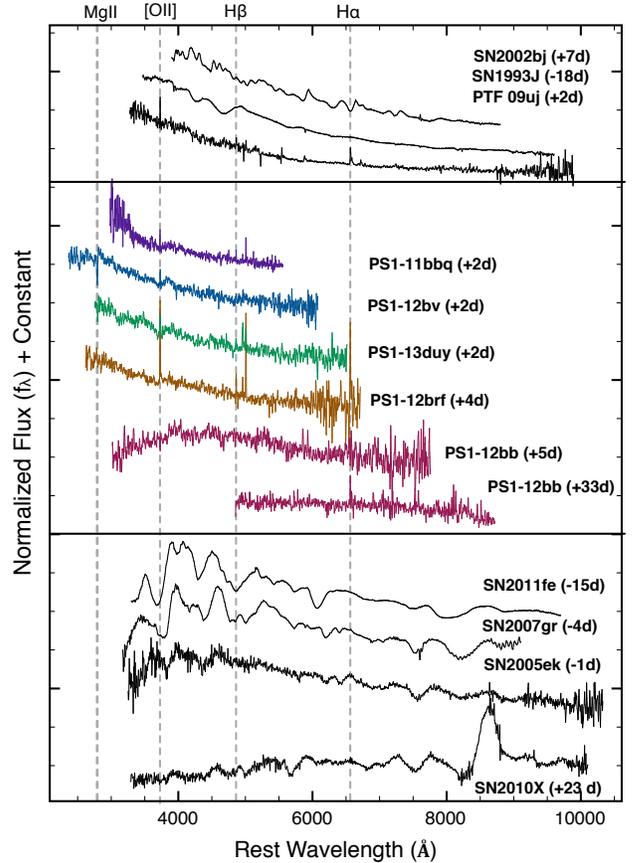}
\caption{Explosion spectra for five PS1-MDS transients (colored) in comparison to events from the literature (black).  With the exception of PS1-12bb our events are dominated by a blue continuum, with a lack of strong P Cygni features.  Some contributions from the host galaxy (e.g.\ nebular emission lines) are still present in these events. PS1-12bb shows a redder continuum and a notable lack of broad nebular features at $+$33 days. The top panel shows literature objects thought to be powered by cooling envelope emission/interaction while the bottom panel shows objects powered by radioactive decay (type Ib SN\,2002bj \citealt{Poznanski2010}; type IIb SN\,1993J \citealt{Barbon1995}; type IIn PTF\,09uj \citealt{Ofek2010}; type Ia SN\,2011fe \citealt{Pereira2013}; type Ic SN\,2007gr \citealt{Valenti2008}; type Ic SN\,2005ek \citealt{Drout2013}; type Ic SN\,2010X \citealt{Kasliwal2010}). \label{fig:ExSpec}}
\end{center}
\end{figure}

For comparison, we also show the pseudo-bolometric light curves of several other rapidly-evolving events (right panel): SN\,2002bj \citep{Poznanski2010}, PTF\,09uj \citep{Ofek2010}, SN\,2010X \citep{Kasliwal2010} and SN\,2005ek \citep{Drout2013}, as well as the double peaked SN\,1993J \citep{Schmidt1993}.  The curves for SN\,2005ek, SN\,2002bj, and SN\,1993J were calculated from multi-band photometry in a manner similar to that utilized here, while the curves for SN\,2010X and PTF\,09uj were derived based on r-band data only.  We note that SN\,2005ek and SN\,2010X, which have been proposed to be powered mainly by radioactive decay \citep{Drout2013,Kasliwal2010,Tauris2013}, are less luminous than the PS1-MDS events, while the more luminous PTF\,09uj and the first peak of SN\,1993J are thought to be powered by cooling envelope emission/interaction. These power sources will be discussed further in Section~\ref{Sec:Discussion}.

\section{Spectroscopic Properties}\label{Sec:Spectra}

We obtained spectra during outburst for five transients: PS1-11bbq, PS1-12bb, PS1-12bv, PS1-12brf, and PS1-13duy.  These spectra are plotted in Figure~\ref{fig:ExSpec}.  One spectrum was obtained between two and four days post maximum for each event and, in the case of PS1-12bb, a second spectrum was obtained at +33 days.  The spectra shown for PS1-11bbq, PS1-12bv, PS1-12brf, and PS1-13duy still contain some contribution from their host galaxies, as evidenced from presence of nebular emission lines.  However, both the lack of a 4000 \AA\ break and the faint apparent magnitude of all four hosts (23 $-$ 25 mag) give us confidence that a majority of the continuum is due to the transient itself.  

\subsection{Basic Properties and Comparison to other Events}

\begin{figure}[ht!]
\begin{center}
\includegraphics[width=\columnwidth]{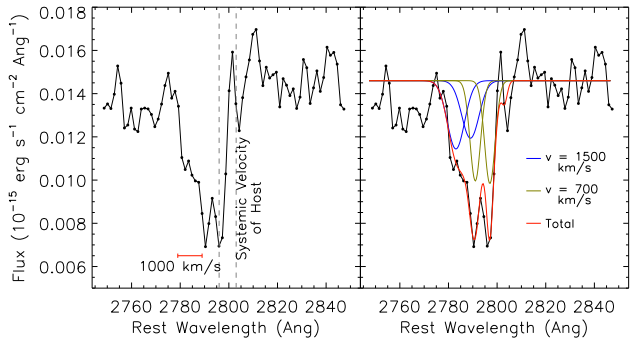}
\caption{Mg II absorption in the spectrum of PS1-12bv.  \emph{Left:} Vertical lines mark the systemic velocities of the host galaxy as measured from [OII] $\lambda$3727 and H$\beta$ in emission.  The red line signifies the extent of the resolved blue ``shelf'', which may be due to absorption in the CSM surrounding the progenitor.  \emph{Right:} An example multiple component fit to the absorption feature.  The green (blue) components are blue shifted $\sim$700 km s$^{-1}$ (1500 km s$^{-1}$) with respect to the host galaxy and have a FWHM of 550 km s$^{-1}$ (900 km s$^{-1}$). The galaxy [OII] $\lambda$ 3727 feature has a FWHM of 300 km s$^{-1}$. \label{fig:MgII}}
\end{center}
\end{figure}

From Figure~\ref{fig:ExSpec} we see that the spectra are dominated by continua as opposed to strong P Cygni features.  Four of the events, PS1-11bbq, PS1-12bv, PS1-12brf and PS1-13duy, show blue continua\footnote{A spectrum of the bronze sample object PS1-13bit was obtained using the OSIRIS instrument at the Gran
Telescopio Canarias under programme GTC2007-12ESO (PI: R. Kotak). Private communication from Rubina Kotak reveals that it is similarly dominated by a blue continuum with no narrow emission or absorption features.}, while PS1-12bb is consistent with a blackbody of 6000 $-$ 7000 K.  PS1-12bb also showed redder photometric colors and a distinct SED evolution (Section~\ref{Sec:Photometry}).  Of the events, PS1-12bv shows the strongest evidence for broad spectral features, with the most evident near 3900 \AA.

Spectra dominated by blue continua have been observed in previous SN.  They are typically found in objects that are hot, optically thick, and powered by cooling envelope emission or interaction.  For instance, although SN\,1993J later went on to develop a plethora of P Cygni features indicative of a type IIb SN, during its initial light curve peak (attributed to cooling envelope emission and recombination) its spectrum was dominated by continua with only a few features between 4000 $-$ 5000 \AA. Hydrogen lines were not evident in the early spectra.  In addition, the rapidly-evolving type IIn SN PTF\,09uj (attributed to shock breakout/interaction) has only weak narrow emission lines in its spectra near maximum, and is dominated by a blue continuum.  Given the signal to noise of our spectra, it is unlikely that we would be able to identify similar emission features in our events if they were present.  Thus, based on the quality and coverage of our spectra, \emph{we are unable to conclude whether the objects in our sample are hydrogen-rich or hydrogen-poor.}  We plot spectra of SN\,1993J and PTF\,09uj, as well as the blue rapidly-evolving SN\,2002bj (whose power source is still under debate), in the top panel of Figure~\ref{fig:ExSpec}.

\begin{figure}[!ht]
\includegraphics[width=\columnwidth]{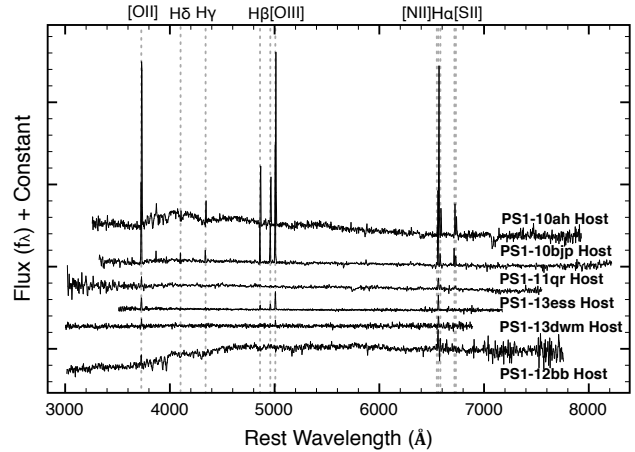}
\caption{Host galaxy spectra for six of the gold/silver transients.  Nebular emission lines are labeled by dashed vertical lines. \label{fig:HostSpec}}
\end{figure}

In contrast, SN that are powered mainly by radioactive decay (type Ia/b/c SN) are typically dominated by P Cygni features formed in the outer ejecta layers.  In particular, they often possess strong line blanketing in the blue due to the presence of iron peak elements.  We plot several examples in the lower panel of Figure~\ref{fig:ExSpec}. We see that our spectra more closely resemble events that are powered by interaction/recombination, as opposed to radioactive decay. 

\subsection{Lack of Nebular Features in PS1-12bb}

We obtained two spectra for PS1-12bb. At the time of the second epoch (+33 days) the $i-$band light curve of PS1-12bb had declined by $\sim$3 mag.  Similar to the spectrum obtained near maximum light, this spectrum is dominated by continuum (with some contribution from its host galaxy's light).  This is in contrast to any other ``late time'' spectra of a rapidly-declining SN that has been obtained to date.  Both SN\,2005ek \citep{Drout2013} and SN\,2010X \citep{Kasliwal2010} were observed between 10 and 35 days post-maximum.  Both events (type Ic) displayed a growing emission component in the Ca II NIR triplet.  By +23 days the spectrum of SN\,2010X was dominated by the Ca II NIR triplet between 8000 and 9000 \AA.  This is not the case for PS1-12bb.  We can place a limit on the luminosity in the Ca II NIR triplet between 8300$-$8700 \AA\ at $+$33 days in PS1-12bb of $<$ 3 $\times$ 10$^{39}$ erg s$^{-1}$.  This is approximately 4 times lower than the feature observed in SN\,2010X at a similar time. PS1-12bb was also our only object whose best fit blackbodies did not show radii that expanded with time.  Thus, this transient may be of a different kind than a SN with an expanding ejecta that eventually becomes optically thin.

\subsection{Potential CSM absorption in PS1-12bv}

The emission lines present in Figure~\ref{fig:ExSpec} are unresolved and attributed to the host galaxies. However, in the spectrum of PS1-12bv there is tentative evidence for CSM absorption in the Mg II ($\lambda \lambda$ 2796,2803) feature.  A view around this feature is shown in the left panel of Figure~\ref{fig:MgII}.  The spectrum of PS1-12bv shows unresolved nebular emission lines of [OII] $\lambda$3727 and H$\beta$ at a redshift of z $=$ 0.405.  In contrast, the two deepest features of the Mg II ($\lambda \lambda$ 2796,2803) absorption feature are blue shifted by $\sim$ 700 km s$^{-1}$ relative to the galaxy emission lines (z $=$ 0.402) and there is a resolved blue wing to the feature.  This blue wing spans an \emph{additional} $\sim$1000 km s$^{-1}$ from the minimum of the absorption feature.  Our resolution corresponds to $\sim$ 400 km s$^{-1}$ in this wavelength region.

\begin{figure*}[ht!]
\begin{center}
\includegraphics[width=\textwidth]{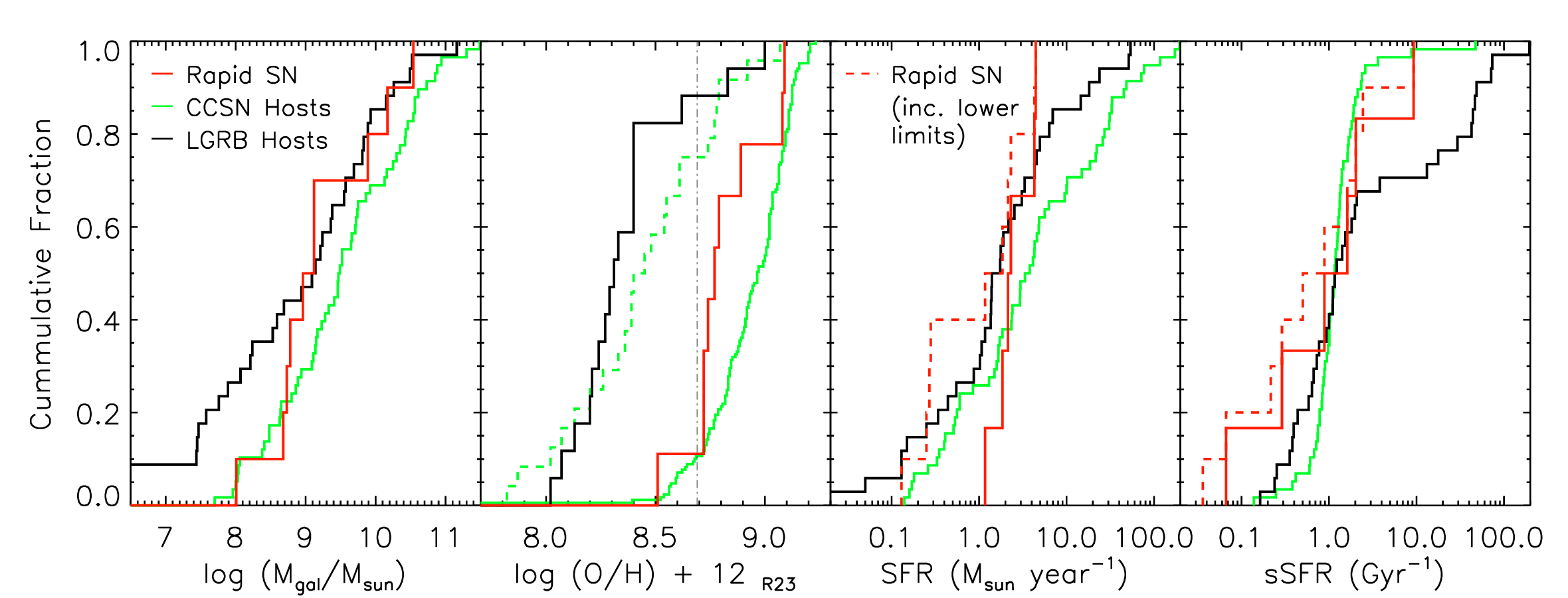}
\caption{Cumulative distributions of various host galaxy properties: stellar mass, metallicity, star formation rate, and specific star formation rate. In all panels the red line(s) represents the rapid evolving SN presented in this work. In panels 3 and 4 the solid red line represents the six gold/silver objects with host galaxy spectra and the dashed red line includes lower limits based on the explosion spectra of the remaining four objects.  Also shown are distributions from LGRBs and core-collapse SN (black and green lines, respectively; see text for references). The dashed green line in panel 2 is the \emph{untargeted} Type Ibc SN from \citet{Sanders2012a}. The vertical dashed line in panel 2 signifies solar metallicity.  \label{fig:HostCums}}
\end{center}
\end{figure*}

\begin{figure}[!ht]
\includegraphics[width=\columnwidth]{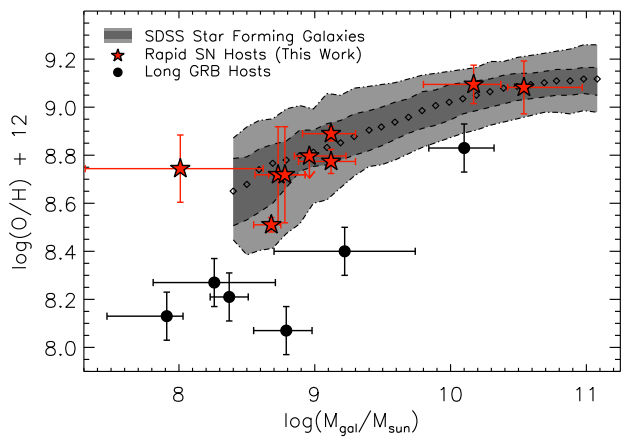}
\caption{Mass-metallicity relation for nine gold/silver transients (red stars).  Also plotted are contours representing the 53000 SDSS star forming galaxies from \citet{Tremonti2004} (shaded region, lines represent the 2.5, 16, 50, 84, and 97.5 percentile of the distribution in each bin) and the long duration GRB hosts from \citet{Levesque2010d}.  Unlike the LGRB hosts, which are offset to lower metallicities, our sample is consistent with being drawn from the greater SDSS population. \label{fig:MassMet}}
\end{figure}

A similar blue absorption wing was seen in the spectra of the type IIn SN\,1998S \citep{Bowen2000}, although in this case the wing spanned only $\sim$ 350 km s$^{-1}$ from the absorption minimum.  Both \citet{Bowen2000} and \citet{Chugai2002} credit this ``shelf'' of absorption to the CSM around the explosion and \citet{Chugai2002} invoked radiative driving to accelerate relatively slow moving CSM material (40 $-$ 50 km s$^{-1}$) to these velocities.  In our case, if the blue component is due to CSM absorption, we would require either stronger radiative driving or a faster CSM wind speed.  The CSM wind speed found for SN\,1998S was typical for a RSG wind, whereas the speeds required for PS1-12bv would be more indicative of a wind from a WR or LBV star.  PS1-12bv is one of our four most luminous objects and is very similar photometrically to PS1-11qr, PS1-11bbq, PS1-13duy and PTF\,09uj (which was hypothesized to be due to the shock breakout from an optically thick wind; \citealt{Ofek2010}). This broad feature may therefore point to a progenitor star that possesses a strong wind in the years before explosion for these transients.

Alternatively, this feature could be explained by either a complex ISM absorption system along the line of sight or an outflow from the host galaxy.  In order to investigate this possibility, we fit pairs of Mg II ($\lambda \lambda$ 2796,2803) doublets with varying redshifts and full-width at half-maximum (FWHM) to this portion of the spectrum. An example fit is shown in the right panel of Figure~\ref{fig:MgII}.  Blue and green lines show the individual Mg II ($\lambda \lambda$ 2796,2803) doublets, and the red line shows their sum.  In all cases, we required absorption components that were both resolved and significantly broader than typical ISM absorptions ($\lesssim$ 100 km s$^{-1}$).

\section{Host Galaxies}\label{Sec:Hosts}

Our method of selecting rapidly-evolving SN was not designed to select transients from a single explosion mechanism/progenitor. However, on the whole, the objects we identified within the PS1-MDS have a similar set of photometric and spectroscopic properties (blue colors, spectra dominated by continua, etc.).  We now turn to the host galaxies of these explosions to see what insight they can provide into their progenitor populations. In Figure~\ref{fig:HostSpec} we display the host galaxy spectra obtained for six of the transients.  These are supplemented by the host galaxy lines present in the explosion spectra (Figure~\ref{fig:ExSpec}).  Nebular emission lines, indicative of active star formation, are found in the spectra of all ten galaxies.  In the sections below we examine the masses, metallicities, star formation rates, and explosion site offsets of the host galaxies in comparison to the hosts of other classes of transients.

\subsection{Mass}

After correcting the host griz-band photometry for Milky Way extinction we use the FAST stellar population synthesis code \citep{Kriek2009} to calculate the total stellar mass contained within our host galaxies.  Our models utilized the \citet{Maraston2005} stellar library and assumed an exponential star formation history and Salpeter IMF.  The total extinction was restricted to $<$ 0.05 (motivated by the Balmer decrement in our highest quality host spectra and the blue colors of our transients; the assumed extinction does not significantly affect the resulting mass).  The results from this analysis are listed in Table~\ref{tab:hosts}.

The total stellar mass in the host galaxies ranges from 8.0 $<$ $\log$ (M/M$_\odot$) $<$ 10.6 with a median value of $\log$ (M/M$_\odot$) $=$ 9.0.  In the first panel of Figure~\ref{fig:HostCums} we plot the cumulative distribution of our host masses in comparison to the core collapse SN (CC-SN) and long gamma-ray burst (LGRB) samples from \citet{Svensson2010}.   A larger fraction of our sample's hosts appear at relatively low masses in comparison to the core collapse SN. A Kolmogorov-Smirnov (KS) test yields a 14\% (37\%) probability that the rapidly-evolving transients hosts are drawn from the same population as the CC-SN (LGRB) hosts. Thus, while our host galaxies skew slightly toward lower masses, there is no statistical evidence for a different parent population from either LGRBs or CC-SN.

\subsection{Metallicity}

Using the Markov Chain Monte Carlo method described in \citet{Sanders2012a} we measure the emission line fluxes and metallicities of nine of the host galaxies.  Due to the varying wavelength coverage and quality of our spectra, it was not possible to use the same strong line diagnostic for all of the galaxies.  In Table~\ref{tab:hosts} we list both the measured value and the diagnostic utilized. For the purposes of comparison, we then convert all of the values to the R23 system of \citet{Kewley2002} using the calibration relations from \citet{Kewley2008}.   

Many of the host galaxies have metallicities that are roughly solar, with a median value of $\log$ (O/H) $+$ 12 $=$ 8.8.  In the second panel of Figure~\ref{fig:HostCums} we plot their cumulative distribution.  Our hosts are offset to a significantly higher metallicity than either LGRB hosts \citep{Svensson2010,Savaglio2009,Levesque2010a,Levesque2010d,Graham2013} or the untargeted type Ibc hosts from \citet{Sanders2012a} (black line and dashed green line, respectively) with a $\lesssim$0.5\% probability of being drawn from the same population.  The entire core-collapse SN sample from \citet{Kelly2011} (solid green line) is shifted to even higher metallicities than our sample, although we caution a fraction of these events were discovered by targeted surveys (which bias towards higher metallicities).

\subsection{Mass-Metallicity Relation}

\begin{figure}[!ht]
\includegraphics[width=\columnwidth]{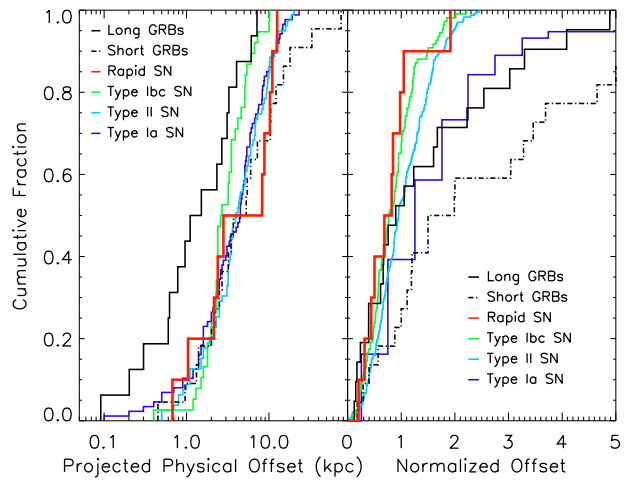}
\caption{Projected host offsets for our sample (red) in comparison to other astronomical transients.  \emph{Left Panel:} physical offsets \emph{Right Panel:} offsets normalized by the g-band half-light radii of the hosts.  Our sample most closely traces the distribution of core-collapse SN offsets. \label{fig:HostOffset}}
\end{figure}

In Figure~\ref{fig:MassMet} we plot the stellar mass versus metallicity for nine of the transients (red stars) versus the $\sim$ 53000 SDSS star forming galaxies from SDSS \citep[shaded regions]{Tremonti2004}.  Also shown are the low redshift LGRB hosts from \citet[black]{Levesque2010d}.  \emph{The hosts of the PS1-MDS rapidly-evolving SN appear to be consistent with the bulk of star-forming galaxies in SDSS.}  This is in contrast to the hosts of LGRBs and type I SLSN, both of which have been shown to obey a relation in the mass-metallicity plane that is offset below the bulk of star-forming galaxies \citep{Levesque2010d,Lunnan2014}.

\subsection{Star Formation Rates}

We estimate host galaxy star formation rates (SFRs) for the six events for which we possess galaxy spectra (Figure~\ref{fig:HostSpec}) by measuring their H$\alpha$ line fluxes and applying the relation of \citet{Kennicutt1998}.  In each case we apply a rough correction for the covering fraction of our spectra by scaling to our PS1 photometry of the hosts.  We do not correct for intrinsic extinction. In the three cases where both H$\alpha$ and H$\beta$ are detected the decrement is reasonably consistent with zero extinction. The resulting SFRs range from 1 $-$ 5 M$_\odot$ yr$^{-1}$ and are listed in Table~\ref{tab:hosts}.  For the four remaining objects we may place lower limits on the SFR by measuring the H$\beta$ emission line flux from the explosion spectra (Figure~\ref{fig:ExSpec}) and assuming zero extinction.  We do not attempt to correct for the covering fraction in these cases.  The lower limits range from 0.1 $-$ 0.3 M$_\odot$ yr$^{-1}$.  In the second and third panels of Figure~\ref{fig:HostCums} we plot the cumulative distribution of our SFRs and specific SFRs in comparison to the CC-SN and LGRB hosts from \citet{Svensson2010}.  Although the small number of events limits the conclusions we can draw, there is no statistical evidence that the samples are drawn from different progenitor populations. The star formation rates measured for our hosts are clustered around the median value observed for both CC-SN and LGRBs.

\subsection{Explosion Site Offsets}

\begin{figure}[ht]
\begin{center}
\includegraphics[width=\columnwidth]{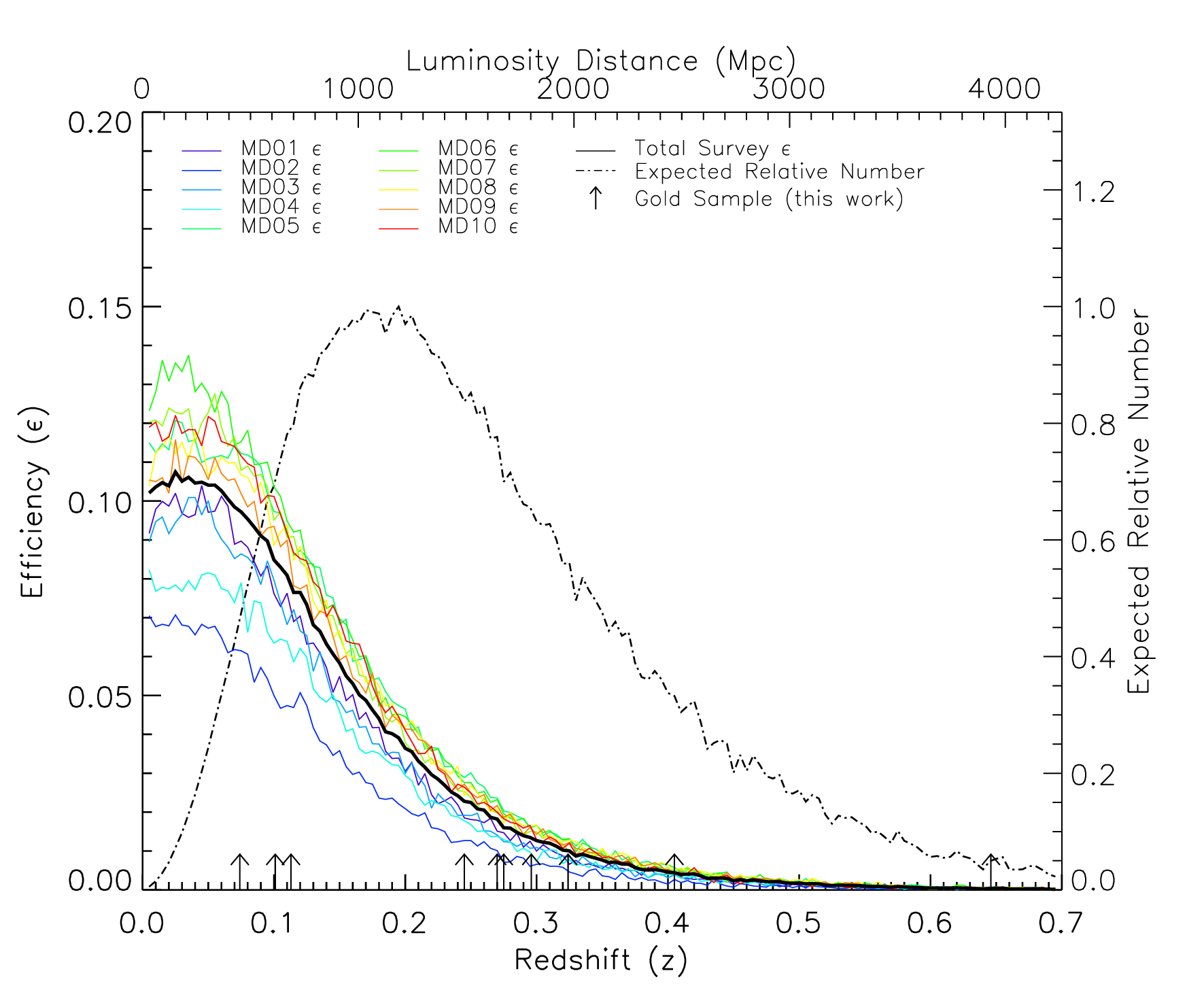}
\caption{PS1-MDS rapid-transient detection efficiencies for our rate calculation.  Solid lines indicate detection efficiencies for individual PS1-MDS fields as a function of redshift (bottom axis) or luminosity distance (top axis).  A black solid line indicates the detection efficiencies for the survey as a whole, which peak around 11\% at z$=$0.1.  The dashed black line indicates the distance at which we expect our detections to be located. This peaks at z$=$0.2.  The actual distances of the PS1-MDS transients are shown as arrows on the bottom axis. \label{fig:RateEff}}
\end{center}
\end{figure}

Using the PS1 centroid positions (good to $\sim$ 0.1$''$) we measure the separation between the location of the transient and the center of its host galaxy.  The cumulative distribution of these offsets (in physical units) are shown in the left panel of Figure~\ref{fig:HostOffset} (red line).  Also shown are distributions of type Ia, type Ibc and type II SN \citep[blue, green, and cyan, respectively]{Prieto2008}, short-duration GRBs \citep[dashed black; SGRBs]{Fong2013}, and LGRBs \citep[solid black]{Bloom2002}.

In the right panel of Figure~\ref{fig:HostOffset} we normalize these offsets by the $g-$band half light radius of the host. Also shown are type Ia SN \citep[blue]{Galbany2012}, type Ibc and type II SN \citep[green and cyan, respectively]{Kelly2011} and the same LGRB and SGRB samples as described above.  For objects with high signal to noise SDSS detections, we use the Petrosian half-light radius from SDSS DR9.  For others (PS1-11bbq, PS1-13duy, and PS1-13ess), we estimate the half light radius based on the light within a 3$''$ radius of the centroid location within the PS1 deep stacks (we caution that these values are more uncertain). Both the physical and normalized offsets are listed in Table~\ref{tab:hosts}.  Our sample most closely resembles the normalized offsets of type Ibc and type II SN, indicating that, at least in this context, our explosion sites are consistent with the environments in which one can expect to find massive stars.

\section{Volumetric and Relative Rates}\label{Sec:Rates}

In this section we use information on the true cadence and sensitivity obtained by the PS1-MDS between Dec.\ 2009 and March 2014 to calculate the volumetric rate of rapidly-evolving transients similar to those presented in this work.  Because these objects were identified based on their light curve morphology, rather than spectroscopic classification, we avoid the observer-dependent bias that results because resources limit our spectroscopic follow-up to roughly 10\% of the transients discovered in the PS1-MDS.  

Because their timescale is rapid, even short observing gaps due to poor weather or maintenance can have a significant effect on the efficiency with which the survey detects these transients.  We therefore use a Monte Carlo approach similar to that utilized in \citet{Quimby2012,Quimby2013} to calculate the efficiency with which each PS1-MDS field can recover these transients as a function of distance.

We begin by constructing light curve and temperature evolution templates based on the gold transients (Figure~\ref{fig:GoldPhot1}).  We then construct a luminosity function which consists of an intrinsic gaussian distribution modified by an exponential function to account for host galaxy absorption.  In following with \citet{Quimby2013} we adopt P(A$_{\rm{V}}$) $\propto$ e$^{\rm{A}_{\rm{V}}/\tau}$ with $\tau = 0.6$ \citep{Hatano1998} for the host galaxy absorption.  As a starting point for the intrinsic distribution we adopt a gaussian with the same mean and variance as the (rest-frame) peak magnitudes of our gold sample after performing a V/Vmax correction for the Malmquist bias.  As  a check we performed several Monte Carlo simulations distributing 500 objects drawn from this distribution evenly in space.  The objects with apparent magnitudes $>$ 1.5 mag above the nominal PS1 detection limit (a rough proxy for detectability) agree well with our observed transients in both luminosity and distance. The nominal values for the intrinsic (i-band) gaussian distribution we adopt below are $\mu = -17.2$ mag and $\sigma = 1.0$ mag.  Ê

For each PS1-MDS field we then determine the detection efficiency within 140 distance bins between z=0.005 and z=0.7. We performed 10,000 iterations of the following process \emph{within each distance bin:}  (i) randomly select one of the six templates; (ii) choose a peak absolute magnitude from the aforementioned luminosity distribution; (iii) choose a random distance \emph{within} the distance bin; (iv) k-correct the template light curves, assuming the spectra can be approximated as a blackbodies described by the temperature evolution template; (v) choose a random explosion epoch between Dec.\ 2009 and March.\ 2014; (vi) map the resulting explosion onto the \emph{actual dates} PS1 observed each filter for that PS1-MDS field; (vii) assign random noise to each epoch that is pulled from the actual distribution of noises obtained by transients found in the PS1-MDS on that epoch in that filter; and finally (viii) run the resulting synthetic data through the same light curve selection criteria described in Section~\ref{Sec:Obs}.  The efficiency within each distance bin is then calculated as the fraction of the 10,000 iterations that were recovered.

In Figure~\ref{fig:RateEff} we plot the resultant efficiencies for each PS1-MDS field (colored lines) as well as the survey as a whole (solid black line).   The survey detection efficiencies range from $\sim$11\% near z $\sim$ 0.1 to $\lesssim$0.1\% at z $\sim$0.7.  The quoted efficiencies are for the entire duration of the PS1 survey (Dec.\ 2009 to March 2014), while any given PS1-MDS field is only observed for roughly half of this time period.  The in-season efficiency at z$\sim$ 0.1 for the survey was approximately 25\%.  This demonstrates how much influence small, in-season, observing gaps can have on the detection of such rapid transients.

Using these efficiencies we calculate the volumetric rate of these rapid and luminous transients as:

\begin{equation}
R = \frac{N}{\sum \epsilon_i V_i t_i}
\end{equation}

\noindent where $\epsilon_i$, V$_i$, and t$_i$ are the efficiency, co-moving volume, and proper time within each distance bin and N is the number of transients actually detected in the PS1-MDS.  We calculate a rapidly-evolving transient volumetric rate of 4800 $-$ 8000 events yr$^{-1}$ Gpc$^{-3}$. The quoted range comes from assuming either 6 detections (the gold transients) or 10 detections (the gold plus bronze transients).  Our rate calculation does not include the silver transients because they would not be selected by the criteria given in Section~\ref{Sec:Obs} (which were the basis for how our efficiencies were derived).  This rate is approximately 24 $-$ 40 times higher than the rate derived by \citet{Quimby2013} for super-luminous SN-like objects and is approximately 4 $-$ 7\% of the core-collapse SN rate at z$\sim$0.2 found by \citet{Botticella2008}.  

In Figure~\ref{fig:RateEff} we also plot the expected relative number of detected events as a function of distance (dashed black line).  This is calculated by multiplying the efficiency at each distance by the relative volume and relative proper time in that bin. We expect that a majority of the PS1-MDS detected transients should be found around z$\sim$0.2.  Black arrows along the bottom axis represent the redshifts of our spectroscopic sample.  A KS test between the observed and expected redshift distributions yields a 50\% probability that they are drawn from the same population.

\begin{figure*}[ht!]
\begin{center}
\includegraphics[width=\columnwidth]{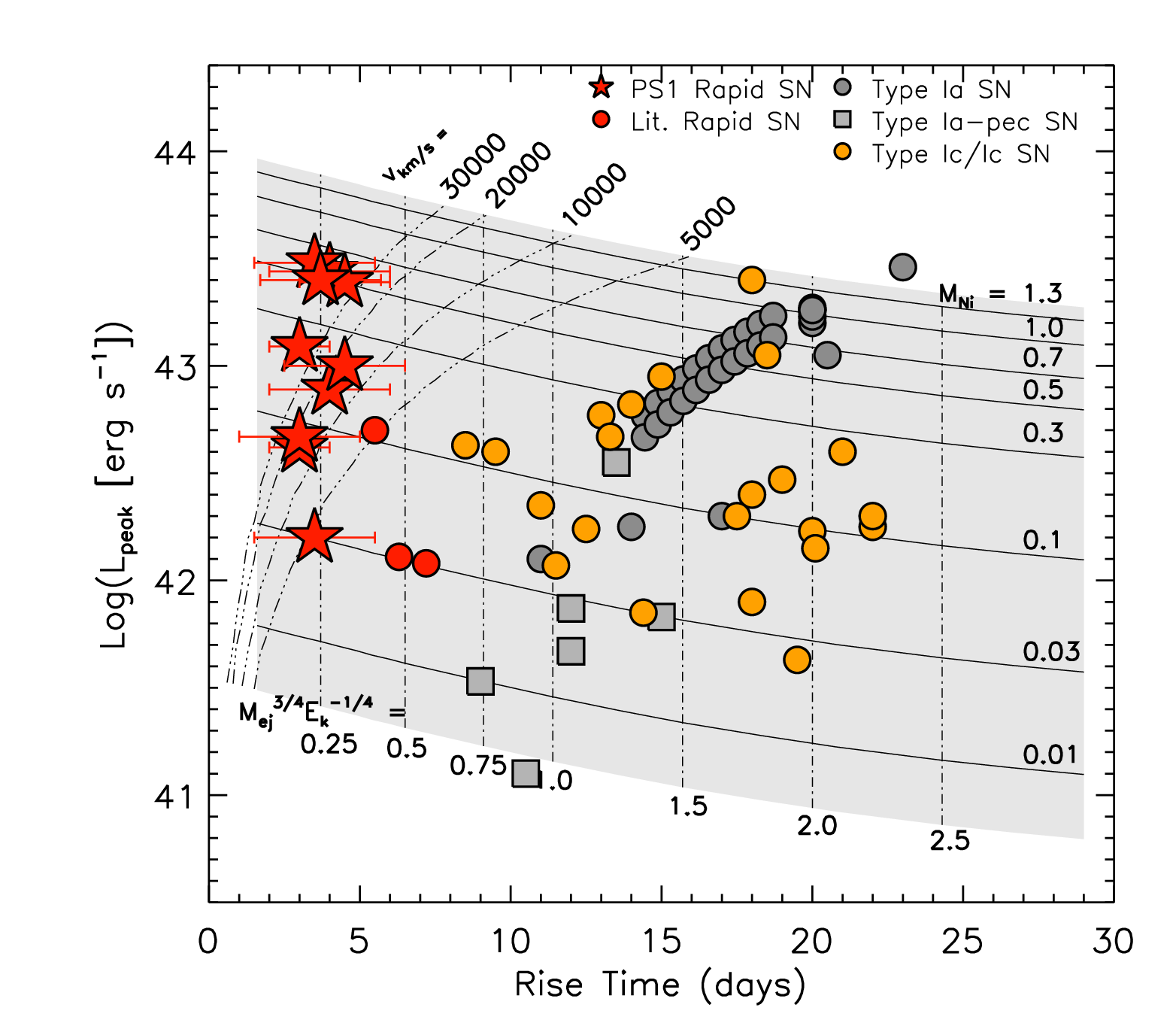}
\includegraphics[width=\columnwidth]{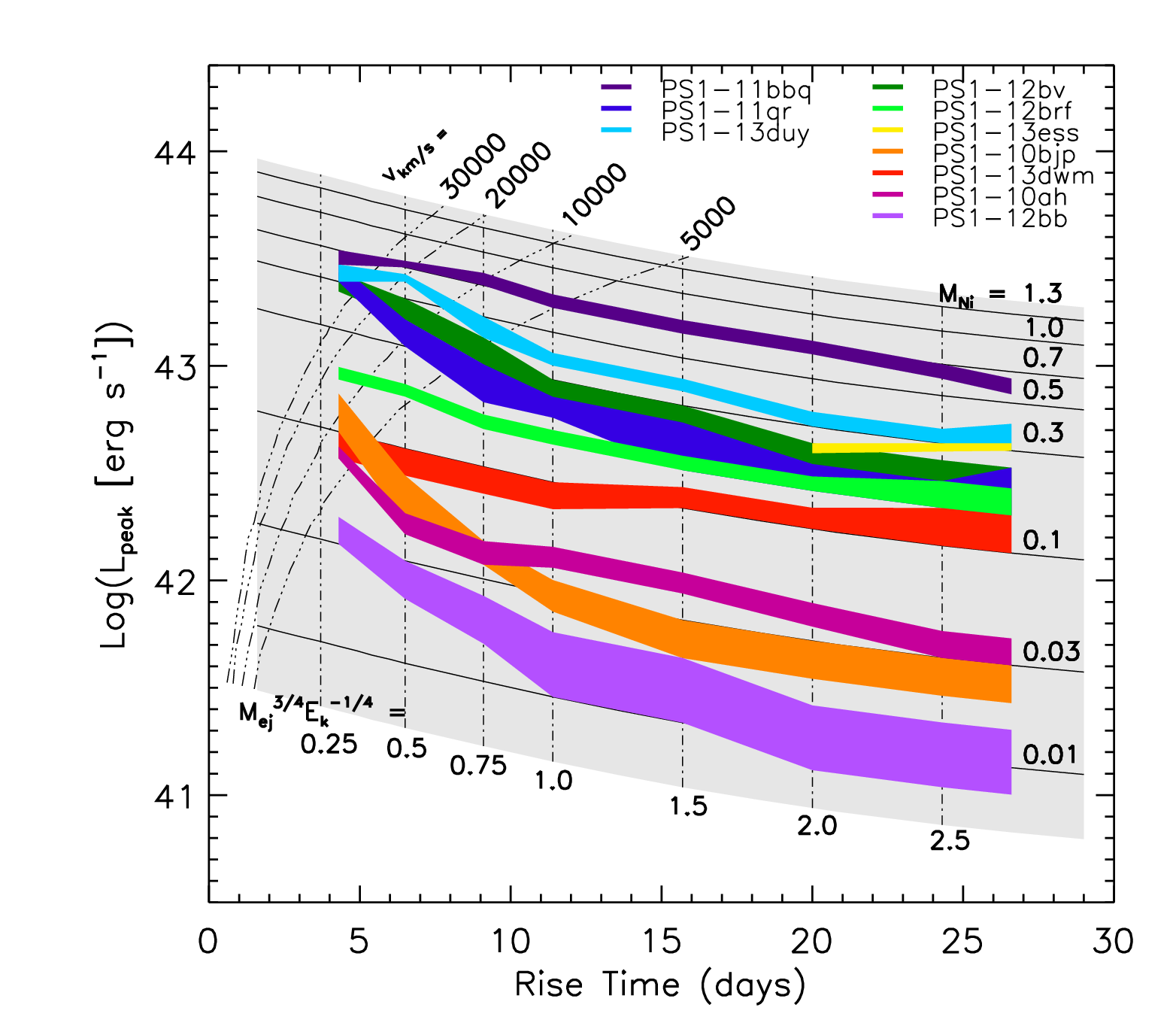}
\caption{Physical constraints for $^{56}$Ni powered explosions. \emph{Left Panel:}  We map the physical parameters M$_{\rm{Ni}}$ and M$_{\rm{ej}}^{3/4}$ E$_{\rm{K}}^{-1/4}$ onto the observable parameters of peak luminosity and rise time.  On this we plot the gold/silver transients (red stars) and objects from the literature that are powered by radioactive decay (type Ia - grey circles; type Ib/Ic - orange circles; type Ia-pec - grey squares; rapid events - red circles).  The curved lines represent locations where M$_{\rm{Ni}}$ $=$ M$_{\rm{ej}}$ for a given ejecta velocity.  Any object above and to the left of a given line is unphysical for that velocity.  \emph{Right Panel:}  Colored regions represent the upper limit on the amount of $^{56}$Ni which could power a second peak (such as that observed in SN\,1993J) in the light curves of our objects. \label{fig:Nickel}}
\end{center}
\end{figure*}

The largest sources of uncertainty in these derived rates are our assumed luminosity function and the effects of host galaxy extinction.  In order to assess these errors, we run additional iterations of procedure described above, varying both the parameters that describe the intrinsic luminosity function and the parameter $\tau$ which describes host galaxy absorption.  Varying $\tau$ between 0.5 and 0.7 produced $\pm 500$ events yr$^{-1}$ Gpc$^{-3}$ in the derived rates (an $\sim$ 8\% variation), while neglecting host extinction entirely yields rates of 3500 $-$ 5800 events yr$^{-1}$ Gpc$^{-3}$.  Introducing small variations to the functional form of the intrinsic luminosity function (high and low luminosity cutoffs, a lognormal distribution, etc.) produces only slightly variations in the derived rates ($\lesssim$8\%).  Increasing the mean luminosity from $-$17.2 to $-$17.5 (which although less ideal, still produces reasonable agreement with our observed sample) lowers our derived rates to 3500 $-$ 5800 events yr$^{-1}$ Gpc$^{-3}$, a factor of 1.4 decrease. 

In addition, the {\tt photpipe} pipeline requires a human to examine every potential detection and select real astronomical events.  As a byproduct of this process, some fraction of real transients will be overlooked. The chance that a given event will be missed is likely higher for transients with shorter overall timescales.  It is beyond the scope of the current paper to address this factor, but any such correction would act to increase our derived rates.

Finally, we address, from a rates perspective, the possibility that no objects of a similar class to the rapidly-evolving SN\,2005ek and SN\,2010X were discovered in the PS1-MDS. Only PS1-12bb possesses a similar luminosity and red colors but its spectra do not exhibit the strong early on-set of the Ca II NIR features that was a distinctive characteristic of those events. Using the data of \citet{Drout2013} we construct light curve and temperature templates for SN\,2005ek and run them through a similar procedure as described above, pulling from a Gaussian luminosity distribution with M$_{\rm{r}}$ $=$ $-$17.5 $\pm$ 0.5.  We find that, due to the lower peak luminosity and red colors of SN\,2005ek, the efficiency with which the PS1-MDS can detect similar transients plummets to zero by z$\sim$0.2.  In addition, the PS1-MDS has a small effective volume at low redshifts. As a result, if the intrinsic rate of SN\,2005ek-like events is similar to that for the bluer, more luminous, objects that make up our sample we would only expect to find $\lesssim$1 over the duration of the PS1-MDS.  \emph{The reason the PS1-MDS sample of rapidly-evolving SN is dominated by luminous, blue events (as opposed to previously discovered, faint, red events) is not due to differences in their intrinsic rates, but rather to the efficiency with which the pencil beam observing strategy of the PS1-MDS detects objects of each class}.

\section{Discussion: Power Sources and Progenitors}\label{Sec:Discussion}

In the sections above, we described the basic properties of the PS1-MDS rapidly-evolving transients, their host galaxies, and their volumetric rates. We now examine the nature of these transients.  We discuss several means by which the optical emission may be powered, and the implications of each on the progenitor systems.  Specifically, we examine SN powered by radioactive decay, SN powered by shock breakout emission/interaction, super-Eddington flares from tidal disruption events, and magnetars formed during NS-NS mergers.

\subsection{Limits on Radioactive Decay}

As described in Section~\ref{Sec:Spectra}, the continuum dominated spectra of our objects more closely resemble those of explosions powered by cooling envelope emission or interaction than those powered by radioactive decay.  \emph{In addition, the rapid timescale and high peak luminosity of several of our events are difficult to reconcile with an explosion powered entirely by $^{56}$Ni}.  This is demonstrated in Figure~\ref{fig:Nickel} where we plot the rise time versus peak luminosity for the gold/silver transients along with SN from the literature that are powered by radioactive decay.  Using the models of \citet{Arnett1982} and \citet{Valenti2008} we map these parameters onto a grid of $^{56}$Ni mass (M$_{\rm{Ni}}$) and a parameter proportional to the total ejecta mass (M$_{\rm{ej}}$) and kinetic energy (E$_{\rm{K}}$) of the explosion.   These results rely on the assumptions of spherical symmetry, centrally localized $^{56}$Ni, and an optically thick ejecta.

The M$_{\rm{Ni}}$ required to reproduce the peak luminosity of our events span a wide range (0.03 M$_\odot$ $<$ M$_{\rm{Ni}}$ $<$ 0.7 M$_\odot$).  This range is similar to that spanned by type Ib/Ic SN, however, the devil is in the details. The degeneracy between M$_{\rm{ej}}$ and E$_{\rm{K}}$ can be broken by using the photospheric velocity of the explosion near maximum light.  For a given velocity, a smaller characteristic time implies a lower ejecta mass.  Thus, for objects with short overall timescales, it is possible to obtain the unphysical result that M$_{\rm{Ni}}$ $>$ M$_{\rm{ej}}$.  

The photospheric velocities of our objects are unknown, so in Figure~\ref{fig:Nickel} we plot curved lines that represent M$_{\rm{Ni}}$ $=$ M$_{\rm{ej}}$ for a variety of velocities.  Any object to the upper left of a given line is physically disallowed for the simplified photospheric model.  While the radioactively powered explosions from the literature fall comfortably in the region where  M$_{\rm{Ni}}$ $<<$ M$_{\rm{ej}}$, several of our transients would require relatively high ejecta velocities ($\gtrsim$ 0.1c) to remain physical.  In all cases the rise times and peak luminosities imply that the required nickel mass makes up a significant fraction of the total ejecta mass. 

One caveat is that the models represented in Figure~\ref{fig:Nickel} assume the ejecta is optically thick.  If this assumption breaks down, incomplete gamma-ray trapping at early times can produce a faster initial light curve evolution \citep{Drout2013}.  For our objects, we can estimate the level of gamma-ray trapping required such that a $^{56}$Co $\rightarrow$ $^{56}$Fe decay tail would fall below our late-time PS1 photometry limits.  For five of our ten gold objects, the late-time PS1 limits \emph{require} incomplete gamma-ray trapping at early times if the peak is truly powered by $^{56}$Ni.  Using the parameterization of \citet{Valenti2008} to relate the level of trapping to M$_{\rm{ej}}$ and E$_{\rm{k}}$ we find that only two of our objects (PS1-10ah and PS1-12bb) have \emph{lower limits} for M$_{\rm{Ni}}$/M$_{\rm{ej}}$ that are less than 0.3.  

Thus, if our objects are powered primarily by $^{56}$Ni, they would require either (1) progenitors that can produce a relatively small ejecta mass composed almost entirely of $^{56}$Ni --- \emph{a scenario that seems inconsistent with the lack of strong UV line blanketing in our spectra} --- or (2) a explosion scenario in which either a significant amount of the radioactive material is mixed into the outer ejecta layers or there is a significant breakdown in spherical symmetry. 

We are therefore converging on a situation where the early emission from many of our rapidly-evolving events is powered by a source other than radioactive $^{56}$Ni.  In this case, it is useful to place limits on the amount of $^{56}$Ni that could power a subsequent peak in the light curve (such as that observed in SN\,1993J or type Ibn SN iPTF13beo; \citealt{Gorbikov2014}) without exceeding our observed light curve points or late-time limits.  SN\,1993J possessed two peaks of similar optical luminosity.  The first was due to cooling envelope emission and the second to the radioactive decay of $\sim$0.07 M$_\odot$ of $^{56}$Ni. Only one of our objects, PS1-13ess, shows evidence for a second peak in its light curve. The M$_{\rm{Ni}}$ required to power this second peak for PS1-13ess is 0.2 $-$ 0.3 M$_\odot$ (depending on the rise time, which is not well constrained). This is approximately four times higher than the amount that powered SN\,1993J.  In Figure~\ref{fig:Nickel} we use our observations to put upper limits on the amount of $^{56}$Ni that could power such a secondary peak in the remainder of our transients.

\begin{figure}[ht!]
\begin{center}
\includegraphics[width=\columnwidth]{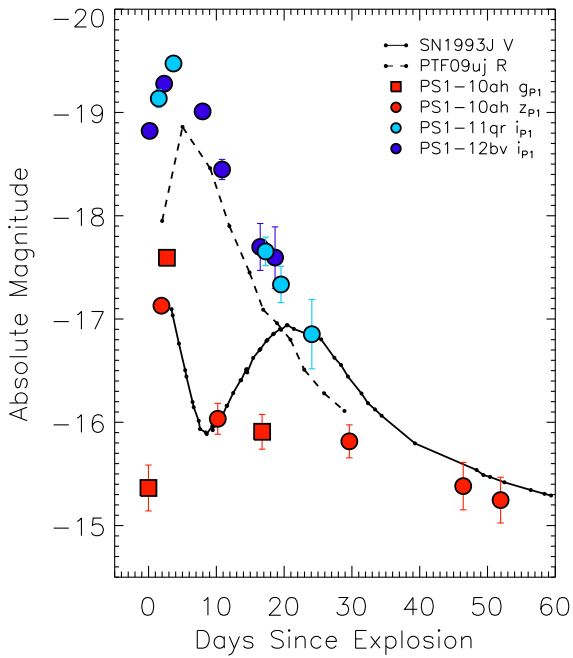}
\caption{A comparison of the light curves of several gold transients to events from the literature thought to be powered by varieties of shock breakout/cooling envelope emission.  PS1-10ah shows a similar initial decline to the type IIb SN\,1993J, but lacks a later peak powered by radioactive decay.  PS1-11qr, PS1-12bv, and a number of the other luminous events are very similar to PTF\,09uj, which was hypothesized to be due to the shock breakout from an optically thick wind \citep{Ofek2010}. \label{fig:CoolingEnv}}
\end{center}
\end{figure}

For each object, we present a colored region which represents the maximum nickel mass that could power an optically thick explosion with a given rise time.  For three of our objects (PS1-10ah, PS1-10bjp, and PS1-12bb) we can put an upper limit on the amount of $^{56}$Ni that could power a secondary peak to $\lesssim$0.03 M$_\odot$.  This is lower than the amount inferred for all but a few type Ibc and peculiar type Ia SN.  Several type II SN have been constrained to have M$_{\rm{Ni}} <$ 0.03 M$_\odot$, although over half of those analyzed in \citet{Hamuy2003} have M$_{\rm{Ni}}$ larger than this.  The limits on the more luminous transients are less constraining in comparison to type Ib/Ic SN.

\subsection{Shock Breakout and Interaction}

The blue colors and spectra of several of our objects resemble those observed in SN powered by cooling envelope emission or interaction with a  dense CSM (e.g.\ type IIb, type IIn and type Ibn SN, several of which have peak luminosities and timescales which also resemble the PS1 transients presented here).  In this section we examine both the shock breakout from the surface of a star and a shock breakout that occurs within an optically thick wind surrounding the progenitor as possible power sources for our transients.

\subsubsection{Shock Breakout from the Surface of a Star}

First, we consider the case where these events are powered by the cooling envelope portion of the explosion of a massive star.  While the shock breakout from a massive star lasts only a matter of seconds (WR stars) to hours (RSGs) the process shock heats the ejecta, which then cools and recombines, giving rise to optical emission which is independent from later emission powered by radioactive decay.  The luminosity reached by the cooling envelope emission can give insight into the radius of the progenitor, with larger progenitors yielding more luminous events (due to both a larger emitting surface and reduced effects of adiabatic cooling), while the timescale of the event is related to the envelope mass that is cooling/recombining (see, e.g., \citealt{Nakar2010,Rabinak2011}).

\begin{figure}[ht!]
\begin{center}
\includegraphics[width=\columnwidth]{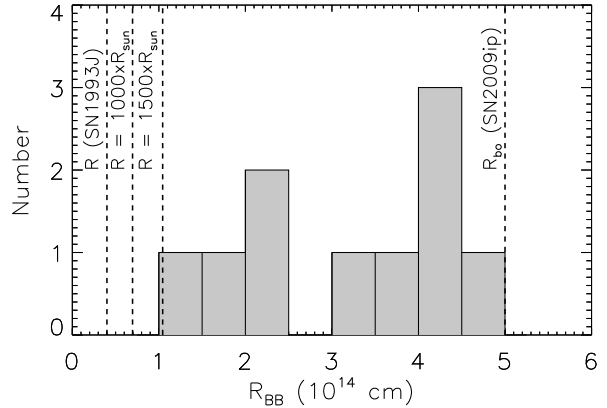}
\caption{Histogram of the first radius constraint obtained for each transient in Section~\ref{Sec:Photometry}.  Vertical lines mark the radius found for the progenitor of SN\,1993J \citep{Woosley1994}, 1000 and 1500 R$_\odot$ (the largest known RSGs) and the radius of shock breakout found for the type IIn SN\,2009ip \citep{Margutti2014}. \label{fig:Rbb}}
\end{center}
\end{figure}

Several of our transients (e.g.\ PS1-10ah, PS1-12brf, PS1-13dwm, PS1-13ess) have peak luminosities and timescales that are similar to those observed for the cooling envelope portion of type IIb SN with extended progenitors.  In Figure~\ref{fig:CoolingEnv} we compare the light curve of SN\,1993J to several of our gold sample objects and see that the initial decline of SN\,1993J is well matched to the initial decline of our object PS1-10ah. This is consistent with the initial light curve signature expected for the explosion of progenitors with extended low-mass envelopes \citep{Nakar2014}. However, with the exception of PS1-13ess, we \emph{do not} observe a second peak due to the decay of $^{56}$Ni as is observed in other type IIb SN.  In fact, these intermediate luminosity objects are the transients for which we can place the strictest limits on the amount of $^{56}$Ni that powers a secondary peak.  For PS1-10ah we constrain M$_{\rm{Ni}}$ $\lesssim$ 0.03M$_\odot$, which is more than a factor of two lower than that found for the second peak of SN\,1993J \citep{Woosley1994}.  This implies that some \emph{stripped envelope} core-collapse SN may eject a very small amount of radioactive isotopes.  A similar point was raised in \citet{Kleiser2014} who modeled the rapidly declining type Ic SN\,2010X as an \emph{oxygen} recombination event.

While this power source seems reasonable for our intermediate luminosity events, a number of our other transients (PS1-11qr, PS1-11bbq, PS1-12bv, and PS1-13duy) possess peak luminosities that are a factor of ten higher than that observed for SN\,1993J (see Figure~\ref{fig:CoolingEnv}).  \citet{Woosley1994} invoked a progenitor radius of $\sim$ 4 $\times$ 10$^{13}$ cm in order to reproduce the light curve of SN\,1993J.  This is only a factor of two below the radii of the largest known RSGs \citep{Levesque2009}.  Thus, if these events can be explained by the cooling envelope emission after shock breakout from the \emph{surface} of a massive star, we would seem to require a progenitor that is both extremely extended, but also possesses a relatively low envelope mass.  Alternatively, these events could represent cases where the shock breakout occurs not directly from a stellar surface, but inside a dense, optically thick, wind surrounding the star.  This hypothesis was also put forth for the rapidly-evolving type IIn event, PFT09uj \citep{Ofek2010}, whose light curve bears striking resemblance to our higher luminosity events (Figure~\ref{fig:CoolingEnv}). 

\subsubsection{Shock Breakout within a Dense Wind}

The case of a shock breakout from a dense wind has been examined from a theoretical perspective by \citet{Chevalier2011}, \citet{Balberg2011}, \citet{Ofek2010} and \citet{Ginzburg2014}. For a wind environment, the density profile around the star goes as $\rho_w \propto r^{-2}$.  While the actual situation for the case of an optically thick mass-loss region may be more complex (e.g. due to shell ejections), this approximation has been made in previous works and will also be utilized here.  These models have been applied to explain a number of super-luminous SN (both type I and type II; \citealt{Chevalier2011}) as well other type IIn SN such as SN2009ip \citep{Margutti2014} and PTF\,09uj \citep{Ofek2010} and the first peak of type Ibn iPTF13beo \citep{Gorbikov2014}. 

\begin{figure*}[ht!]
\includegraphics[width=\textwidth]{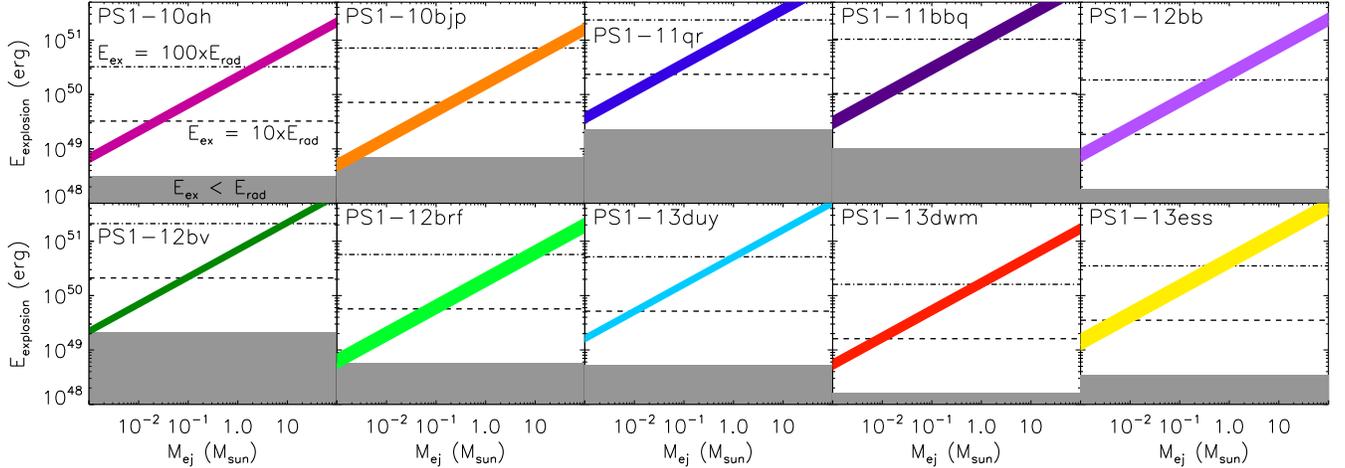}
\caption{Values of M$_{\rm{ej}}$E$^{-1/2}$ derived for the gold/silver transients for the case of a shock breakout from a dense, optically thick, wind.  The width of the colored region gives the error inferred for M$_{\rm{ej}}$E$^{-2}$ based on the errors for each of our observables. Grey shading represents an excluded region where the explosion energy is less than the observed radiated energy.  Horizontal lines in each panel represent where E$_{\rm{explosion}}$ $=$ 10$\times$E$_{\rm{rad}}$ and 100$\times$E$_{\rm{rad}}$ for each object. \label{fig:DenseShell}}
\end{figure*}

\begin{figure*}[ht!]
\includegraphics[width=\textwidth]{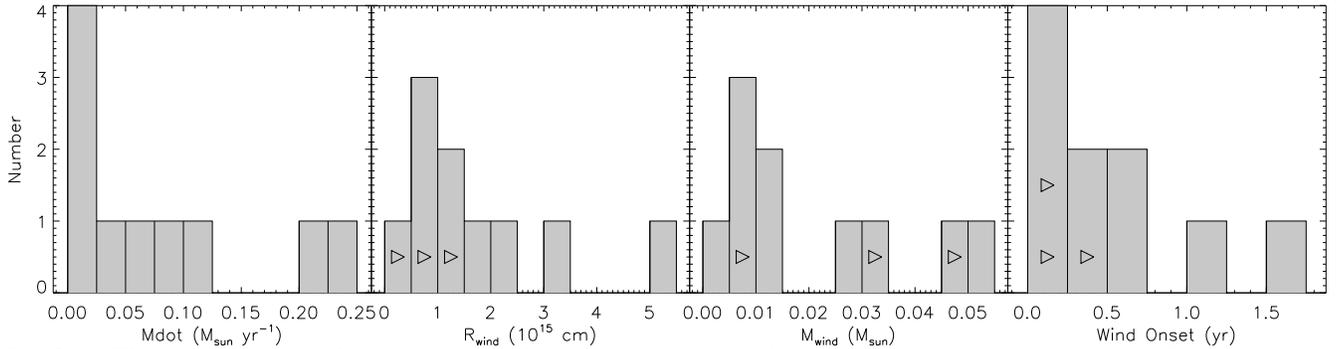}
\caption{Histograms of the physical parameters derived for the mass-loss region surrounding each object in the case of a shock breakout from a dense, optically thick, wind.  Panels 1,2,3 and 4 display the mass-loss rates inferred from the explosion, the outer radius of the region of dense mass-loss, the total mass contained in the region of dense mass-loss, and the inferred time for the onset of the mass-loss (prior to the main explosion), respectively.  Both the mass-loss rates (panel 1) and wind onset times (panel 4) listed assume a wind velocity of 1000 km s$^{-1}$ (as measured from a potential CSM absorption in the spectrum of PS1-12bv).  The results scale linearly for other assumed velocities. \label{fig:DenseShellHist}}
\end{figure*}

In their appendix, \citet{Margutti2014} solve the equations given in \citet{Chevalier2011} for the case where the wind radius, R$_w$, is larger than the radius at which diffusion becomes important, R$_d$.  These equations give the ejecta mass and kinetic energy of the explosion, as well as the opacity and density of the mass-loss region (as parameterized by a mass-loss rate \.{M}) in terms of three observables: the rise time of the explosion (t$_{\rm{rise}}$), the energy radiated on the rise (E$_{\rm{rad}}$), and the radius at breakout (R$_{\rm{bo}}$).  Using our first measured blackbody radius for each transient (Figure~\ref{fig:Rbb}) as an order of magnitude estimate for R$_{\rm{bo}}$, we can utilize these models to estimate the physical parameters of the explosion and mass-loss region surrounding the progenitor.  From Figure~\ref{fig:Rbb} we see that the breakout radius for our events is relatively small.  All the events (including those whose first radius constraint was measured at maximum light) have values below that measured for SN\,2009ip and an order of magnitude smaller than those measured for several SLSN \citep{Chevalier2011}.  

These models are degenerate in M$_{\rm{ej}}$E$^{-2}$ and we find values for the ejecta mass ranging from 1.0 $(\frac{\rm{E}}{10^{51} \rm{ergs}})$ M$_\odot$  $\lesssim$ M$_{\rm{ej}}$ $\lesssim$ 45.0 $(\frac{\rm{E}}{10^{51} \rm{ergs}})$ M$_\odot$.  In comparison, \citet{Margutti2014} found M$_{\rm{ej}} =$ 50.5 $(\frac{\rm{E}}{10^{51} \rm{ergs}})$ M$_\odot$ for the 2012b outburst of SN2009ip.  In Figure~\ref{fig:DenseShell} we plot diagonal lines which represent the value of M$_{\rm{ej}}$E$^{-2}$ obtained for each object.  The shaded gray region in each panel represents an excluded region where the explosion energy (vertical axis) is lower than the total radiated energy calculated in Section~\ref{Sec:Photometry}.  Horizontal lines in each panel then represent explosion energies that are 10 and 100 times the radiated energy.  

For these models we infer mass-loss rates ranging from 2 $\times 10^{-3} (\frac{\rm{v}_w}{1000 \rm{km s}^{-1}})$ M$_\odot$ yr$^{-1}$ $\lesssim$  \.{M} $\lesssim$ 2 $\times 10^{-1} (\frac{\rm{v}_w}{1000 \rm{km s}^{-1}})$ M$_\odot$ yr$^{-1}$.  A histogram of these values are shown in the first panel of Figure~\ref{fig:DenseShellHist}. We have normalized to a wind speed of 1000 km s$^{-1}$ based on the tentative CSM absorption observed in PS1-12bv (Section~\ref{Sec:Spectra}). These values exceed the mass-loss rates of WR stars (typically $\lesssim$10$^{-4}$ M$_\odot$ yr$^{-1}$) and may hint at a more complex or eruptive mass-loss history.

In order to obtain more information about the mass-loss environment implied by these models, we next use the time when the bolometric luminosity calculated in Section~\ref{Sec:Photometry} falls below that expected from continued interaction (\citealt{Chevalier2011} Eqn.\ 9; \citealt{Margutti2014} Eqn.\ A6) to estimate the time at which the shock reaches the outer radius of the dense wind region.  For all of our objects, with the exception of PS1-10ah, this occurs within a few days after maximum light.  For PS1-10ah the flattening of the light-curve at +10 days is consistent with the emission expected for continued interaction.  Using this information and Eqn.\ A7 from \citet{Margutti2014}, we find that the dense wind extends to a few $\times$ 10$^{15}$ cm for all of our objects (second panel, Figure~\ref{fig:DenseShellHist}).  For three events (PS1-11bbq, PS1-12brf, and PS1-13duy) we were only able to put lower limits on the extent of the mass-loss region.  These values are comparable to the dense wind radius of 1.2 $\times$10$^{15}$ cm found by \citet{Margutti2014} for the 2012b explosion of SN\,2009ip.

From this we find that the total mass contained within the region of dense mass-loss is only a few hundredths of a solar mass for most of our objects (third panel, Figure~\ref{fig:DenseShellHist}). This is an order of magnitude lower than the CSM mass found by \citet{Margutti2014} for the 2012b explosion of SN\,2009ip. In addition, the timescale over which the ejection of this mass from the progenitor would have occurred is very short. For wind speeds of 1000 km s$^{-1}$ the mass-loss would have occurred in the $\lesssim$ 1 year prior to explosion (fourth panel, Figure~\ref{fig:DenseShellHist}).

\citet{Chevalier2011} predict that the temperature of the explosion should \emph{rise} to maximum in the case of shock breakout within a region of dense mass-loss.  Unfortunately, for many of our objects we do not robustly constrain their temperature evolution on the rise. However, one of our best observed objects on the rise, PS1-10ah, does show evidence for a rise in temperature to maximum.  Our best fit blackbody for this object at maximum light is $\sim$ 18,000 K, but several days before max is only 10,000 K.  This trend is also reflected in the observed g$-$r colors, which get bluer at maximum light. 

If this physical scenario is an accurate representation of our transients, then they (along with PTF09uj) add to the interesting picture painted by type IIn, type Ibn, SLSN-II and other transients about the (possibly eruptive) mass-loss that some massive stars undergo in the year(s) immediately prior to explosion.  The rates of these short duration transients are not insignificant compared to the rates of longer duration type IIn SN (8$-$9\% of core-collapse SN; \citealt{Smith2011}).  While the mass-loss rates and dense wind radii required for our transients are similar to those derived for other type IIn events, the total mass contained in the dense CSM is a factor of ten lower than that inferred for a majority of of type IIn SN (see \citealt{Smith2014} and references therein).  This, in turn, leads to a smaller break-out radius and shorter transient duration.

\subsection{Super-Eddington Tidal Disruption Flares}

We now turn our attention to a set of non-explosive transients whose predicted properties bear some resemblance to several of our objects: the super-Eddington phase of a tidal disruption flare.  To date, a majority of tidal disruption event candidates have been discovered in the UV and X-rays. This is expected given the high predicted blackbody temperature of material close to the SMBH ($\sim$3$\times$10$^5$ K; see, e.g., \citealt{Lodato2011}).  However, for SMBHs with M $\lesssim$ 10$^8$ M$_\odot$, there may be a period of time after the initial disruption when accretion is super-Eddington, which can drive an outflow of material.  \citet{Strubbe2009} model the light curves and spectra for this type of emission.  They find that the transients are short-lived ($\lesssim$ 30 days), luminous (L $\sim$ 10$^{43}$ ergs s$^{-1}$), and peak in the optical/UV.

\citet{Cenko2012} argue that a super-Eddington outflow may explain the event PTF\,10iya, which has several properties in common with the PS1-MDS transients. PTF\,10iya has a rapid timescale ($\sim$ 10 days), high peak luminosity ($\sim$10$^{44}$ ergs s$^{-1}$) and a blackbody temperature of $\sim$20,000 K.  However, there are several difficulties in interpreting our objects similarly.  First, and most obviously, PTF\,10iya was coincident with the nucleus of its host galaxy to within 350 mas.  The same is not the case for many of the transients presented here.  Only PS1-10ah lies in the very central region of its host.  While, in principle, a non-nuclear event could be due to a tidal disruption by an intermediate-mass black hole (IMBH), in the models of \citet{Strubbe2009} a characteristic timescale of $\sim$10 days requires a SMBH with a mass between 10$^6$ and 10$^8$ M$_\odot$.  The timescale for an IMBH would be $\lesssim$ 1 day.  

In addition, \citet{Strubbe2009} predict that the observed temperature should \emph{increase} slightly as the transient declines.  In contrast, our objects show ejecta which cool with time.  Finally, \citet{Cenko2012} argued that the blackbody radii measured for PTF\,10iya ($\sim$ 10$^{15}$ cm) are too large for SN ejecta traveling at typical velocities to reach in such a short period of time.  The same argument does not hold for our objects, whose blackbody radii are an order of magnitude smaller.  Thus, despite sharing several observational properties with PTF\,10iya, it is not obvious that any of our events are consistent with the emission predicted for the super-Eddington phase of a tidal disruption flare.  

In contrast, \citet{Strubbe2009} predicted that the PS1-MDS should detect approximately 20 tidal disruption flares in their super-Eddington phase \emph{per year}.  Using their models and detection rates as a function of BH mass and the black hole mass function from \citet{Hopkins2007} we estimate that roughly 40\% of these should fall in the rapidly-evolving parameter space that was systematically examined in this paper.  In addition, \citet{Strubbe2009} calculate their rates using a limiting flux of 25 AB mag for the PS1-MDS, which is 1 $-$ 1.5 mag fainter than what the survey actually achieved.  Using Equation 32 in \citet{Strubbe2009} to scale their derived rates to the true sensitivity obtained by PS1-MDS we produce a modified predicted detection rate of 1 $-$ 2 events per year within the parameter space we have examined. Thus, given the stringent detection criteria we utilized in Section~\ref{sec:selection} it is possible that our lack of detections in four years of the PS1-MDS is not in conflict with the models of \citet{Strubbe2009}.

\subsection{Magnetars formed during NS-NS Mergers}

In some cases, the merger of two low mass neutron stars (NS) may produce a stable massive NS as opposed to collapsing to a black hole. Both \citet{Yu2013} and \citet{Metzger2014} investigate the electromagnetic counterparts that would be produced from the formation of such a system if the magnetar spin-down energy is channeled into the ejecta from the merger, as opposed to a collimated jet.  They predict optical/UV transients with peak bolometric luminosities of 10$^{43}$ $-$ 10$^{44}$ erg s$^{-1}$, rise timescales of $\lesssim$ 1 day and overall timescales of $\sim$10 days.  Although these predicted luminosities and timescales are slightly higher/shorter than those observed for the PS1-MDS transients presented here, their properties are broadly comparable.

However, both the derived rates for the PS1-MDS transients and their explosion site offset distribution disfavor a significant fraction originating from this channel.  The ``realistic'' NS-NS merger rate from \citet{Abadie2010} is 1000 events yr$^{-1}$ Gpc$^{-3}$ and the ``high'' rate is 10000 events yr$^{-1}$ Gpc$^{-3}$.  The rates we derive in Section~\ref{Sec:Rates} are 4800 $-$ 8000 events yr$^{-1}$ Gpc$^{-3}$. Thus, if \emph{all} of our events originated from this channel we would require a high NS-NS merger rate of which a significant fraction produce stable NS remnants. In addition, the offset distribution observed for our transients is consistent with that observed for CC-SN, rather than SGRBs (which are also thought to originate from NS-NS mergers).  However, while this progenitor channel cannot explain the entire population of rapidly-evolving objects we identify in the PS1-MDS, we cannot rule it out for some individual objects.  To do so, we would require models with more detailed radiation transport to determine if the blue colors and continuum dominated spectra we observe are consistent with predictions.

\section{Summary and Conclusions}

{\bf The Search:} We have performed a search for transients with rapid timescales (t$_{1/2} \lesssim$ 12 days) and high peak luminosities ($-$15 $>$ M $>$ $-$21) within the $\sim$5000 transients discovered in the PS1-MDS.  We identify 14 light curves of interest, 10 of which have confirmed peak magnitudes brighter than $-$16 mag.  This sample increases the known number of events in this region of transient phase space by approximately a factor of three.  The median redshift of our confirmed sample is 0.275.

{\bf Basic Properties:} The PS1-MDS rapidly-evolving transients span a wide range peak magnitudes ($-$16.8 mag $<$ M $<$ $-$20 mag) and an order of magnitude in peak pseudo-bolometric luminosity (2 $\times$ 10$^{42}$ erg s$^{-1}$ $<$ L$_{\rm{peak}}$ $<$ 3 $\times$ 10$^{43}$ erg s$^{-1}$).  In general, they possess much faster rise than decline timescales. Our data for six objects cannot exclude a transient that rose on a timescale of $\lesssim$ 1 day.  With the exception of PS1-12bb, all of the transients possess blue colors near maximum light ( g$_{\rm{P1}}$ $-$ r$_{\rm{P1}}$ $\lesssim$ $-$0.2).  Best fit blackbodies reveal photospheric temperatures/radii which cool/expand with time, consistent with explosions that involve expanding ejecta.  The explosion spectra obtained near maximum light are dominated by a blue continuum, consistent with a hot optically thick ejecta. Based on these spectra we cannot conclude if the explosions are hydrogen-rich or hydrogen-poor.

{\bf Power Sources and Progenitors:} The short timescales and high peak luminosities of the PS1-MDS transients are difficult to reconcile with explosions powered by the radioactive decay of $^{56}$Ni.  Either extreme ejecta velocities or significant outward mixing of $^{56}$Ni would be required to avoid a situation where the required nickel mass is a significant fraction ($>$50\%) of the total ejecta mass. However, the blue, continuum dominated, spectra are consistent with early spectra observed for transients powered by shock breakout cooling envelope emission or interaction with a dense CSM.  

The luminosity and duration of several transients are consistent with the cooling envelope emission observed in type IIb SN with extended low-mass envelopes (e.g.\ SN\,1993J).  However, in all but one case we see no evidence for a second peak in the optical powered by radioactive decay.  For the others, we can restrict the amount of $^{56}$Ni which could power a subsequent peak to $\lesssim$0.03 M$_\odot$. In contrast, several of the more luminous transients may be produced by a shock breakout within a region of dense mass-loss surrounding the progenitor star.  The mass-loss rates and dense CSM radii required are comparable to those derived for normal type IIn SN.  However, the total mass contained within the dense CSM is an order of magnitude lower ($\sim$0.01 M$_\odot$).

{\bf Host Galaxies: } The host galaxies of the PS1-MDS rapidly-evolving transients are consistent with the interpretation of massive star progenitors.  They are all star-forming, possess roughly solar metallicity, and are consistent with the mass-metallicity relation observed for SDSS star forming galaxies \citep{Tremonti2004}. The explosion sites span a wide range of offsets from the galaxy centers, and their distribution most closely resembles that observed for core-collapse SN.

{\bf Volumetric Rates: } Using information on the true cadence and sensitivity achieved by the PS1-MDS, we use a Monte Carlo code to calculate the efficiency of the PS1-MDS in detecting rapidly-evolving transients as a function of distance.  For luminous blue transients (such as those presented here) we find in season efficiencies that peak around $\sim$25\% at z$=$0.1. This demonstrates the large effect that small observing gaps (due to weather, maintenance, etc.) can have on the ability of optical SN searches to detect rapidly-evolving events.  Using these efficiencies we calculate a volumetric rate for these transients of 4800 $-$ 8000 events Gpc$^{-3}$ yr$^{-1}$.  This is 4 $-$ 7\% of the core-collapse SN rate at z=0.2 found by \citet{Botticella2008}.  In addition, we find that the efficiency with which PS1-MDS detects faint, red, transients (similar to the rapidly-declining events presented in the literature; SN\,2010X, SN\,2005ek) plummets to zero by z$\sim$0.2.  The PS1-MDS has a small survey volume at such low redshifts. Thus, the lack of similar objects in our sample does not reflect their intrinsic rate, but rather the efficiency with which the PS1 observing strategy can recover events of each subclass.

{\bf Implications and Future Directions: } Both the event rates we derive for these transients and our interpretation for their progenitors have significant implications for our understanding of the final stages of stellar evolution.  The rates are not insignificant compared to the core-collapse SN rate, and both the progenitor systems we favor imply non-standard final configurations of the star.  Our interpretation of some events as cooling envelope emission from a progenitor with a low-mass extended envelope implies that some stripped envelope core-collapse SN eject very little radioactive material.  This could either be due to fallback within a high mass progenitor whose outer envelope is stripped by winds, or due to binary stripping of the outer envelope in a lower mass progenitor.  Our interpretation of some other events as a shock breakout within an optically thick wind adds to the complex picture painted by type IIn SN of the eruptive mass loss that seems to occur for some massive stars in the year(s) immediately prior to explosion.  In view of these results, there are several future directions for both theorists and observers that will aid our ability to interpret these transients:

\emph{Theory:} (1) Detailed models of these fast, blue, luminous transients. In particular, when detailed models (which treat outward mixing of $^{56}$Ni, departures from spherical symmetry, and time varying opacity) are performed, do they confirm that the luminosities, timescales, and spectra are inconsistent with a transient powered mainly by radioactive decay? What stellar envelope parameters does a cooling envelope interpretation imply? Is it possible to reproduce the most luminous transients with a shock breakout from the surface of a star, or are the implied stellar radii unphysical? Can a model other than those investigated here reproduce all of the observed properties? (2) Updated models of the final stages of stellar evolution.  What final state do stars which are stripped by binary interaction actually take? Can detailed models reproduce the mass-loss in the final year(s) before explosion implied by these events and other type IIn SN? 

\emph{Observation:} (1) Multi-band data taken on the same epoch with a $\sim$1 day cadence.  This data will allow us to constrain both the true rise time of the transients ($\lesssim$ 1 day versus 2$-$3 days) and their temperature evolution on the rise.  The shock breakout from the surface of a star and a shock breakout from within a region of dense mass loss give different predictions for these parameters. (2) Multiple epochs of spectroscopy, especially at late times.  The spectra obtained for all of our blue transients were at maximum light.  While spectra dominated by blue continua are consistent with a cooling envelope or shock breakout interpretation at these epochs, late time spectra should display features from either the SN ejecta or from interaction with the CSM.

\vspace{2 mm}

M.R.D thanks Hagai Perets, Jerod Parrent, and Luke Zoltan Kelley for useful conversations regarding this manuscript. M.R.D. is supported in part by the NSF through a Graduate Research Fellowship. Support for this work was provided by the David and Lucile Packard Foundation Fellowship for Science and Engineering awarded to A.M.S. E.K. acknowledges financial support from the Jenny and Antti Wihuri Foundation. S.J.S. acknowledges European Research Council funding under the European Union's Seventh Framework Programme (FP7/2007-2013)/ERC Grant agreement n$^{\rm o}$ [291222]

The Pan-STARRS1 Surveys (PS1) have been made possible through contributions of the Institute for Astronomy, the University of Hawaii, the Pan-STARRS Project Office, the Max-Planck Society and its participating institutes, the Max Planck Institute for Astronomy, Heidelberg and the Max Planck Institute for Extraterrestrial Physics, Garching, The Johns Hopkins University, Durham University, the University of Edinburgh, Queen's University Belfast, the Harvard-Smithsonian Center for Astrophysics, the Las Cumbres Observatory Global Telescope Network Incorporated, the National Central University of Taiwan, the Space Telescope Science Institute, the National Aeronautics and Space Administration under Grant No. NNX08AR22G issued through the Planetary Science Division of the NASA Science Mission Directorate, the National Science Foundation under Grant No. AST-1238877, the University of Maryland, and Eotvos Lorand University (ELTE)

This paper includes data gathered with the 6.5m Magellan Telescopes located at Las Campanas Observatory, Chile. Some observations reported here were obtained at the MMT observatory, a joint facility of the Smithsonian Institution and the University of Arizona. Some observations were obtained under Program IDs GN-2011B-Q-3 (PI: Berger) and GS-2012A-Q-31 (PI: Berger) at the Gemini Observatory, which is operated by the Association of Universities for Research in Astronomy, Inc., under a cooperative agreement with the NSF on behalf of the Gemini partnership: the National Science Foundation (United States), the National Research Council (Canada), CONICYT (Chile), the Australian Research Council (Australia), Ministerio da Ciencia, Tecnologia e Inovacao (Brazil) and Ministerio de Ciencia, Tecnologia e Innovacion Productiva (Argentina). The data presented here were obtained in part with ALFOSC, which is provided by the Instituto de Astrofisica de Andalucia (IAA) under a joint agreement with the University of Copenhagen and NOTSA. Some of the computations in this paper were run on the Odyssey cluster supported by the FAS Science Division Research Computing Group at Harvard University.

\emph{Facilities:} PS1 (GPC1), Magellan Baade (IMACS), Magellan Clay (LDSS3), MMT (Blue Channel spectrograph, Hectospec), Gemini North (GMOS-N),Gemini South (GMOS-S), NOT (ALFOSC)


\appendix

\section{Most Likely Host Analysis of Bronze Sample Objects}

Our bronze sample is composed of four objects that passed all of the photometric selection criteria described in Section~\ref{sec:selection}, but whose distance we were not able to constrain spectroscopically.  As a result, we cannot confirm that they are ``high luminosity'' transients (M $<$ $-$ 15 mag).  We now attempt to assess this likelihood by examining the possible hosts for each event.

\begin{figure}[ht!]
\begin{center}
\includegraphics[width=0.9\columnwidth]{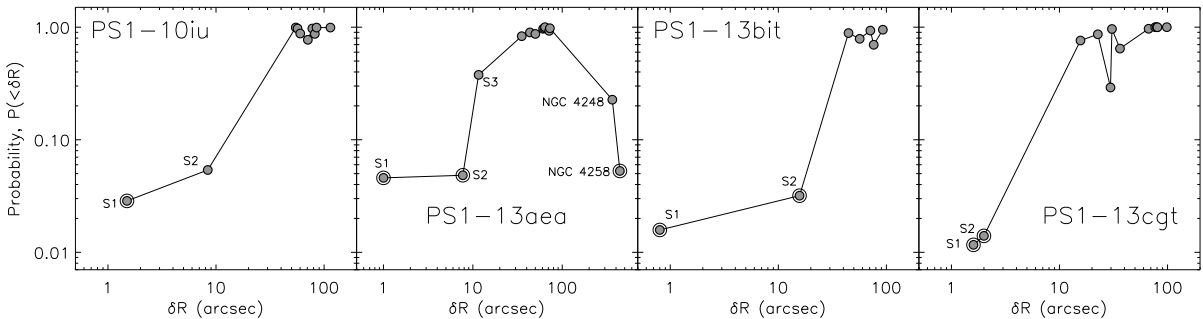}
\end{center}
\caption{Most probable host analysis for the bronze transients.  Vertical axis gives the probability of chance alignment with a given galaxy as a function of $\delta$R.  Galaxies of particular note are labeled. \label{fig:HostProb}}
\end{figure}

The region around each transient is shown in Figure~\ref{fig:SilverStamp}, and we begin by assessing the most likely host for all four objects via the method outlined in \citet{Bloom2002} and \citet{Berger2010}.  We calculate the probability of chance alignment, P($\delta$R), with various galaxies in the surrounding area (labeled in red in Figure~\ref{fig:SilverStamp}) based on their i$-$band magnitudes and $\delta$R from the explosion site.  $\delta$R for underlying versus distant galaxies are calculated as described in \citet{Bloom2002}.  The results from this analysis are shown in Figure~\ref{fig:HostProb}.

There are faint (25  mag $<$ m$_i$ $<$ 22 mag) sources located within 1$''$ of the explosion site for all four bronze objects, as well as more luminous galaxies at slightly larger separations.  In all four cases we find that the underlying, faint, source has the lowest probability of chance alignment with the transient ($\lesssim$ 4\%).  However, all four events have at least one other source with a $<$10\% probability of chance alignment.  We now examine, in turn, the consequences if each of these sources were the true host of the transient. 

The additional sources with a $<$10\% probability of chance alignment are bright (m$_i$ $>$ 19.5) and can morphologically be identified as galaxies.  Thus, if our transients are associated with these sources, they would be analogous to our gold/silver transients, with larger offsets from their hosts.  PS1-10iu S2, PS1-13aea S2, and PS1-13bit S2 possess SDSS photometric redshifts of 0.136, 0.290, and 0.134, respectively.  If these are their true hosts, the transients would possess peak absolute magnitudes between $-$16.8 and $-$19.5, well in line with the range exhibited by the gold/silver samples.  For PS1-13aea, the situation has the added complication that the transient exploded in the region between NGC 4258 and NGC 4248 (See Figure~\ref{fig:SilverStamp}), with distances of 7.2 and 7.3 Mpc, respectively.  If the transient were associated with either of these galaxies (NGC 4258 has roughly the same probability of chance alignment as PS1-13aea S2) then the peak magnitude of  PS1-13aea would only be $-$8.5, and likely an extragalactic nova.

We now turn to the possibility that the hosts of the bronze transients are the faint sources located within $\sim$ 1$''$ of the explosion site.  If this is the case, then two possibilities exist.  Either these sources are galaxies, and the bronze events represent additional rapidly-evolving and luminous transients (likely at higher redshifts), or these sources are stellar and these events are less luminous (and likely known) classes of transients.  Not one of the transients shows signs of prior or later outbursts in the four years of PS1-MDS data.  Unfortunately, the signal to noise of the detections are not sufficient to do a definitive morphological assessment of the sources.  However, for both PS1-10iu and PS1-13bit, the location of the transient (accurate to approximately 0.1$''$) appears offset from the centroid of the underlying source.  If the sources were located within the Milky Way (at distances $< $3 kpc) then the absolute magnitude of the \emph{quiescent} sources would be only $\sim$ 10 mag.  While this is in line with magnitudes expected for quiescent cataclysmic variables, several of the sources also possess red colors, which is not.  

Based on these arguments, we find it likely that at least some subset of our bronze objects represent additional members of this class of rapidly-evolving and luminous transients.

\begin{deluxetable*}{l c c c cc cccc c}
\tabletypesize{\tiny}
\tablecaption{Basic Observational Information\label{tab:basic}}
\tablewidth{0pt}
\tablehead{
\colhead{Event} &
\colhead{Disc.\ Date} &
\colhead{R.A.} &
\colhead{Dec.} &
\colhead{Redshift} &
\colhead{d$_{\mathrm{lum}}$} &
\multicolumn{4}{c}{Effective Pivot Wavelength (\AA)} &
\colhead{E(B-V)$_{\mathrm{MW}}$} \\
\cline{7-10}
& 
(UT) &  
(2000) & 
(2000) & 
& 
(Mpc) & 
\colhead{g$_\mathrm{P1}$} &
\colhead{r$_\mathrm{P1}$} &
\colhead{i$_\mathrm{P1}$} &
\colhead{z$_\mathrm{P1}$} &
}
\startdata
\multicolumn{11}{c}{Gold Sample} \\
\cline{1-11}
PS1-10ah & 2010 Feb.\ 21  & 10:48:15.784 &  +57:24:19.48& 0.074 & 330.5 & 4540.7 & 5772.9 & 7001.9 & 8068.3 & 0.008 \\
PS1-10bjp& 2010 Oct.\ 24  & 23:26:21.402 & $-$01:31:23.11& 0.113 & 518.6 &  4381.6   &   5570.6  &    6756.5   &   7785.5  & 0.049 \\
PS1-11qr & 2011 April 2  &09:56:41.767 & +01:53:38.25 & 0.324 & 1685.3 & 3683.3  &    4682.9 &     5679.8  &    6544.8 & 0.017 \\
PS1-12bb & 2012 Jan.\ 20  & 09:57:23.866 & +03:11:04.47 & 0.101 & 459.7 & 4429.3  &    5631.3  &    6830.2  &    7870.4 & 0.026\\
PS1-12bv & 2012 Jan.\ 20 & 12:25:34.602 & +46:41:26.97& 0.405 & 2192.2 &3471.0  &    4412.9 &     5352.3  &    6167.5 & 0.010\\
PS1-12brf & 2012 Oct.\ 10 & 22:16:06.892 & $-$00:58:09.81& 0.275 & 1393.6 &  3824.9  &    4862.8  &    5898.0 &     6796.3 & 0.084\\
\cline{1-11}
\multicolumn{11}{c}{Silver Sample} \\
\cline{1-11}
PS1-11bbq & 2011 Dec.\ 1  &08:42:34.733 & +42:55:49.61& 0.646 & 3857.8 &  2962.8   &  3766.8   &   4568.7  &   5264.5 & 0.026 \\
PS1-13duy & 2013 Nov.\ 1  & 22:21:47.929 & $-$00:14:34.94 & 0.270 & 1364.5 & 3839.9 & 4882.0 & 5921.9 & 6823.1  & 0.053  \\
PS1-13dwm & 2013 Nov.\ 5  & 22:20:12.081 & +00:56:22.35 & 0.245 & 1220.8 & 3917.0 & 4980.0 & 6040.8 & 5960.1& 0.054 \\
PS1-13ess & 2013 Dec.\ 26  & 02:22:09.428 & $-$03:03:00.51 & 0.296 & 1517.2 & 3762.9 & 4784.0 & 5803.1 & 6686.2 & 0.025 \\
\cline{1-11}
\multicolumn{11}{c}{Bronze Sample} \\
\cline{1-11}
PS1-10iu & 2010 June 6  & 16:11:34.886 & +55:08:47.91 & \nodata & \nodata & \nodata& \nodata& \nodata & \nodata & 0.006 \\
PS1-13aea & 2013 April 7  & 12:18:14.320 & +47:20:12.60&  \nodata & \nodata & \nodata& \nodata& \nodata & \nodata & 0.014 \\
PS1-13bit & 2013 May 12  & 16:12:00.765 & +54:16:08.16& \nodata & \nodata & \nodata& \nodata& \nodata & \nodata &  0.008 \\
PS1-13cgt & 2013 July 17  & 16:18:56.245 & +54:19:33.71 & \nodata & \nodata & \nodata& \nodata& \nodata & \nodata & 0.010 
\enddata              
\end{deluxetable*}

\begin{deluxetable*}{l l l r r r c}
\tabletypesize{\scriptsize}
\tablecaption{Optical Photometry \label{tab:Photom}}
\tablewidth{0pt}
\tablehead{
\colhead{Event} &
\colhead{filter} &
\colhead{MJD} &
\colhead{Phase\tablenotemark{a}} &
\colhead{AB mag\tablenotemark{b}} &
\colhead{Error} &
\colhead{Instrument}
 }
\startdata
PS1-10ah & g$_{\rm{P1}}$ &    55230.5 &  -19.5  &   $>$22.38  &  \nodata & PS1 \\
PS1-10ah & g$_{\rm{P1}}$ &    55233.5 &  -16.7  &   $>$22.60  &   \nodata &  PS1 \\
PS1-10ah & g$_{\rm{P1}}$ &    55236.6 &  -13.8  &   $>$22.40  &  \nodata & PS1 \\
PS1-10ah & g$_{\rm{P1}}$ &    55239.5  & -11.1  &   $>$22.75  &   \nodata &  PS1 \\
PS1-10ah & g$_{\rm{P1}}$ &    55242.5  &  -8.3   &  $>$22.70  &   \nodata &  PS1 \\
PS1-10ah & g$_{\rm{P1}}$ &    55248.5  &  -2.7   &   22.18   &   0.22 & PS1 \\
PS1-10ah & g$_{\rm{P1}}$ &    55251.4  &   0.0   &  19.95    &  0.05 &  PS1 \\
PS1-10ah & g$_{\rm{P1}}$ &    55266.5  &  14.1  &   21.64   &   0.17 & PS1 \\
PS1-10ah & g$_{\rm{P1}}$ &    55281.3  &  27.8  &   $>$22.02  &   \nodata &  PS1 \\
PS1-10ah & g$_{\rm{P1}}$ &    55293.4  &  39.1  &   $>$22.26  &   \nodata & PS1
\enddata
\tablenotetext{a}{Rest-frame days since observed $g-$band maximum.}
\tablenotetext{b}{Upper limits presented are 3$\sigma$.}
\tablecomments{Table \ref{tab:Photom} is published in its entirety in the electronic edition. A portion is shown here for guidance regarding its form and content.}
\end{deluxetable*}

\begin{deluxetable*}{l l l l c l l }
\tabletypesize{\scriptsize}
\tablecaption{Optical Spectroscopy \label{tab:Spectra}}
\tablewidth{0pt}
\tablehead{
\colhead{UT Date} &
\colhead{MJD} &
\colhead{Event} &
\colhead{Target} &
\colhead{Phase\tablenotemark{a}} &
\colhead{Telescope} &
\colhead{Instrument\tablenotemark{b}}
 }
\startdata
2010 Mar.\ 08 & 55263 & PS1-10ah & Host & \nodata & NOT & ALFOSC \\ 
2011 Apr.\ 06 &  55657 & PS1-10ah & Host & \nodata&  MMT & Hectospec\\
2011 Jun.\ 08 & 55720 & PS1-10ah & Host & \nodata & MMT & Hectospec  \\
2012 Jan.\ 17 & 55943 & PS1-10ah & Host & \nodata & MMT & Blue Channel \\
2010 Nov.\ 28 & 55528 & PS1-10bjp & Host & \nodata & MMT & Hectospec \\
2011 Sep.\ 25 & 55829 & PS1-10bjp & Host &\nodata & MMT & Hectospec \\
2011 Jun.\ 26  & 55738 & PS1-11qr & Host & \nodata & Magellan-Clay & LDSS3 \\
2011 Dec.\ 04 & 55899 & PS1-11bbq & Transient & +2 &Gemini-N & GMOS  \\ 
2012 Jan.\ 27 & 55953 & PS1-12bb & Transient & +5 & MMT & Blue Channel \\ 
2012 Feb.\ 27 & 55984 & PS1-12bb & Transient & +33 & Gemini-S &  GMOS \\ 
2012 Jan.\ 27 & 55953 & PS1-12bv & Transient & +2 & MMT & Blue Channel \\ 
2012 Oct.\ 14  & 56214 & PS1-12brf & Transient & +4 & MMT & Blue Channel \\  
2013 Nov.\ 07 & 56603 & PS1-13duy & Transient & +2 & MMT & Blue Channel \\
2013 Nov.\ 10 & 56606 & PS1-13dwm & Host & \nodata & MMT & Blue Channel \\
2013 Dec.\ 29 & 56655 & PS1-13ess & Host & \nodata & Magellan-Clay & LDSS3

\enddata
\tablenotetext{a}{Rest-frame days since observed $g-$band maximum.}
\tablenotetext{b}{ALFOSC = Andalucia Faint Object Spectrograph and Camera; Hectospec \citep{Fabricant2005}; Blue Channel \citep{Schmidt1989}; LDSS3 = Low Dispersion Survey Spectrograph-3 \citep{Allington-Smith1994}; GMOS = Gemini Multi-Object Spectrograph \citep{Hook2004} }
\end{deluxetable*}

\begin{deluxetable*}{l l ccc c cc c}
\tabletypesize{\scriptsize}
\tablecaption{Measured Photometric Properties (Gold and Silver Samples) \label{tab:PhotomProps}}
\tablewidth{0pt}
\tablehead{
\colhead{Event} &
\colhead{Band} &
\colhead{m$_{\mathrm{obs,max}}$} &
\colhead{M$_{\mathrm{obs,max}}$} &
\colhead{M$_{\mathrm{rest,max}}$} &
\colhead{t$_{\mathrm{rise}}$}&
\colhead{t$_{\mathrm{1/2,rise}}$} & 
\colhead{t$_{\mathrm{1/2,decline}}$} & 
\colhead{$\Delta m_{15}$} \\
& & \colhead{(mag)} & \colhead{(mag)}& \colhead{(mag)} & \colhead{(day)} & \colhead{(day)} & \colhead{(day)} & \colhead{(mag)} 
 }
\startdata
PS1-10ah & $g_{\mathrm{P1}}$ & 19.95 (0.05)  & -17.59 (0.11) & -17.5 & 1 $-$ 3 &  1.0 (0.1) & 6.3 (0.6) & 1.7 (0.2) \\
PS1-10ah & $r_{\mathrm{P1}}$ & 20.15 (0.04) & -17.38 (0.10) & -17.2 & 1 $-$ 3 & 1.1 (0.1) & 7.4 (0.8) & 1.4 (0.1) \\
PS1-10ah & $i_{\mathrm{P1}}$ & 20.56 (0.05) & -16.97 (0.11) & -17.0 & 1 $-$ 3 & $<$ 0.9 & 10.6 (1.0) & 1.0 (0.1) \\
\vspace{0.5mm}
PS1-10ah & $z_{\mathrm{P1}}$ & 20.40 (0.04) & -17.13 (0.11) & -16.8 & 1 $-$ 3 & $<$ 2.7 & 5.7 (0.7) & 1.2 (0.1) \\
\cline{1-9}
PS1-10bjp & $g_{\mathrm{P1}}$ & 20.29 (0.02) & -18.34 (0.11) & -18.2 & 1 $-$ 4 & 1.0 (0.1) & 7.7 (0.6) & 1.7 (0.2) \\
PS1-10bjp & $r_{\mathrm{P1}}$ & 20.45 (0.03) & -18.14 (0.11) & -18.0 & 3 $-$ 5 &  3.4 (0.1) &  6.0 (0.4) & 1.8 (0.1) \\
PS1-10bjp & $i_{\mathrm{P1}}$ & 20.46 (0.01) & -18.10 (0.11) & -17.8 & 3 $-$ 6 & 1.8 (0.1) & 10.0 (0.5) & 1.1 (0.1) \\
\vspace{0.5mm}
PS1-10bjp & $z_{\mathrm{P1}}$ & 20.75 (0.03) & -17.79 (0.11) & -17.6 & 2 $-$ 6 & 3.1 (0.1) & 11.1 (0.7) & 1.0 (0.1)  \\
\cline{1-9}
PS1-11qr & $g_{\mathrm{P1}}$ & 21.05 (0.03) & -19.84 (0.06)& -19.5 & 2 $-$ 5 & 2.8 (0.1) & 6.0 (0.5) & 2.1 (0.2)  \\
PS1-11qr & $r_{\mathrm{P1}}$ & 21.31 (0.06) & -19.56 (0.08) & -19.3 & 2 $-$ 5  & 2.9 (0.1) & 8.7 (0.4) & 1.7 (0.1)\\
PS1-11qr & $i_{\mathrm{P1}}$ & 21.39 (0.04) & -19.47 (0.07) & -19.0 & 3 $-$ 6  & 3.0 (0.2) & 5.6 (0.5) & 2.0 (0.1) \\
\vspace{0.5mm}
PS1-11qr & $z_{\mathrm{P1}}$ & 21.78 (0.17) & -19.08 (0.18) & -18.8 & 1 $-$  6 & 3.0 (0.4) & 10.3 (2.1) & 1.1 (0.3)  \\
\cline{1-9}
PS1-11bbq & $g_{\mathrm{P1}}$ & 22.53 (0.09) & -19.96 (0.10) & -19.6 & 1 $-$ 5& $<$1.7 & $<$ 14.4 & $>$ 0.7 \\
PS1-11bbq& $r_{\mathrm{P1}}$ & 22.73 (0.09) & -19.73 (0.10) & -19.4 & 1 $-$ 6 & $<$3.3 & $<$ 9.5 & $>$ 0.9 \\
PS1-11bbq & $i_{\mathrm{P1}}$ & 22.64 (0.10) & -19.80 (0.11) & -19.1 &  1 $-$ 6& $<$5.4 & $<$ 12.7 & $>$ 0.9 \\
\vspace{0.5mm}
PS1-11bbq & $z_{\mathrm{P1}}$ & 22.61 (0.20) & -19.83 (0.19) & -18.9 & 1 $-$ 4& 2.5 (0.7) & $<$ 15.0 & $>$ 0.7\\
\cline{1-9}
PS1-12bb & $g_{\mathrm{P1}}$& 21.33 (0.03) & -16.97 (0.12) & -17.0 & 1 $-$ 5 & $<$ 1.8 &  6.3 (0.3) & $\sim$ 1.8 \\
PS1-12bb & $r_{\mathrm{P1}}$ & 21.27 (0.04) & -17.00 (0.12) & -17.0 &  1 $-$ 5 & $<$ 1.7 & 8 -- 11.0 & $\sim$ 1.6 \\
PS1-12bb & $i_{\mathrm{P1}}$ & 21.12 (0.02) & -17.14 (0.12) & -16.9 & 1 $-$ 3 & $<$ 1.0 & 9.6 (0.5) & 1.0 (0.07) \\
\vspace{0.5mm}
PS1-12bb & $z_{\mathrm{P1}}$ & 21.25 (0.04) & -16.99 (0.13) & -16.8 & 1 $-$ 4 & $<$1.3 & 8.2 (0.5) & 1.8 (0.2) \\
\cline{1-9}
PS1-12bv & $g_{\mathrm{P1}}$ & 21.69 (0.05) & -19.68 (0.07) & -19.4 & 3 $-$ 7 & $<$ 3.4 & 4 -- 8 & $>$ 1.6\\
PS1-12bv & $r_{\mathrm{P1}}$ & 21.87 (0.05) & -19.49 (0.07) & -19.1 & 3 $-$ 7 & $<$ 2.2 & 3 -- 9 & $>$ 1.5 \\
PS1-12bv & $i_{\mathrm{P1}}$ & 22.07 (0.06) & -19.27 (0.08) & -18.8 & 2 $-$ 6 & $<$ 3.4 & 8.1 (0.4) & 1.6 (0.2) \\
\vspace{0.5mm}
PS1-12bv & $z_{\mathrm{P1}}$ & 22.12 (0.11) & -19.12 (0.12) & -18.6 & 3 $-$ 7  & $<$ 4.0 & 9.0 (2.0) & $>$1.3 \\
\cline{1-9}
PS1-12brf & $g_{\mathrm{P1}}$ & 21.93 (0.05) & -18.84 (0.07) & -18.4 & 1 $-$ 3 & $<$ 0.9 & 6.8 (0.5) & $>$ 1.1 \\
PS1-12brf & $r_{\mathrm{P1}}$ & 22.24 (0.06) & -18.43 (0.08) & -18.3 & 1 $-$ 3 & $<$ 1.0  & 8.8 (0.6) & $>$ 0.7 \\
PS1-12brf & $i_{\mathrm{P1}}$ & 22.11 (0.11) & -18.50 (0.12) & -18.3 & 4 $-$ 6 & 3.3 (0.2) & 10.5 (0.7) & 0.9 (0.2) \\
\vspace{0.5mm}
PS1-12brf & $z_{\mathrm{P1}}$ & 22.24 (0.22) & -18.34 (0.23) & -18.2 & 5 $-$ 7 & 5.0 (0.2) & $\sim$10.0 & $>$ 0.6 \\
\cline{1-9}
PS1-13duy & $g_{\mathrm{P1}}$ & 20.84 (0.10) & -19.75 (0.12) & -19.4 & 1 $-$ 5 & $<$ 4.0 & $<$ 16.4 & $>$ 0.7 \\
PS1-13duy & $r_{\mathrm{P1}}$ & 21.46 (0.04) & -19.08 (0.08) & -19.1 &  \nodata & \nodata & $<$ 17.3 & $>$ 0.7 \\
PS1-13duy & $i_{\mathrm{P1}}$ & 21.25 (0.05) & -19.26 (0.09) & -18.9 & 1 $-$ 3 & $<$ 2.5 & $<$ 12.9 & $>$ 0.9 \\
\vspace{0.5mm}
PS1-13duy & $z_{\mathrm{P1}}$ & 21.49 (0.10) & -19.00 (0.12) & -18.6 & 1 $-$ 4 & 1.8 (0.3) & 6.5 (1.0) & $>$ 1.3 \\
\cline{1-9}
PS1-13dwm & $g_{\mathrm{P1}}$ & 22.56 (0.12) & -17.80 (0.14)& -17.7 & 1 $-$ 4 & $<$ 3.0 & 2 $-$ 5 & $>$ 0.5 \\
PS1-13dwm & $r_{\mathrm{P1}}$ & 22.69 (0.11) & -17.63 (0.13) & -17.5 & 1 $-$ 4 & $<$ 3.0  & 3 $-$ 7 & $>$ 0.5 \\
PS1-13dwm & $i_{\mathrm{P1}}$ & 23.09 (0.16) & -17.19 (0.17) & -17.2 & 1 $-$ 6 & \nodata & \nodata & $>$ 0.5 \\
\vspace{0.5mm}
PS1-13dwm & $z_{\mathrm{P1}}$ & 22.78 (0.22) & -17.49 (0.23) & -17.0 & 1 $-$ 3 & $<$ 1.8 & 3 $-$ 6 & $>$ 0.5 \\
\cline{1-9}
PS1-13ess & $g_{\mathrm{P1}}$ & 22.02 (0.12) & -18.68 (0.14) & -18.5 & 1 $-$ 6 & $<$ 7.6 & 4.7 (0.5) & $>$ 1.0 \\
PS1-13ess & $r_{\mathrm{P1}}$ & 22.69 (0.11) & -18.43 (0.18) & -18.3 & 1 $-$ 6 & $<$  8.9 & 5.9 (0.3) & $>$ 1.2 \\
PS1-13ess & $i_{\mathrm{P1}}$ & 22.96 (0.12) & -17.70 (0.13) & -18.0 & \nodata & \nodata & \nodata & $>$ 0.7 \\
PS1-13ess & $z_{\mathrm{P1}}$ & 22.34 (0.19) & -18.30 (0.20) & -17.8 & 1 $-$ 4 & $<$ 6.1 & 10.6 (1.4) & $>$ 1.0 
\enddata
\end{deluxetable*}

\begin{deluxetable}{lc cc c cc ccc}
\tabletypesize{\scriptsize}
\tablecaption{Host Galaxy Properties \label{tab:hosts}}
\tablewidth{0pt}
\tablehead{
\colhead{Event} &
\colhead{m$_{\rm{obs,i}}$} &
\colhead{log (O/H) + 12\tablenotemark{a}} &
\colhead{Method\tablenotemark{b}} &
\colhead{log(M$_{\rm{gal}}$/M$_\odot$)}&
\colhead{SFR} &
\colhead{sSFR} &
\colhead{Offset} & 
\colhead{Offset} & 
\colhead{Norm.\ Offset\tablenotemark{c}} \\
& (mag) & \colhead{(measured)} &  & &\colhead{M$_\odot$ yr$^{-1}$} & \colhead{Gyr$^{-1}$} & \colhead{(arcsec)} & \colhead{(kpc)} &
 }
\startdata
PS1-10ah & 18.08 (0.02)& 8.49 (0.03) & PP04N2 & 9.12$^{+0.18}_{-0.21}$& $\sim$1.2 & $\sim$0.9 & 0.42 & 0.68 & 0.21 \\
PS1-10bjp & 19.22 (0.03)  & 8.38 (0.05) & PP04N2 & 9.12$^{+0.18}_{-0.24}$ & $\sim$2.1 & $\sim$1.6 & 0.95 & 2.39 & 0.85 \\
PS1-11qr & 19.80 (0.06) & 8.85 (0.08) & PP04N2 & 10.17$^{+0.20}_{-0.37}$ & $\sim$4.3 & $\sim$0.3 & 1.54 & 12.58 & 0.98 \\
PS1-11bbq & 24.40 (0.15) & 8.67 (0.14) & Z94 &  8.01$^{+0.61}_{-0.70}$ &  $>$0.3 & $>$2.4 & 0.47 & 8.81 & 1.05 \\
PS1-12bb & 16.59 (0.01) & 8.79 (0.11) & PP04N2 & 10.54$^{+0.43}_{-0.12}$ & $\sim$2.3 & $\sim$0.06  & 4.93 & 10.99 & 1.92 \\
PS1-12bv & 22.01 (0.20) & \nodata  & \nodata  & 9.89$^{+0.62}_{-0.52}$ & $>$0.3 & $>$0.03 & 0.97 & 10.29 & 0.50 \\
PS1-12brf & 22.04 (0.21) & 8.6 (0.2) & KD02 &  8.73$^{+0.20}_{-0.17}$ & $>$0.3 & $>$0.5  & 0.42 & 2.90 & 0.69 \\
PS1-13duy & 22.20 (0.04) & 8.6 (0.2) & KD02 & 8.78$^{+0.15}_{-0.12}$ & $>$0.1 & $>$0.2 & 0.33 & 2.16 & 0.44 \\
PS1-13dwm & 21.38 (0.08) & $<$ 8.4  & PP04N2  &  8.96$^{+0.27}_{-0.01}$ &  $\sim$1.9 & $\sim$2.0  & 1.40 & 8.28 & 0.82 \\
PS1-13ess & 22.54 (0.08) & 8.43 (0.03) & M91 & 8.68$^{+0.07}_{-0.13}$ &$\sim$4.5 &$\sim$9.3 & 0.14 & 1.04 & 0.32 
\enddata
\tablenotetext{a}{Metallicity measured from host galaxy spectra.  Diagnostics utilized varied between objects}
\tablenotetext{b}{Diagnostic used to obtain column 2. PP04N2 = \citet{Pettini2004}; Z94 = \citet{Zaritsky1994}; KD02 = \citet{Kewley2002}; M91 = \citet{McGaugh1991}}
\tablenotetext{c}{Offset normalized by the g$-$band half light radius of the host galaxy.}
\end{deluxetable}

\end{document}